\documentclass{aastex61}
\pdfoutput=1 
\usepackage{amsmath,amstext}
\usepackage[T1]{fontenc}
\usepackage{apjfonts} 
\usepackage[figure,figure*]{hypcap}
\usepackage{colortbl}
\usepackage{comment}

\usepackage{color}  
\usepackage[figuresright]{rotating}

\newcommand{\oiii}{[\ion{O}{3}]}
\newcommand{\oii}{[\ion{O}{2}]}

\newcommand{\esc}{{\rm erg\,s^{-1}\,cm^{-2} } }

\newcommand{\kms}{{\rm km\,s^{-1}}}

\newcommand{\iscea}{\textit{ISCEA}}

\newcommand{\wise}{{\it WISE}}

\def\la{\mathrel{\mathpalette\fun <}}
\def\ga{\mathrel{\mathpalette\fun >}}
\def\fun#1#2{\lower3.6pt\vbox{\baselineskip0pt\lineskip.9pt
        \ialign{$\mathsurround=0pt#1\hfill##\hfil$\crcr#2\crcr\sim\crcr}}}

\shorttitle{ISCEA: Infrared Satellite for Cosmic Evolution Astrophysics}
\shortauthors{Wang et al.}

\begin{document}

\title{Illuminating Galaxy Evolution at Cosmic Noon with ISCEA:\\
the Infrared Satellite for Cosmic Evolution Astrophysics}

\author{Yun Wang}\footnote{*wang@ipac.caltech.edu}
\affiliation{IPAC, California Institute of Technology, Mail Code 314-6, 1200 E. California Blvd., Pasadena, CA 91125, USA}
\author{Lee Armus}
\affiliation{IPAC, California Institute of Technology, Mail Code 314-6, 1200 E. California Blvd., Pasadena, CA 91125, USA}
\author{Andrew Benson}
\affiliation{Carnegie Observatories, 813 Santa Barbara Street, Pasadena, CA 91101}
\author{Emanuele Daddi}
\affiliation{CEA, IRFU, DAp, AIM, Universit\'e Paris-Saclay, Universit\'e de Paris,  Sorbonne Paris Cit\'e, CNRS, F-91191 Gif-sur-Yvette, France}
\author[0000-0002-9382-9832]{Andreas Faisst}
\affiliation{IPAC, California Institute of Technology, Mail Code 314-6, 1200 E. California Blvd., Pasadena, CA 91125, USA}
\author[0000-0002-0933-8601]{Anthony Gonzalez}
\affiliation{Department of Astronomy, University of Florida, Gainesville, FL 32611-2055, USA}
\author[0000-0001-7503-8482]{Casey Papovich}
\affiliation{Department of Physics and Astronomy, Texas A\&M University, College
Station, TX, 77843-4242 USA}
\affiliation{George P.\ and Cynthia Woods Mitchell Institute for
 Fundamental Physics and Astronomy, Texas A\&M University, College
 Station, TX, 77843-4242 USA}
\author{Zoran Ninkov}
\affiliation{Chester F. Carlson Center for Imaging Science,
Rochester Institute of Technology, USA}
\author[0000-0002-9573-3199]{Massimo Robberto}
\affiliation{Space Telescope Science Institute, 3700 San Martin Drive, Baltimore, MD 21218, USA}
\affiliation{Johns Hopkins University, 3400 N. Charles Street, Baltimore, MD 21218, USA}
\author{Randall J. Rose} 
\affiliation{Southwest Research Institute, 6200 Culebra Road, San Antonio, TX 78238, USA}
\author{Thomas (TJ) Rose}
\affiliation{Southwest Research Institute, 6200 Culebra Road, San Antonio, TX 78238, USA}
\affiliation{Ann and HJ Smead Department of Aerospace Engineering Sciences, University of Colorado, Boulder, Colorado, 80309, USA}
\author{Claudia Scarlata}
\affiliation{Minnesota Institute for Astrophysics, University of Minnesota, 116 Church St SE, Minneapolis, MN 55455, USA}
\author{S. A. Stanford}
\affiliation{Department of Physics and Astronomy, University of California, Davis, CA 95616, U.S.A.}
\author{Todd Veach}
\affiliation{Southwest Research Institute, Space Science and Engineering Division, 6220 Culebra Road, San Antonio, TX 78238, USA}
\author{Zhongxu Zhai}
\affiliation{Waterloo Center for Astrophysics, University of Waterloo, Waterloo, ON N2L 3G1, Canada}
\affiliation{Department of Physics and Astronomy, University of Waterloo, Waterloo, ON N2L 3G1, Canada}
\author{Bradford Benson}
\affiliation{Department of Astronomy and Astrophysics, University of Chicago, 5640 South Ellis Avenue, Chicago, IL, 60637, USA]
\affiliation{}Fermi National Accelerator Laboratory, MS209, P.O. Box 500, Batavia, IL, 60510, USA}
\author[0000-0001-7665-5079]{L.~E.~Bleem}
\affiliation{High Energy Physics Division, Argonne National Laboratory, Argonne, IL, 60439, USA}
\affiliation{Kavli Institute for Cosmological Physics, University of Chicago, Chicago, IL, 60637, USA}
\author{Michael W. Davis}
\affiliation{Southwest Research Institute, Space Science and Engineering Division, 6220 Culebra Road, San Antonio, TX 78238, USA}
\author{George Helou}
\affiliation{IPAC, California Institute of Technology, Mail Code 314-6, 1200 E. California Blvd., Pasadena, CA 91125, USA}
\author{Lynne Hillenbrand}
\affiliation{Cahill Center for Astronomy and Astrophysics, California Institute of Technology, 1200 E. California Blvd., Pasadena, CA 91125, USA}

\begin{abstract}
 
\iscea\  (Infrared Satellite for Cosmic Evolution Astrophysics) is a small astrophysics mission whose Science Goal is to discover how galaxies evolved in the cosmic web of dark matter at cosmic noon. 
The \iscea\  Science Objective is to determine the history of star formation and its quenching in galaxies as a function of local density and stellar mass when the Universe was $3-5$ Gyrs old ($1.2<z<2.1$).
\iscea\  is designed to test the following Science Hypothesis: During the period of cosmic noon, at $1.7 < z < 2.1$, environmental quenching is the dominant quenching mechanism for typical galaxies not only in clusters and groups, but also in the extended cosmic web surrounding these structures.
\iscea\  meets its Science Objective by making a 10\% shot noise measurement of star formation rate down to $6\,{\rm M_\odot\,yr^{-1}}$ using H$\alpha$ out to a radius $>$ 10 Mpc in each of 50 protocluster (cluster and cosmic web) fields at $1.2 < z < 2.1$. \iscea\  measures the star formation quenching factor in those fields, and galaxy kinematics with a precision $< 50\,\kms$  to deduce the 3D spatial distribution in each field.
\iscea\  will transform our understanding of galaxy evolution at cosmic noon.

\iscea\  is a small satellite observatory with a 700 cm$^2$ (30~cm equivalent diameter) aperture telescope with a field of view (FoV) of 0.32 deg$^2$, and a multi-object spectrograph with a digital micro-mirror device (DMD) as its programmable reflective slit mask. Using the approach pioneered by the DMD-based Infrared Multi-Object Spectrograph (IRMOS) on Kitt Peak, \iscea\  will obtain spectra of $\sim$ 1000 galaxies simultaneously at an effective resolving power of $R=1000$, with $2.8^{\prime\prime}\times 2.8^{\prime\prime}$ slits, over the near-infrared wavelength range of 1.1 to 2.0${\rm \mu m}$, a regime not accessible from the ground without large gaps in coverage and strong contamination from airglow emission. \iscea\  will achieve a pointing accuracy of $\leq 2^{\prime\prime}$ FWHM over $200\,{\rm s}$. \iscea\  will be launched as a small complete mission into a Low Earth Orbit, with a prime mission of 2.5 years. \iscea's space-qualification of DMDs opens a new window for spectroscopy from space, enabling revolutionary advances in astrophysics.

\end{abstract}

\section{Introduction}
\label{sec:intro}

Galaxies form and assemble within the cosmic web, with evolutionary histories that are determined by a combination of internal and environmental processes. Galaxy clusters emerge from the cosmic web at the intersections of filaments, and it has long been known that the properties of galaxies in these densest environments are influenced by a host of processes. These environmental processes include abrupt mechanisms like ram pressure stripping, mergers, and tidal interactions \citep[e.g.,][]{Toomre1972,Gunn1972,Mihos1994,Mihos1996,Moore1996,Hopkins2009,Haas2013,Schawinski2014,Faisst2017} as well as the slower process of strangulation and starvation \citep[e.g.,][]{Larson1980,Balogh2000,Feldmann2011,McGee2014}, with the different processes varying in importance as a function of environmental density and galaxy mass. Together these different mechanisms yield the long-observed star formation-density relation at low redshift \citep{Dressler1980,Balogh1998} in which star formation is quenched in dense environments. Subsequent work in the intervening decades has demonstrated that the star formation-density relation extends over a wide range of densities \citep[e.g.,][]{Balogh2004,Tanaka2004} and is already in place by $z\sim1$ \citep[e.g.,][]{Cooper2010}. Studies of cluster environments at low redshift have also shown that environmental factors influence galaxy properties even out well beyond the virial radius where the local densities are only moderately enhanced relative to field levels \citep[e.g.,][]{Chung2011}. Environment however is not the only factor that influences the star formation history of a galaxy. Mass-quenching, which is related to the dark matter mass of galaxies, occurs when gas falling onto galaxies is shock heated, hence prevented from cooling and forming stars. It is understood to be an important quenching mechanism for massive galaxies \citep[e.g.,][]{Croton2006,Cattaneo2008,Peng2010,Woo2013,Carollo2013,Woo2014,Schawinski2014}. 

From a theoretical perspective, the onset of this star-formation vs. density relation is expected to occur at $z\sim2$, with the precise timing being a function of mass. For example, the IllustrisTNG simulations predict that there already exists a substantial (factor of 2.5) decline in the specific star formation rates (sSFRs) for galaxies with mass $>10^{10.5}\,{\rm M_\odot}$ in clusters relative to the field, with this transition being delayed for lower mass galaxies \citep{Harshan2021}. The details of this transition, including both the timing and how it depends on local density, are however dependent on the relative importance of the various environmental mechanisms. Understanding this physics is critical to our general picture of galaxy evolution, and yet remains relatively poorly constrained -- especially during this critical epoch during which star formation in the Universe peaks and massive galaxies and nascent (proto-)clusters are rapidly assembling.

Existing observations present a complex and incomplete picture. Many studies have found evidence for a reversal of the star formation-density relation at $z>1$, with enhanced sSFRs relative to field levels in overdense environments \citep{Cooper2008,Tran2010,Brodwin2013,Alberts2016,Hatch2017}. Meanwhile, there is also evidence for molecular gas deficits in star-forming galaxies within $z\sim1.5$ galaxy clusters, indicating that environmental factors are actively depleting their gas supply, and tentative evidence for environmental quenching at $z\sim2$ based upon clustering of quiescent galaxies \citep{Ji2018}. Together, these results paint a picture of accelerated evolution in the densest environments, with both star formation and its quenching starting earlier than in lower densities. The detailed dependence of star formation and its quenching upon environment is however unclear during the $z\sim2$ epoch when the most massive galaxies are expected to feel the onset of quenching. Recently, \cite{Harshan2021} conducted a pioneering investigation of the star formation histories of galaxies in a low-mass ($M\sim5\times 10^{13}$~M$_\odot$) protocluster using spectral energy distribution (SED) fitting. They found that the most massive galaxies ($\log (M_*/{\rm M_\odot})>10.5$) showed a slight 1$\sigma$ decrement in sSFRs relative to the field, which is suggestive of the onset of quenching but at odds with the star formation reversal seen by other groups.

At a fundamental level, one of the key outstanding questions in our picture of galaxy formation is the relative importance of environmental and mass-quenching as a function of local density and redshift, and the need for data to resolve this question is most acute at $z\sim2$.
 Many of the trends identified at lower redshift should first manifest at this epoch.  This is where the cosmic star formation rate (SFR) density peaks \citep{Madau2014}, and we expect galaxy groups and clusters to be assembling rapidly (based on simulations).  By $z\sim1-1.5$, many of the most massive clusters have already formed their cores and quenched their massive galaxies.  To understand these mechanisms we must study and characterize the star-forming properties of the galaxies in overdensities at $z\sim2$.  
 Such an investigation has not been possible previously either from the ground, because of limitations for ground-based observatories, or from space because current space-based facilities do not have a sufficiently large field-of-view (FoV) to cover galaxies from the cores of protoclusters, to the filaments, to the field.   
Next generation facilities like {\it JWST} will provide high-fidelity information on the field population and can also target the very cores of clusters; however the FoV of {\it JWST} is very limited and covering the full cluster environment requires tiling together $>100$ {\it JWST} fields. On the other hand, wide area slitless spectroscopic surveys, such as those planned for {\it Euclid} and {\it Roman}, do not have the necessary spectroscopic resolution to measure accurately the velocity structure of the protoclusters. The ideal means of addressing this question is to look at a statistical sample of protocluster environments at this epoch, with sufficient spatial coverage to study lower density environments as well as dense cluster cores.

In the recent years infrared galaxy cluster searches have begun to extend to $z>1.5$, driven by improved data and selection methods \citep[e.g.][]{Papovich2008,eisenhardt2009,Muzzin2013b,noirot2018,gonzalez2019,Wen2020}. 
At the same time, the Atacama Cosmology Telescope (ACT, see \citealt{Hilton2020}) and South Pole Telescope (SPT, see \citealt{Bocquet2019}) surveys have identified a handful of clusters at $z>1.5$, providing a sample of very high mass clusters that can be used to probe the most extreme central densities.  In all cases, high-redshift clusters are expected to be rapidly growing within the larger protcluster environment.
These protoclusters span the full range of local densities from field levels to dense cluster cores, providing an opportunity to assess the impact of environmental factors on galaxy evolution at all density levels -- including in the filamentary structures surrounding the forming cluster cores in which "pre-processing" of future cluster galaxies may occur. It has also recently been argued that simulations indicate that gas accretion is less efficient within filaments at $z\sim2$ due to high angular momentum, and that this may contribute to the large fraction of passive galaxies in filaments \citep{Song2021}. The new samples of high-redshift cluster candidates are now sufficiently large that with deep studies over the entire protocluster regions statistically robust analysis of environmental quenching as a function of local density and stellar mass becomes possible.
Protoclusters are also the unambiguous formation sites of the first massive galaxies, and hence the best locations in which to constrain the elusive processes leading to the formation of passive, early-type galaxies. Within the protocluster regions, which have angular extents of tens of arcmin at $z=2$ \citep{Overzier2016}, addressing this science requires high-fidelity SFR estimates (e.g. via H$\alpha$), stellar masses, and robust local density determinations.

To find definitive answers to the outstanding questions on galaxy evolution at cosmic noon, we propose the Infrared Satellite for Cosmic Evolution Astrophysics (\iscea). \iscea\  is a space mission with a 700 cm$^2$ (27cm~$\times$~27cm square or 30cm diameter equivalent) aperture telescope in Low Earth Orbit (LEO), to obtain 1.1-2$\mu$m near-infrared (NIR) slit spectra at $R=1000$ of galaxies with SFR as low as $6\,{\rm M_\odot\,yr^{-1}}$ over a FoV 218 times that of {\it HST} and 128 times that of {\it JWST}. It has superior spectroscopic coverage and resolution compared to {\it HST}, {\it SPHEREx}, {\it Euclid} and {\it Roman} (Table~\ref{tab:missions}).

\begin{figure}
    \centering
    \includegraphics[width = 3.5in]{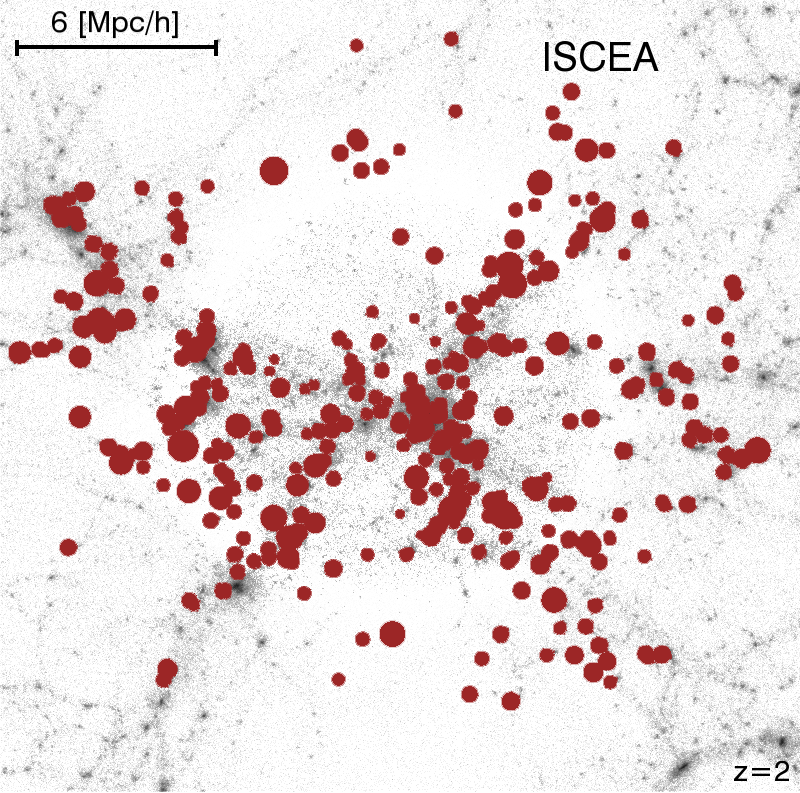}
    \includegraphics[width = 3.5in]{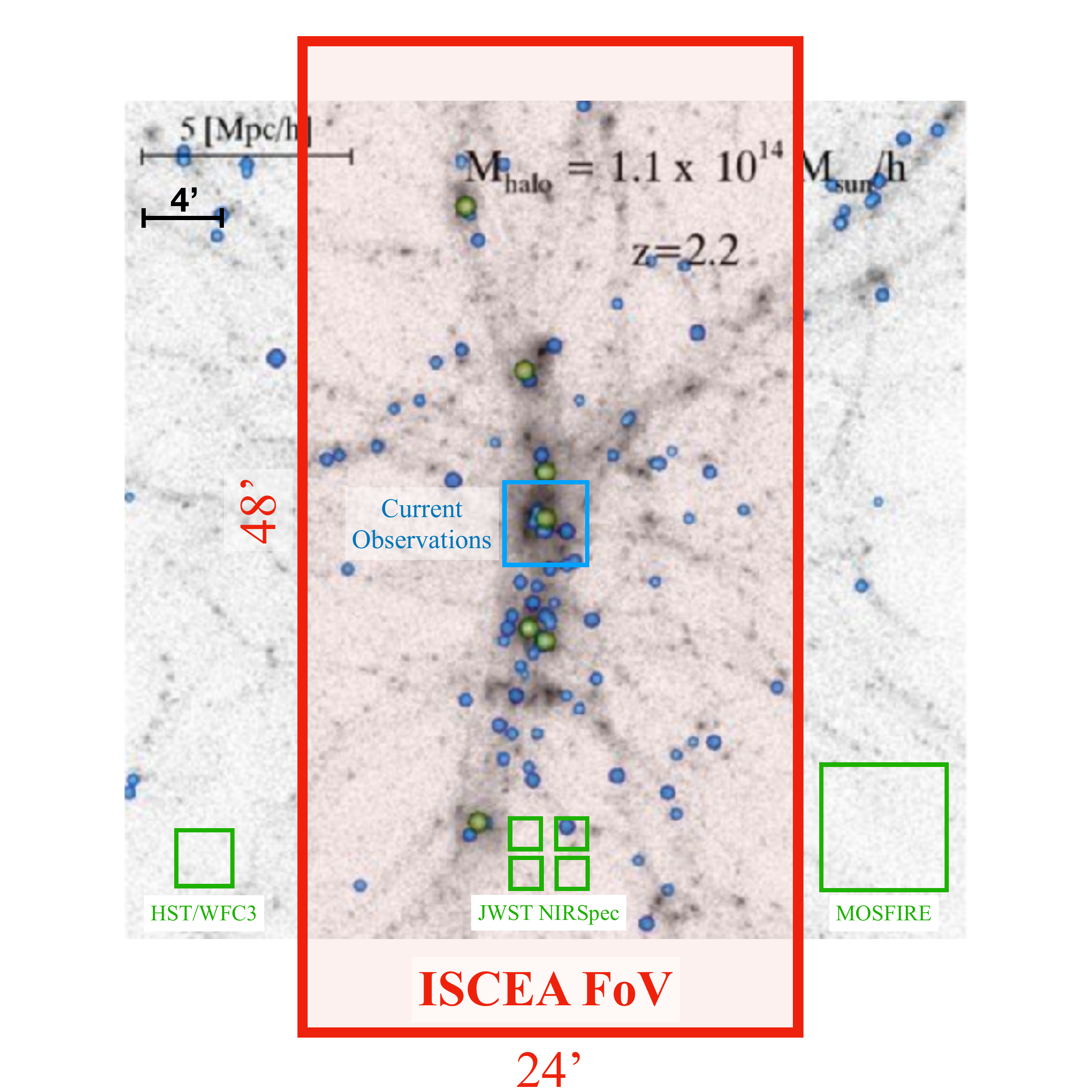}
    \caption{\textit{Left}: Simulated \iscea\  galaxies (red) tracing the cosmic web of dark matter (grayscale) in a protocluster at $z = 2$ (courtesy of Alvaro Orsi). \iscea\  will map 50 protoclusters, each containing a cluster and its cosmic web environment.
    \textit{Right}: The \iscea\  FoV (red rectangle) compared to the scope of current observations (small blue square), overlaying a simulated protocluster at $z = 2.2$ \citep{Orsi2016}. The background grayscale shows the network of dark matter filaments that comprise the protocluster. The blue and green filled circles highlight star-forming and elliptical galaxies in the protocluster, respectively. Protoclusters are very extended $-$ the cluster in each protocluster contains less than 20 per cent of all protocluster galaxies at $z \sim 2$ \citep{Muldrew2015}. 
    \iscea\  spectroscopically studies 50 protoclusters at $1.2 < z < 2.1$ using wide-field, multi-object spectroscopy to discover how environment affects galaxy formation and evolution at cosmic noon. 
    }
    \label{fig:cosmic-web}
\end{figure}

\iscea\  is optimally matched to the sizes of the high redshift protocluster regions (Fig.~\ref{fig:cosmic-web}), enabling study of the distribution of star-formation and Active Galactic Nuclei (AGN, another indicator of galaxy evolution) not only in cluster cores, but also the infalling galaxy population in filaments and associated substructures that will accrete onto the cluster at later times. We will quantify the distribution of star formation and AGN activity over entire protocluster regions surrounding collapsed cluster cores, including regions that span three orders of magnitude in local density, $\log(1+\delta)=-1$ to 2, where $\delta=(\rho-\langle \rho\rangle)/\langle \rho\rangle$.

\newcommand{\mygoodcolor}{green!25}
\newcommand{\mybadcolor}{red!25}
\begin{table}[]
    \centering
    \begin{tabular}{|c|c|c|c|c|c|c|}
    \hline
        Mission & \cellcolor{\mygoodcolor}ISCEA & HST & JWST & Euclid & SPHEREx & Roman \\
         \hline
      Slit spectroscopy   & \cellcolor{\mygoodcolor}Yes & No & \cellcolor{\mygoodcolor}Yes & No & No & No\\
      \hline
      Spectral resolution & \cellcolor{\mygoodcolor}1000 & 130 & \cellcolor{\mygoodcolor}100-2700 & 380 & 35-130 & 460\\\hline
      FoV (deg$^2$) & \cellcolor{\mygoodcolor}0.32 & 0.00147 & 0.0025 & \cellcolor{\mygoodcolor}0.53 & \cellcolor{\mygoodcolor}39.6 & \cellcolor{\mygoodcolor}0.281  \\\hline
      Depth ($\esc$) & \cellcolor{\mygoodcolor}$3\times 10^{-17}$ & $5\times 10^{-17}$ & \cellcolor{\mygoodcolor}$3.5\times 10^{-19}$ & $2\times 10^{-16}$ & N/A & $10^{-16}$ \\
      & \cellcolor{\mygoodcolor}(5$\sigma$) & (5$\sigma$) & (10$\sigma$) & (3.5$\sigma$) & & (6.5$\sigma$)\\
      Integration time per target & 668ks & 5ks & 100ks& 4.32ks&  & 2.4ks \\\hline
      Launch date & $\sim$ 2027 & 1990 & 2021 & $\sim$ 2022 & $\sim$ 2024 & 2027 \\\hline
    \end{tabular}
    \caption{\iscea\  will provide an unprecedented combination of spectral resolution R=1000, ultra depth, and wide FoV for procluster studies, unmatched by current or planned space-based IR projects. \iscea\  depth assumes the typical 668ks (including 80\% additional time for pointing jitter compensation) of observing time per protocluster field. {\it HST} depth is the 3-D {\it HST} Treasury Program 5$\sigma$ line flux limits for "typical objects" \citep{Brammer2012,VanDokkum2013,Momcheva2016}. {\it Euclid} and {\it Roman} depths correspond to their wide-area surveys. {\it SPHEREx} will be limited by its large pixel size (6$^{\prime\prime}$ pixel scale, vs. 1.4$^{\prime\prime}$ for \iscea) in crowded fields such as clusters at $z\sim2$.
    Fields shaded in green satisfy the observational requirements for the science case here.  Only \iscea\  meets all of these requirements. 
    }
    \label{tab:missions}
\end{table}

In this paper, we present an overview of the science investigation with \iscea.
The \iscea\  instrument and mission design will be presented elsewhere.
\S\ref{sec:background} discusses the \iscea\  Science Goal \& Objective.
\S\ref{sec:req} contains the detailed derivations of the \iscea\  science, instrument, and high-level mission requirements.
\S\ref{sec:baseline} describes the \iscea\  Baseline Mission.
\S\ref{sec:sim} presents the simulations of \iscea\  data.
\S\ref{sec:data_acq}, \S\ref{sec:data_ana}, \S\ref{sec:data-prod} discuss \iscea\  data acquisation, data analysis, and data products respectively.
\S\ref{sec:sys} discusses the modeling and mitigation of systematic effects that affect the \iscea\  science investigation.
We discuss how \iscea\  tests its Science Hypothesis and meet its Science Objective in \S\ref{sec:test-hypo}, and the \iscea\  secondary science program in \S\ref{sec:stars}.
\S\ref{sec:summary} contains the summary and conclusion.

\section{\iscea\  Science Goal \& Objective}
\label{sec:background}

\begin{figure}
    \centering
    \includegraphics[width = 3.5in]{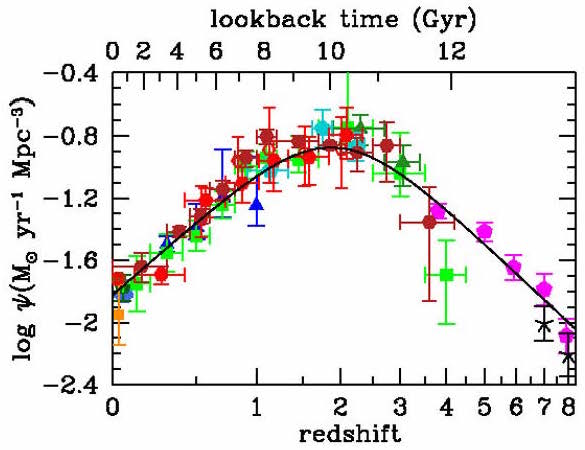}
    \caption{The volume density of cosmic star formation rate (SFR), the key indicator for galaxy evolution, as a function of redshift $z$ and lookback time \citep{Madau2014}. Colors denote data sets from different observational data sets. \iscea\  will explore the peak epoch in galaxy evolution, which occurred during the peak in the cosmic SFR density ($1.2<z<2.1$). }
    \label{fig:SFR}
\end{figure}

{\bf The \underline{Science Goal} of \iscea\  is to discover how galaxies formed and evolved within the cosmic web of dark matter at cosmic noon}, in particular, the extent to which local density is destiny for these galaxies. The redshift regime $z\sim2$ is not only the peak era of star formation in the Universe (see Fig.\ref{fig:SFR}), but also an epoch at which galaxies and galaxy groups are undergoing rapid assembly and the influence of environment is beginning to be keenly felt in the most overdense regions. In cluster and group mass halos various processes such as strangulation, starvation, ram pressure stripping, and galaxy interactions are expected to truncate accretion and quench satellite galaxies, with some of these processes also acting within the filaments of the cosmic web. 

To meet its Science Goal, {\bf the \iscea\  \underline{Science Objective} is to determine the history of star formation and its quenching in galaxies as a function of local density and stellar mass when the Universe was $3-5$ Gyrs old ($1.2<z<2.1$)}, in order to understand the importance of environmental influences versus mass quenching in establishing the galaxy populations we see today. 

The \iscea\  mission is designed through the flow-down of the science objective to the science requirements to the technical requirements, and implemented on a small complete mission that meets all its requirements. \emph{\iscea\  advances NASA astrophysics science objectives, and addresses the Astro2020 key science question "How do the histories of galaxies and their dark matter halos shape their observable properties?", in the decadal survey priority area "Unveiling the Drivers of Galaxy Growth" \citep{Astro2020}}.
During the 2 months of each year when the \iscea\  protocluster fields are not accessible due to orbital constraints, we will carry out a secondary science program to discover and study low mass stars and brown dwarfs in young stellar clusters (see \S\ref{sec:stars}). 

To meet its Science Goal and Objective, \iscea\  is designed to test the following {\bf \underline{Science Hypothesis}: During the period of cosmic noon, at $1.7 < z < 2.1$, 
environmental quenching is the dominant quenching mechanism for typical galaxies not only in clusters and groups, but also in the extended cosmic web surrounding these structures.} 
We discuss the \iscea\  requirement flow-down in detail in the following section. 

\begin{figure}
    \centering
    \includegraphics[width = 7in]{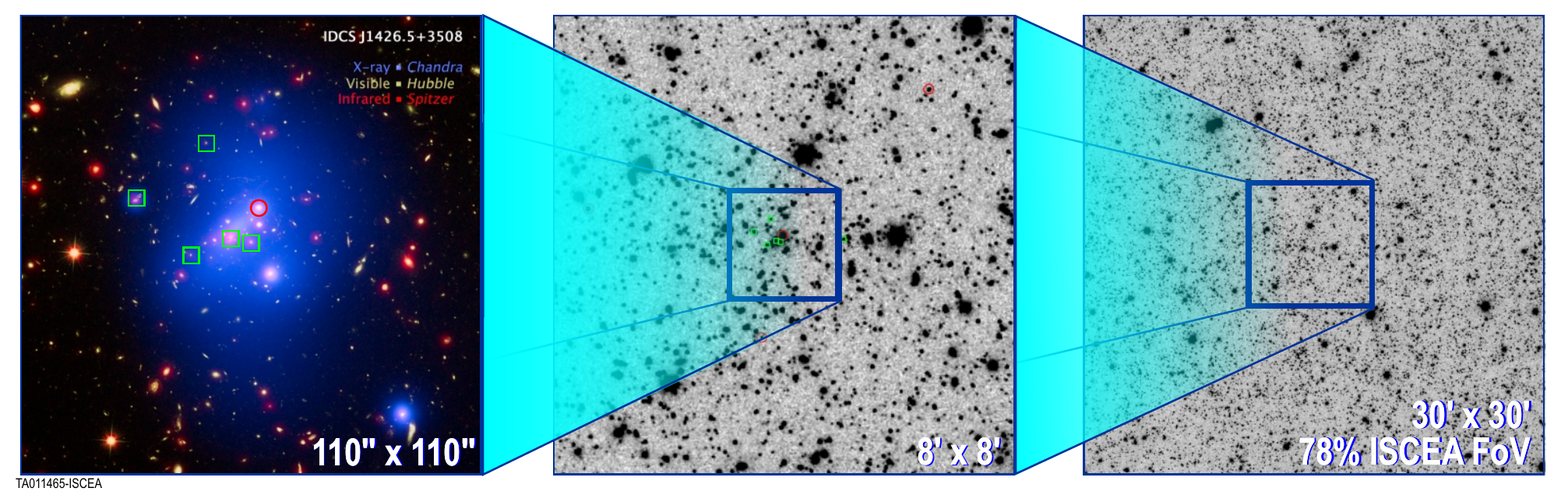}
    \caption{\iscea\  increases the scale and resolution of 3D galaxy maps of clusters by $>10\times$ current observations. \textit{Left}: A Great Observatories image of the $1.9^\prime\times 1.9^\prime$ core of $z = 1.75$ cluster IDCS J1426.5+3508, with the few galaxies with spectroscopic redshifts indicated with green squares (from HST/WFC3 grism) and red circles (from \textit{Keck} LRIS \& MOSFIRE). \textit{Middle}: Image zoom-out illustrating the limited spectroscopic sampling currently possible (with the same green squares and red circles as in the left panel). \textit{Right}: Spitzer IRAC $3.6\,{\rm \mu m}$ image, showing 78\% of the large \iscea\  FoV.  \citep{Stanford2012,Brodwin2012,Brodwin2016,Gonzalez2012,Mo2016} }
    \label{fig:cluster_zoom-out}
\end{figure}

\section{\iscea\  Requirements}
\label{sec:req}

\subsection{\iscea\  Science Requirements}
\label{sec:science_req}

\iscea's  Science Objective, “Determine the history of star formation and its quenching in galaxies as a function of local density and stellar mass when the Universe was 3-5 Gyrs old ($1.2~<~z~<~2.1$)”, flows down to four {\bf \iscea\  Science Requirements}:
\begin{itemize}
    \item[(1)] Measure Star Formation Rate (SFR) using H$\alpha$ emission lines at $S/N \geq 5$ for $\geq 100$ galaxies out to a radius of 10 Mpc in each of 50 protocluster fields at $1.2~<~z~< 2.1$;
    \item[(2)] Measure Star Formation Quenching Factor (SFQF) out to a radius of 10 Mpc in each of 50 protocluster fields at $1.2~<~z~<~2.1$;
    \item[(3)] Measure 3D galaxy distribution out to a radius of 10 Mpc in each of 50 protocluster fields at $1.2~<~z~< 2.1$;
    \item[(4)] Measure radial velocity with $\Delta v~<~50\,\kms$ for each target galaxy.
\end{itemize}

Galaxy evolution peaks at $1.7 < z < 2.1$ (Fig.~\ref{fig:SFR}). 
To meet the \iscea\  science objective, we require 45 bins (5 bins in stellar mass $M_*$, times 9 bins in local mass density $\delta$), similar to the low redshift zCOSMOS 20k study at $0.4<z<0.7$ \citep{Kovac2014}, with the same binsize in $M_*$ of 0.25 dex, and $\sim$3 bins in $\delta$ per order of magnitude. \iscea\  probes the key range of  $M_*=10^{10}-10^{11.25} {\rm M_\odot}$, resulting in 5 bins of 0.25 dex.  \iscea\  covers three orders of magnitude in $\log(1+ \delta)$, $-1$ to 2, resulting in 9 bins with three bins per order of magnitude. 
In each bin, we require 100 galaxies to measure SFR with 10\% shot noise (sufficient statistical precision) since the shot noise is $1/\sqrt{N}$, where $N$ is the number of objects used in the measurement. \iscea\  measures the H$\alpha$ line flux and its intrinsic scatter to derive the SFR using a detailed physical model. At \iscea’s H$\alpha$ line flux limit of $3\times 10^{-17}\esc$ (chosen to optimize the study of environmental effects on galaxy evolution, see \S\ref{sec:instru_req}), the average number of galaxies is $\sim 210$ inside a protocluster region observed by \iscea, based on the \iscea\  mock (see \S\ref{sec:count-gal}). Assuming an overall observing efficiency of 48\% (80\% per field, see \S\ref{sec:efficiency}, and a factor of $\sim$1.7 margin to allow for instrument and spacecraft constraints), 45 protoclusters at $1.7<z<2.1$ are required to yield 4500 galaxies. \iscea\  will also observe five protoclusters at $1.2<z<1.7$, to calibrate and tie into the existing extensive data on clusters at $z<1.7$. 

The \iscea\  target list includes the five highest redshift Sunyaev-Zel’dovich (SZ) clusters at $z>1.5$ observed by SPT \citep{Bocquet2019} and ACT \citep{Hilton2020}. The SZ signal \citep{Sunyaev1972} results from the scattering of the Cosmic Microwave Background photons off the free electrons in the hot ionized gas that permeates each cluster. 
SZ clusters are generally more massive than the clusters selected using photometry \citep{Wen2020}, and provide unique insight into the physical mechanism for star formation quenching in massive clusters at cosmic noon (see \S\ref{sec:intro}).

\begin{figure}
    \centering
    \includegraphics[width = 3.2in]{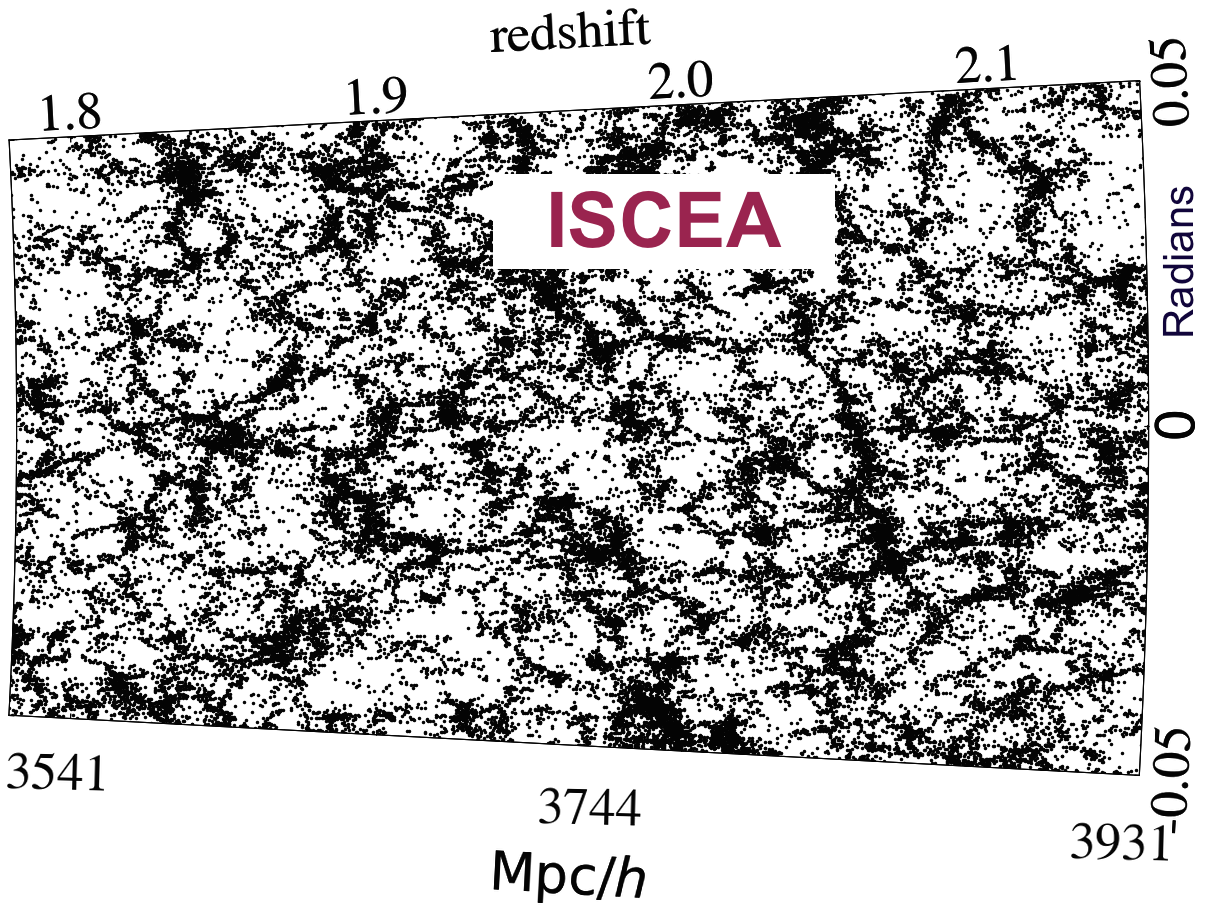}\hskip 0.1in
        \includegraphics[width = 3.2in]{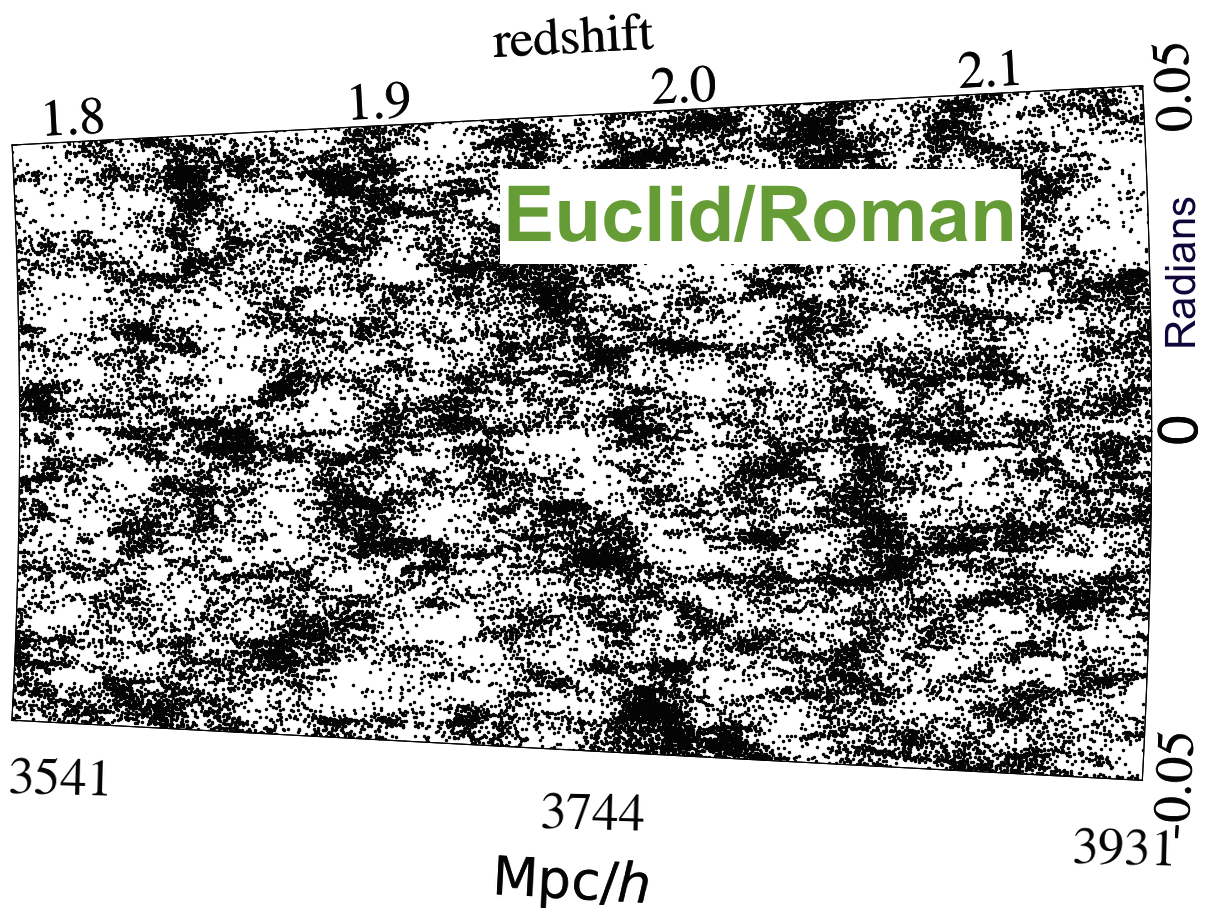}
    \caption{\iscea\  maps the cosmic web with $R = 1000$ slit spectroscopy (left), which is beyond the capabilities of the future missions \textit{Euclid} and \textit{Roman}, both with slitless spectroscopy only (right). See also Table~\ref{tab:missions} for a comparison of different current and future missions. (Courtesy of Alvaro Orsi)
    }
    \label{fig:ISCEA-z}
\end{figure}

Galaxy velocity measurements with $\Delta v \sim 50\,\kms$  will enable us to definitively associate any given galaxy with a cluster, group or filament, and assess whether any observed groups and filamentary structures are truly associated with the protocluster, or lie outside the turnaround region for the protocluster. This requires a redshift precision only available from R $\sim$ 1000 slit spectroscopy (Fig.~\ref{fig:ISCEA-z}).

The cosmic web out to a radius 10 Mpc around a cluster contains all relevant information regarding galaxy evolution as a function of local density. This requires a FoV $> 24^\prime \times 40^\prime$ centered on each cluster, to reach the radius of 10 Mpc, while providing sufficient statistics ($\geq 100$ galaxies per protocluster) to meet the science requirements (see Fig.~\ref{fig:cosmic-web}).
To meet all Science Requirements, \iscea\  will observe 50 protocluster fields at $1.2<z<2.1$ (including 90\% at $z>1.7$) to measure galaxy spectra with $S/N \geq 5$ (on the strongest emission line or absorption feature) at $R = 1000$ (effective) using slit spectroscopy for $\geq 100$ galaxies on average inside each protocluster.

\subsection{\iscea\  Instrument Requirements}
\label{sec:instru_req}

\iscea\  Science Requirements flow down to {\bf \iscea\  Instrument Requirements}:

{\bf Wavelength Range}: $1.1-2.0\,{\rm \mu m}$, to enable the detection of the two strongest emission lines per galaxy, including H$\alpha$ as tracer of star formation activity over $1.2 < z < 2.1$. H$\alpha$ emission line flux is considered the ideal tracer of SFR \citep{Kennicutt1998}. The minimum wavelength is set by requiring the observation of \oiii~ emission line at $z\geq 1.2$.
 
{\bf Spectral Resolving Power}: $R = \lambda/\Delta\lambda \geq 1000$ with slit spectroscopy, to measure galaxy velocities. This is required to map the cosmic web in sufficient details to study environmental effects on galaxy evolution (see Fig.\ref{fig:ISCEA-z}).
 
\begin{figure}
    \centering
    \includegraphics[width = 4in]{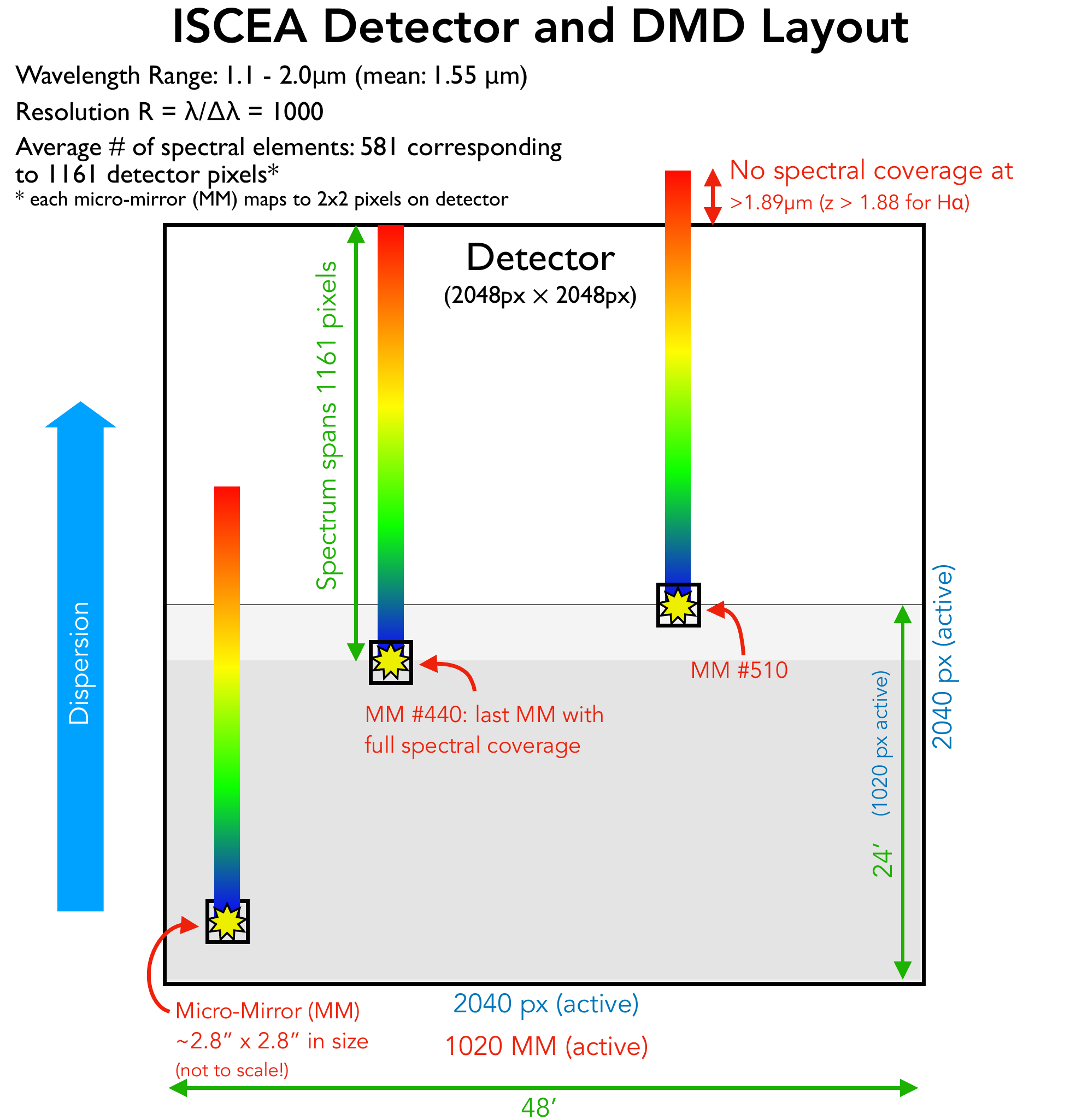}
    \caption{Illustration of how \iscea\  will select galaxy targets for multi-slit spectroscopy by using 1020$\times$510 micro-mirrors (MM) on the Texas Instrument DMD DLP7000 (TRL 6) with 1024$\times$768 MM. The top 258 rows of unused MM are not shown. 
    Only one MM per column is used to select a target object, removing the problem of overlapping spectra by design. With this conservative choice, \iscea\ can obtain $\sim$ 1000 non-overlapping spectra simultaneously. 
    }
    \label{fig:DMD_layout}
\end{figure}

{\bf Spectroscopic Multiplex Factor} (number of spectra obtained simultaneously): $\geq$ 300, to observe all cluster galaxy and cosmic web galaxy candidates in a single observation. 
Fig.~\ref{fig:DMD_layout} shows that by using a DMD as the spectroscopic target selector, \iscea\  can obtain $\sim$ 1000 non-overlapping spectra simultaneously.

{\bf Flux Limit}: $3\times 10^{-17}\esc$. 
\iscea’s flux limit is set by the SFR measurement requirement. The SFR vs. galaxy position (with respect to other nearby galaxies and to the protocluster as a whole) can tell us how SFR tracks with density and radius, which allows us to predict how a galaxy might evolve, i.e., if it will end up as a spiral or elliptical. \iscea\  requires SFR measurements to the limit of $6\,{\rm M_\odot\,yr^{-1}}$, corresponding to a stellar mass of $\la 10^{10}\,{\rm M_\odot}$ on the star-forming main sequence at $1.2<z<2.1$ \citep[see, e.g.,][]{Daddi2007,Koyama2013,Shivaei2015,Valentino2017}, optimal for studying environmental effects on galaxy evolution (see Fig.~\ref{fig:sfrDistribution}).  A SFR of $6\,{\rm M_\odot\,yr^{-1}}$ roughly translates to the H$\alpha$ emission line flux limit of $3\times 10^{-17}\esc$ \citep{Kennicutt1998}, \iscea’s 5$\sigma$ line flux limit. The \iscea\  3$\sigma$ line flux limit of $1.8\times 10^{-17}\esc$ enables definition of quiescent galaxies as galaxies with no detectable emission lines at this limit.

{\bf Spectroscopic "Slit” Size (DMD pixel scale)}: $2.8^{\prime\prime}\times 2.8^{\prime\prime}$, to ensure that each “slit” (i.e., DMD micro-mirror) is small enough to select $\sim$1 galaxy and minimize its sky background, but large enough to capture most of the light from the galaxy in the presence of pointing jitter. 
 
{\bf FoV}: $\geq \sim 24^\prime \times 40^\prime \sim 0.27$ deg$^2$, centered on a cluster, to reach the radius of 10Mpc, while providing sufficient statistics ($\geq 100$ galaxies per protocluster field) to meet the science requirements, see Fig.~\ref{fig:cosmic-web} (right panel).

\subsection{\iscea\  Mission Functional Requirements}
\label{sec:mission_req}

\iscea\  Instrument Requirements flow down to the high-level {\bf \iscea\  Mission Functional Requirements} as follows.

{\bf Telescope Aperture:}  $\ge 700$ cm$^2$ (27cm $\times$ 27cm square), in order to meet the flux limit and the signal-to-noise ratio ($S/N$) requirements for \iscea\  in 2.5 years, based on our exposure time estimates. The \iscea\  exposure time calculator (ETC) assumes a telescope aperture area of 700 cm$^2$, $R=1000$, total system throughput 0.2 (expected instrument performance is 0.21), effective wavelength 1.55$\mu$m (mean of $1.1-2\mu$m), thermal noise of 0.011 phot/s/pix (met at $T_{\rm optics}=181$K, expected $T_{\rm optics} = 177$K), dark current of 0.01 $e^-$/s/pix (met at $T_{\rm detector} = 100$K; dark current is 0.002 e/s/pix at the expected $T_{\rm detector} = 90$K, see \citealt{Blank2002}), effective read noise 5$e^-$ via FOWLER8 sampling (total read noise is 10$e^-$ for 2$\times$2 pixels of extraction area per spectral resolution element), $F_{\rm sky} = 0.14\,{\rm MJy\,sr^{-1}}$, and slit size $2.8^{\prime\prime}\times 2.8^{\prime\prime}$. We divide long science observations into 200s exposures to reject cosmic rays. 
\iscea\  reaches the 5$\sigma$ H$\alpha$ line flux depth of $3\times 10^{-17}\esc$ in 668ks including additional time of 80\% to compensate for pointing jitter of $\leq 2^{\prime\prime}$ full-width-at-half-maximum (FWHM) over 200 sec (see also \S\ref{sec:jitter}). Table \ref{tab:obs-time} shows that \iscea\  meets its Science Requirements with significant resiliency.
\begin{table}
\begin{center}
\begin{tabular}{|l|l|l|}\hline
 Scenario &	Parameters &	Observing Time \\\hline
Reference & $T_{\rm optics}= 181$K, $T_{\rm detector}=100$K, no pointing jitter &	371ks  \\\hline
{\bf Requirement} & {\bf $T_{\rm optics}= 181$K, $T_{\rm detector}=100$K, 2$^{\prime\prime}$ pointing FWHM over 200s} &	 {\bf 371ks$\times$1.8=668ks}   \\\hline
Variation 1 & $T_{\rm optics}= 181$K, $T_{\rm detector}=90$K, 2$^{\prime\prime}$ pointing FWHM over 200s &	 310ks$\times$1.8=558ks   \\\hline
Variation 2 & $T_{\rm optics}= 181$K, $T_{\rm detector}=90$K, 2.2$^{\prime\prime}$ pointing FWHM over 200s &	 310ks$\times$2.1=651ks  \\\hline
Variation 3 & $T_{\rm optics}= 185$K, $T_{\rm detector}=90$K, 2$^{\prime\prime}$ pointing FWHM over 200s &	 430ks$\times$1.8=774ks  \\\hline
Variation 4 & $T_{\rm optics}= 185$K, $T_{\rm detector}=90$K, 2.2$^{\prime\prime}$ pointing FWHM over 200s &	 430ks$\times$2.1=903ks  \\\hline
\end{tabular}
\caption{\iscea\  meets science requirements with significant resiliency to pointing performance and thermal control.
The available observing time per target field is assumed to be 1Ms (maximum time per target is 1.2Ms).}
\end{center}
\label{tab:obs-time}
\end{table}
 
{\bf Observing Strategy:} Each protocluster will be observed multiple times, This enables us to access the faintest galaxies repeatedly, maximize the number of galaxies targeted, and resolve any blended spectra of galaxies occasionally selected by the same slit (i.e., DMD micro-mirror), using the slight offset in roll angles from different visits. Since emission line galaxies (ELGs) have a distribution in line flux, \iscea\  will observe 50 protoclusters in sequence multiple times, with priority given to those containing highest redshift clusters and SZ clusters at $z > 1.5$. The galaxies with $S/N \geq 5$ (on the strongest emission line or absorption feature) in their spectra will be removed from the target list and replaced with new targets in each protocluster field. The faintest cluster ELGs and most elliptical galaxies (characterized by absorption features) will remain on the target list for all \iscea\  visits to a cluster.
 
{\bf Launch Window:} No constraints.
 
{\bf Mission Life:}  2.5 years, to meet science requirements, based on exposure time calculations (Table 2).
 
{\bf Extended Mission Life:}  1-2 years, to enable the Guest Observer (GO) program.
 
\begin{figure}
    \centering
    \includegraphics[width = 6in]{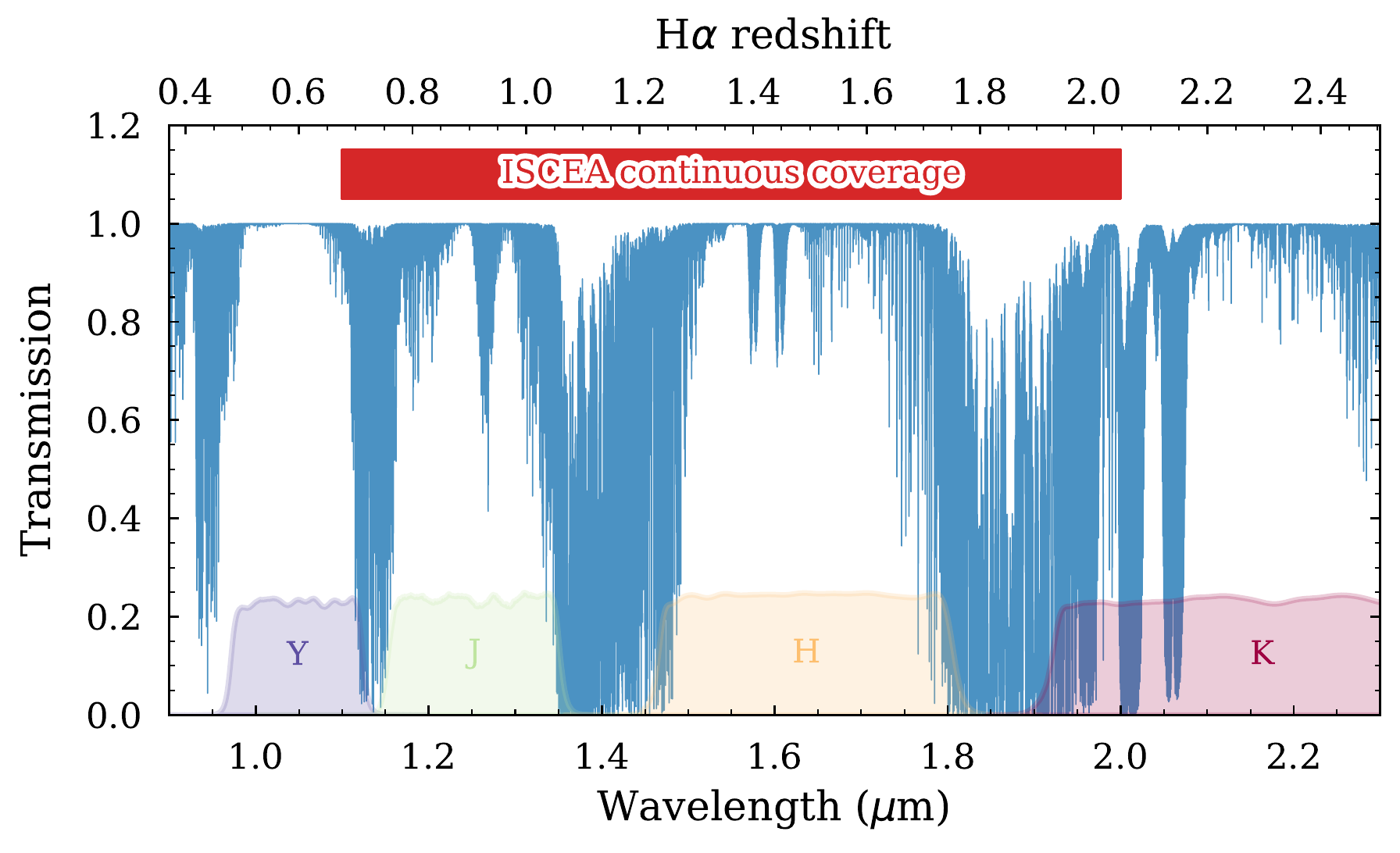}
    \caption{The atmospheric transmission at Mauna Kea for a typical water vapor column of 1.6mm and an airmass of 1 (see \url{http://www.gemini.edu/observing/telescopes-and-sites/sites}). 
    The \iscea\ wavelength range cannot be accessed from the ground without large gaps in redshift and thus galaxy evolution (see Table~\ref{tab:gaps}).}
    \label{fig:air-trans}
\end{figure}

{\bf Observatory Orbit:}  Sun Synchronous LEO.  A space platform is required to explore the peak epoch of galaxy evolution robustly and without gaps. The pervasive presence of strong and highly variable hydroxyl (OH) lines in the Earth’s atmosphere, and the substantial opacity gaps (especially at $\sim 1.4\mu$m and $\sim 1.9\mu$m) limit the possibility of obtaining homogeneous high $S/N$ spectra of faint sources in the NIR from the ground. Fig.\ref{fig:air-trans} shows the atmospheric transmission at Mauna Kea, representative of the best observing conditions from the ground. The gaps in atmospheric transmission correspond to redshift ranges for H$\alpha$, H$\beta$, \oiii~and \oii~(key spectroscopic features for galaxies) inaccessible from the ground, see Table \ref{tab:gaps}. These introduce large gaps in the galaxy evolution history during the critical epoch of mass assembly in protoclusters. Ground-based facilities, even those with NIR coverage, e.g. MOSFIRE at Keck on Mauna Kea\footnote{\url{https://www2.keck.hawaii.edu/inst/mosfire/home.html}}, and ESO’s MOONS in Chile\footnote{\url{https://www.eso.org/sci/facilities/develop/instruments/MOONS.html}}, are inadequate for meeting the \iscea\  Science Objective. \iscea\  requires the continuous coverage of $1.1-2\,{\rm \mu m}$ for any target in the sky throughout the year over a wide FoV to track SFR with H$\alpha$. It will detect H$\alpha$ at $z=0.7-2.1$, H$\beta$ at $z=1.3-3.1$,  \oiii$\lambda$5007 at $z = 1.2-3.0$, and \oii$\lambda$3727 at $z = 2.0-4.4$. 
{\bf \iscea\  will detect H$\alpha$ and \oiii~ for any galaxy with $1.2 < z < 2.1$ to robustly determine redshift and SFR.} \iscea\  will also detect both \oiii~and \oii~emission lines at $2<z<3$, with discovery potential for protoclusters up to $z=3$.
 
\begin{table}
\begin{center}
\begin{tabular}{|l|l|l|}
\hline
Redshift Gap &	$1.3-1.5\,{\rm \mu m}$ & $1.8-2\,{\rm \mu m}$ \\\hline
$z$(H$\alpha$) &	1.0-1.3	 & \cellcolor{\mygoodcolor}{1.7-2.1} \\\hline
$z$(H$\beta$) &	\cellcolor{\mygoodcolor}{1.7-2.1}	& 2.7-3.1 \\\hline
$z$(\oiii)	& \cellcolor{\mygoodcolor}{1.6-2.0} &	2.6-3.0 \\\hline
$z$(\oii) &	2.5-3.0	& 3.8-4.4 \\\hline
\end{tabular}
\caption{Redshift gaps corresponding to atmospheric gaps in Fig.~\ref{fig:air-trans}. \iscea\  observes star-forming galaxies at cosmic noon ($1.7<z<2.1$) not accessible from the ground.}
\label{tab:gaps}
\end{center}
\end{table}

{\bf Observatory will accommodate a spectrograph and an imaging channel:} needed for \iscea\  science data acquisition via spectroscopy, and calibration, target selection and verification via imaging.
 
\begin{figure}
    \centering
    \includegraphics[width = 7in]{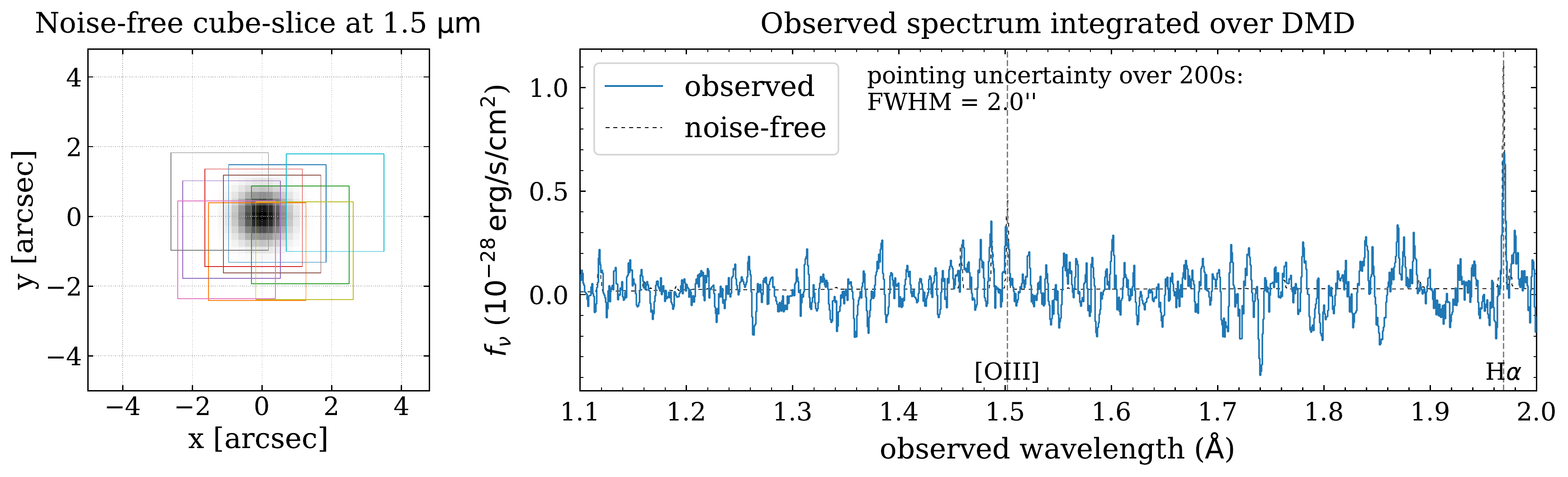}
\caption{\textit{Left}: Each box shows a micro-mirror position for $10$ randomly picked jitters assuming a pointing uncertainty with a FWHM of 2$^{\prime\prime}$ during a $200\,{\rm s}$ exposure. The background shows a model sources at $1.5\,{\rm \mu m}$ convolved with the optical PSF. See \S\ref{sec:jitter} for more details.
\textit{Right}: Resulting simulated \iscea\  spectrum with pointing uncertainty (``jitter'') FWHM of 2$^{\prime\prime}$ for an ELG with H$_{AB}=25$, and H$\alpha$ line flux of $3\times 10^{-17}\esc$ at the $5\sigma$ detection limit. Pointing jitter leads to random positions of the DMD micro-mirrors during the 200s exposure time and the observed spectrum results from adding contributions from all exposures. 
}
\label{fig:jittered}
\end{figure}

{\bf Observatory will have a pointing jitter $\leq 2^{\prime\prime}$ full-width-at-half-maximum (FWHM) over 200 sec:} 
The observatory must have sufficient pointing precision to differentiate individual galaxies in order to measure their spectra. The pointing jitter of $2^{\prime\prime}$ FWHM is well matched to the slit size (i.e., DMD micro-mirror scale) of $2.8^{\prime\prime}\times 2.8^{\prime\prime}$, without being overly stringent and becoming a mission cost driver.
The pointing jitter has two effects on the observed spectrum of a galaxy: (1) a reduction in the effective spectral resolution $R$ (“spectral blurring”), due to the galaxy’s position being jittered in the dispersion direction, and (2) a reduction of $S/N$ (“aperture effect”), due to the decreased coverage of the galaxy when the DMD micro-mirror misses the galaxy (see illustration from our simulation in Fig.~\ref{fig:jittered}). Our simulations (see also \S\ref{sec:jitter}) show that a pointing jitter of $2^{\prime\prime}$ FWHM leads to $<10$\% degradation in $R$ from $R=1000$, which requires that the \iscea\  spectrograph is designed with $R=1100$ to compensate this, and a loss in $S/N$ of $<15$\%, which needs to be compensated by increasing the observing time on the faintest galaxies.  
\S\ref{sec:jitter} contains a detailed discussion of the DMD micro-mirror "slit" loss and pointingjitter modeling.
 
\section{\iscea\  Baseline Investigation}
\label{sec:baseline}

The \iscea\  Baseline Mission is 2.5 years in duration, and observes 50 protoclusters in sequence. These are selected from 100 protocluster candidates with imaging and quick spectroscopic observations. In each protocluster field, we target $>$ 1000 galaxies with photometric redshifts (photo-$z$'s) closest to the confirmed BCG (Bright Cluster Galaxy) spectroscopic redshift. 
Each protocluster field will be observed for a total of 668ks (including the average additional time of 80\% to compensate for $S/N$ loss due to pointing jitter, see \S\ref{sec:jitter}), to reach the \iscea\  flux limit of $3\times 10^{-17}\esc$ at 5$\sigma$ (see Table \ref{tab:obs-time}). Each observation is divided into multiple visits, each with a slightly different roll angle to optimize access to targets and resolve the occasional blended spectra of galaxies selected by the same slit (i.e., DMD micro-mirror). ELGs with $S/N \geq 5$ will be replaced with new galaxies for the next protocluster visit ($>$ 2 weeks later). The faintest galaxies will remain on the target list for all visits to the field, reaching a line flux limit of $3\times 10^{-17}\esc$ at 5$\sigma$. At least $\sim$1000 galaxy spectra will be obtained per cluster field. Our Monte Carlo simulations indicate that \iscea\  science requirements (see \S\ref{sec:req}) are met with $>30$\% margin (see \S\ref{sec:efficiency}).

{\bf \iscea\  meets science requirements with significant resiliency to pointing performance and thermal control (see Table \ref{tab:obs-time}), since the available observing time per protocluster is at least 1Ms (with a maximum of 1.2Ms).}
{\bf If the \iscea\  instrument and mission requirements are fully met, i.e., $T_{\rm optics} \leq 181$K and pointing jitter $\leq 2^{\prime\prime}$ FWHM over 200s, these margins will be used to enhance \iscea\  science}, as 668ks of observing time per target is sufficient to reach the H$\alpha$ line flux limit of $3\times 10^{-17}\esc$ at 5$\sigma$.
Since (1000ks~$-$~668ks)$\times$50~=~16.6Ms~=~24$\times$668ks~+~568ks, {\bf \iscea\  can use these margins to observe 24 additional confirmed protoclusters at $1.7<z<2.1$ to the H$\alpha$ line flux limit of $3\times 10^{-17}\esc$ at 5$\sigma$ (for a total of 69 at $z>1.7$), using 568ks for additional target selection if needed. }
Alternatively, \iscea\  can observe the same 50 protoclusters to the fainter H$\alpha$ line flux limit of $\sim 2.4\times 10^{-17}\esc$ (corresponding to the SFR of $\sim 4.8\,{\rm M_\odot\,yr^{-1}}$). We will do a detailed trade study during Phase A comparing the scientific gains from these two options.  

The Baseline Mission design is supported by the following sections as follows:
\begin{itemize}
    \item{\S\ref{sec:target_sel} describes the \iscea\  protocluster target selection, and \S\ref{sec:sel-gal} the \iscea\ galaxy target selection in each protocluster, respectively.} 
\item{\S\ref{sec:mock} present the \iscea\ mock.}
\item{\S\ref{sec:photo-z} discusses the validation of the galaxy photo-$z$'s. }
\item{\S\ref{sec:count-gal} predicts the number of \iscea\ galaxies per protocluster field using the \iscea\ mock.}
\item{\S\ref{sec:jitter} models the effects due to pointing jitter.}
\item{\S\ref{sec:vpec} discusses \iscea\ velocity measurements.}
\item{\S\ref{sec:efficiency} presents details of the derivation of the \iscea\  observing efficiency.}
\item{\S\ref{sec:data_acq} describes \iscea\  data acquisition.}
\item{\S\ref{sec:data_ana} describes how the \iscea\  Baseline Mission is supported by the \iscea\  Science Operations Center (SOC).}
\item{\S\ref{sec:data-prod} describes the expected data products.}
\item{\S\ref{sec:sys} discusses systematic effects relevant to the \iscea\ science investigation and their mitigation.}
\item{ \S\ref{sec:test-hypo} shows how the \iscea\  Baseline Mission tests the \iscea\  Science Hypothesis to meet the \iscea\  Science Goal and Objective.}
\item{\S\ref{sec:stars} describe the \iscea\ secondary science program to make full use of the two winter months each year when the protocluster targets have limited visibility due to spacecraft orbital constraints.}
\end{itemize}

\subsection{Protocluster Target Selection}
\label{sec:target_sel}

\begin{figure}
    \centering
    \includegraphics[width=\linewidth]{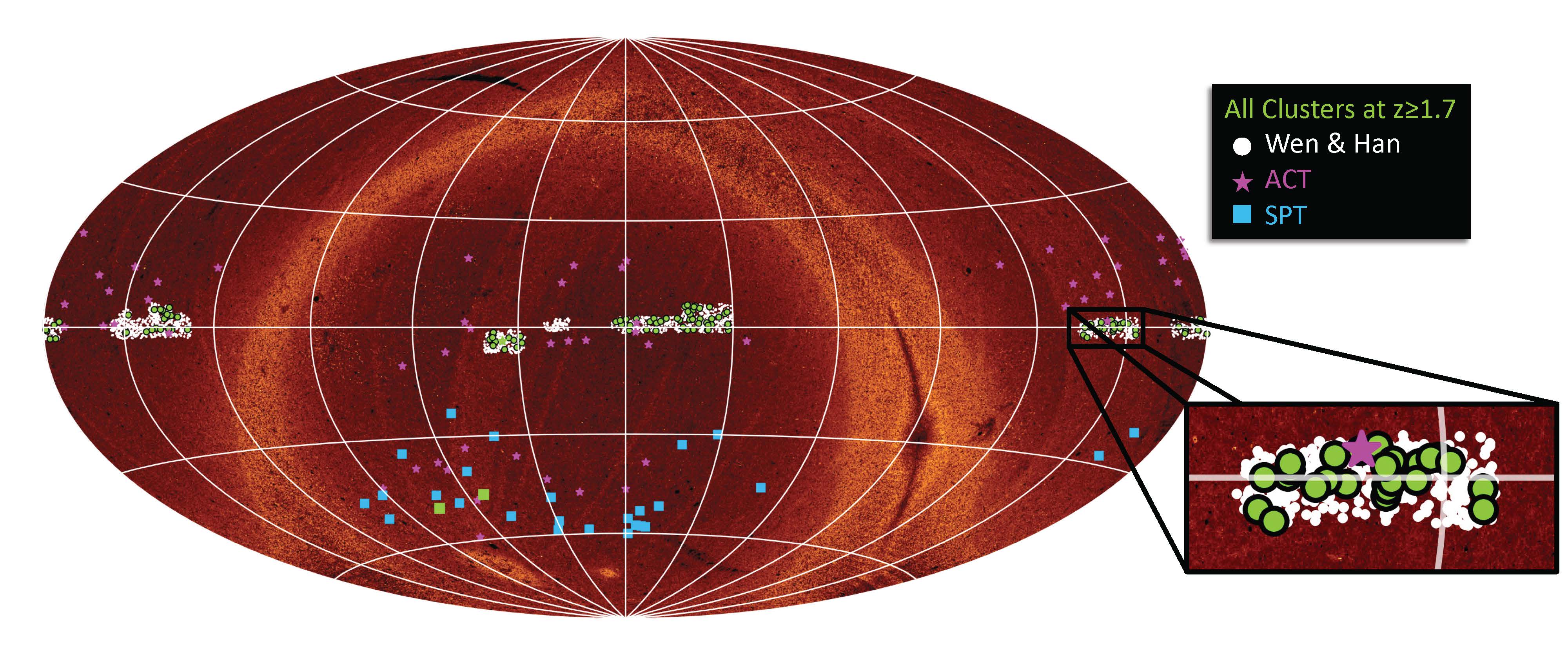}
    \caption{Sky distribution of high-redshift ($z>1.2$) clusters accessible to \iscea, from ACT, SPT, and the equatorial survey fields from the  \citet{Wen2020} cluster catalog, with a density map from All\wise\ shown as the background. All clusters at $z>1.7$ are plotted in green regardless of the source catalog. The inset panel shows the density of high-redshift clusters within one of the \citet{Wen2020} cluster fields.
    The \iscea\  preliminary target list consists of 100 cluster candidates selected from this distribution, including 90 clusters at $1.7 < z < 2.1$, and 10 clusters at $1.2 < z < 1.7$. The \iscea\  confirmed target list will consist of 50 clusters (including 90\% at $1.7 < z < 2.1$).
    }
    \label{fig:clusterdist}
\end{figure}

The \iscea\  preliminary target list consists of 100 protocluster candidates, including 90 protocluster fields at $1.7 < z < 2.1$, and 10 at $1.2 < z < 1.7$, each centered on a cluster candidate.
We have selected 90 fields at $1.7 < z < 2.1$ to prioritize highest redshift clusters from the \cite{Wen2020} cluster catalog.
These are supplemented by  five massive SZ clusters at $z > 1.5$ from ACT and SPT, in order to improve statistics at the highest local densities to gain insight into the physical mechanism of star formation quenching in clusters at this epoch 
(see \S1). The \cite{Wen2020} catalog currently provides the largest sample of $z>1.8$ cluster candidates with photometric redshifts, and hence is the best available sample for this program. Fig. \ref{fig:clusterdist} shows the spatial distribution of \iscea\  preliminary cluster targets in the sky, with the zoomed inset illustrating the cluster target density within one of the fields of \citet{Wen2020}. 
The \iscea\  confirmed target list will consist of 50 protoclusters (each a cluster with its adjacent cosmic web environment), with 90\% of the protoclusters at $1.7 < z < 2.1$. 

\begin{figure}
    \centering
    \includegraphics[width = 3.5in]{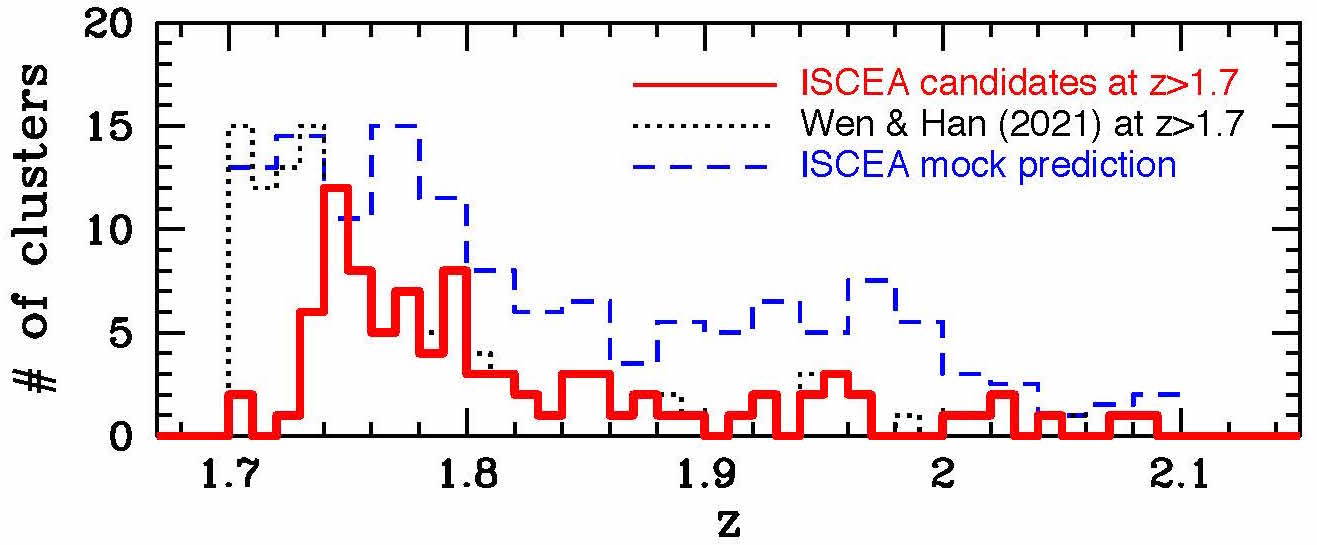}
    \includegraphics[width = 3.3in]{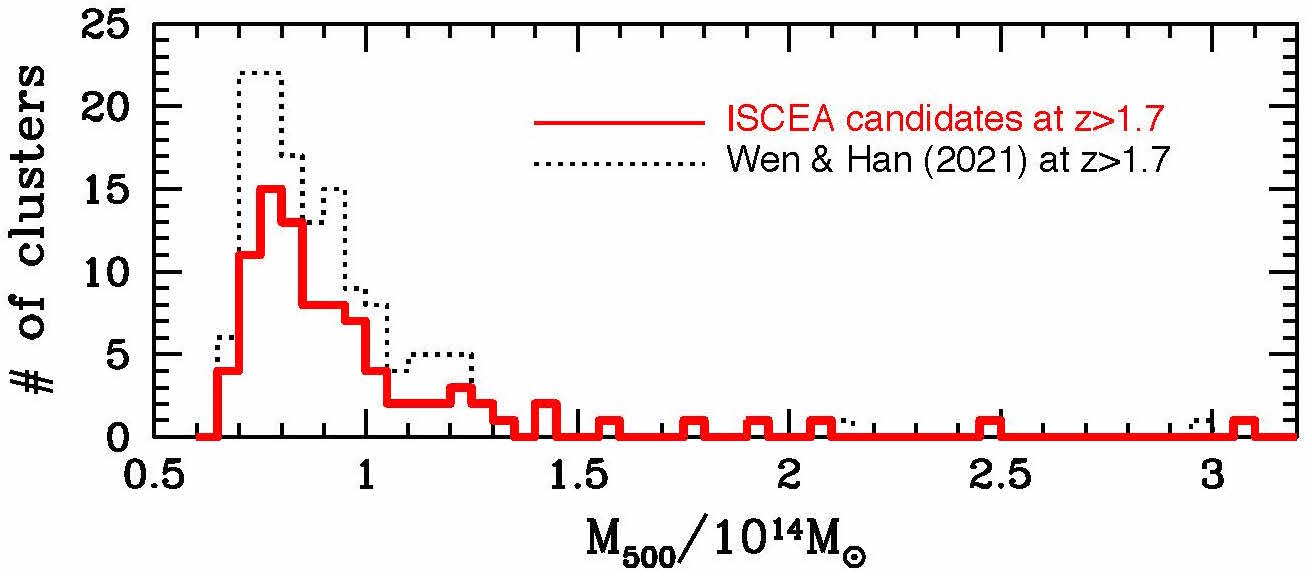}\\
    \caption{\iscea\  protocluster candidates at $1.7 < z < 2.1$, binned in redshift (left) and cluster mass $M_{500}$ (right). The \iscea\  mock catalog predicts 69 protoclusters at $z>1.8$, many more than included in \cite{Wen2020}. This increases the likelihood of the highest $z$ protocluster candidates being at $z>1.8$.
}
    \label{fig:target_dist}
\end{figure}

The \cite{Wen2020} cluster galaxy photo-$z$’s are computed using a nearest-neighbor algorithm based on the 7-band photometry from HSC SSP ($grizy$) and \wise\ (W1, W2) over $\sim$ 800 deg$^2$. The training sample for the algorithm contains 554,996 galaxies with spectroscopic redshifts, of which 240,409 are matched with the HSC-SSP $\times$ unWISE galaxies and have magnitudes in the $grizyW_1$ bands.
This extensive set of spectroscopic data includes the data from SDSS DR14 \citep{Abolfathi2018}, DEEP3 \citep{Cooper2011}, PRIMUS DR1 \cite{Cool2013}, VIPERS PDR1 \citep{Garilli2014}, VVDS \citep{LeFevre2013}, GAMA DR2 \cite{Liske2015}, WiggleZ DR1 \citep{Drinkwater2010}, zCOSMOS DR3 \citep{Lilly2009}, UDSz \citep{Bradshaw2013,McLure2013}, FMOS-COSMOS \citep{Silverman2015,Kashino2019} and 3DHST \citep{Skelton2014,Momcheva2016}. To have enough data for training at $z > 1$, their spectroscopic training set is supplemented by accurate photo-$z$'s from the COSMOS2015 catalogue, which are based on 30-band photometry with an accuracy of 0.021 \citep{Laigle2016}. 
This results in photo-$z$'s with impressive precision and accuracy.
\citet{Wen2020} define the photo-$z$ uncertainty as
\begin{equation}
\sigma_{\Delta z}= 1.48 \times {\rm median}\left(
\frac{\left|z_p-z_s\right| }{1+z_s} \right)
\simeq 0.055 z_s - 0.0145, 
\label{eq:zphot}
\end{equation}
where $z_p$ is photo-$z$, and $z_s$ is spectroscopic redshift. 
They estimated an outlier fraction of $\sim$ 6\%, with outliers defined as photo-$z$'s with a deviation larger than 3$\sigma_{\Delta z}$ or 0.15, compared to spectroscopic redshifts.
They found that $\sigma_{\Delta z} \sim 0.11$ at $z \sim 2$ (see the lower panel of their Figure 1). The addition of \iscea's $1.1-2\,{\rm \mu m}$ broadband imaging will break common degeneracies of the photo-$z$ estimation method, hence result in a further improvement of the photo-$z$ estimates (see \S\ref{sec:photo-z}).
Fig.\ref{fig:target_dist} shows the distribution of the 90 \iscea\  protocluster candidates at $1.7<z<2.1$ in redshift and cluster mass $M_{500}$ (mass contained within radius $r_{500}$, where $\rho_{\rm local}=500\rho_c$, with $\rho_c$ denoting the critical density).

We estimate that $>50$\% of the \iscea\  cluster candidates to be \textit{bona fide} clusters at $1.7 < z < 2.1$. There are many hundreds of cluster candidates to choose from at $1.2 < z < 1.7$, with most of those expected to be \textit{bona fide} clusters in that redshift range, due to the increasing photo-$z$ accuracy and precision with decreasing redshift (see Eq.~\ref{eq:zphot}). We will prioritize the 90 highest redshift cluster candidates at $1.7 < z < 2.1$, and choose 10 cluster candidates at $1.2 < z < 1.7$ to tie in with existing observational data.  
Our \iscea\  mock catalog (\S\ref{sec:mock}) indicates that the \citealt{Wen2020} catalog is complete at $z\sim 1.7$, but highly incomplete at $z > 1.76$ (Fig.~\ref{fig:target_dist}, left panel), increasing the likelihood of the highest $z$ cluster candidates being at $z>1.8$, since we expect 69 clusters at $z > 1.8$ over the \iscea\  survey area of $\sim 800$ deg$^2$. Since  \iscea\  focuses on protoclusters to probe environmental effects on galaxy evolution over three orders of magnitude in local mass density, the cluster mass in each protocuster is of secondary importance. Note that \iscea\ provides follow-up opportunities for future studies to derive cluster masses and enable additional science through studies of lensing (e.g., from {\it Roman} observations) or targeted SZ studies from CMB telescopes.  

\iscea\  will carry out quick spectroscopy (4 hours) of each of the 100 candidate protocluster fields (see \S\ref{sec:obs}), to obtain spectroscopic redshifts for $>10$ galaxies (expected to be met with a large margin since the \iscea\ 5$\sigma$ H$\alpha$ line flux limit for spectroscopy in 4 hours is $2.2\times 10^{-16}$erg/s/cm$^2$, see Fig.\ref{fig:1000nearest-z} left panel), including the Brightest Cluster Galaxy (BCG) in each field. This will result in the \iscea\  confirmed target list of 50 protoclusters to meet the \iscea\  Science Requirements (see \S\ref{sec:science_req}). 

We also have the flexibility to adjust the \iscea\ target list as additional high-redshift clusters from ACT, SPT, and newer, ongoing surveys are 
confirmed. Doing so can further increase the galaxy yield by enabling inclusion of the protoclusters associated with the most massive clusters at that epoch.

\subsection{Galaxy Target Selection}
\label{sec:sel-gal}

In each BCG-confirmed protocluster field, \iscea\  will obtain $\sim$ 1000 spectra simultaneously (see Fig.~\ref{fig:DMD_layout}). 
Since each protocluster field will be visited multiple times, separated by more than 2 weeks, we will update the galaxy target list by removing bright galaxies with high S/N spectra, and replace them with new galaxy targets. Thus \iscea\  will obtain spectra of $>1000$ galaxies per field (see \S\ref{sec:data_acq}).
\iscea\  will target 1000 galaxies with photo-$z$'s closest to that of the BCG in the first visit, and expand the target list in subsequent visits as needed, by selecting the next galaxies with photo-$z$'s closest to the BCG not yet targeted.

\begin{figure}
    \centering
    \includegraphics[width = 6in]{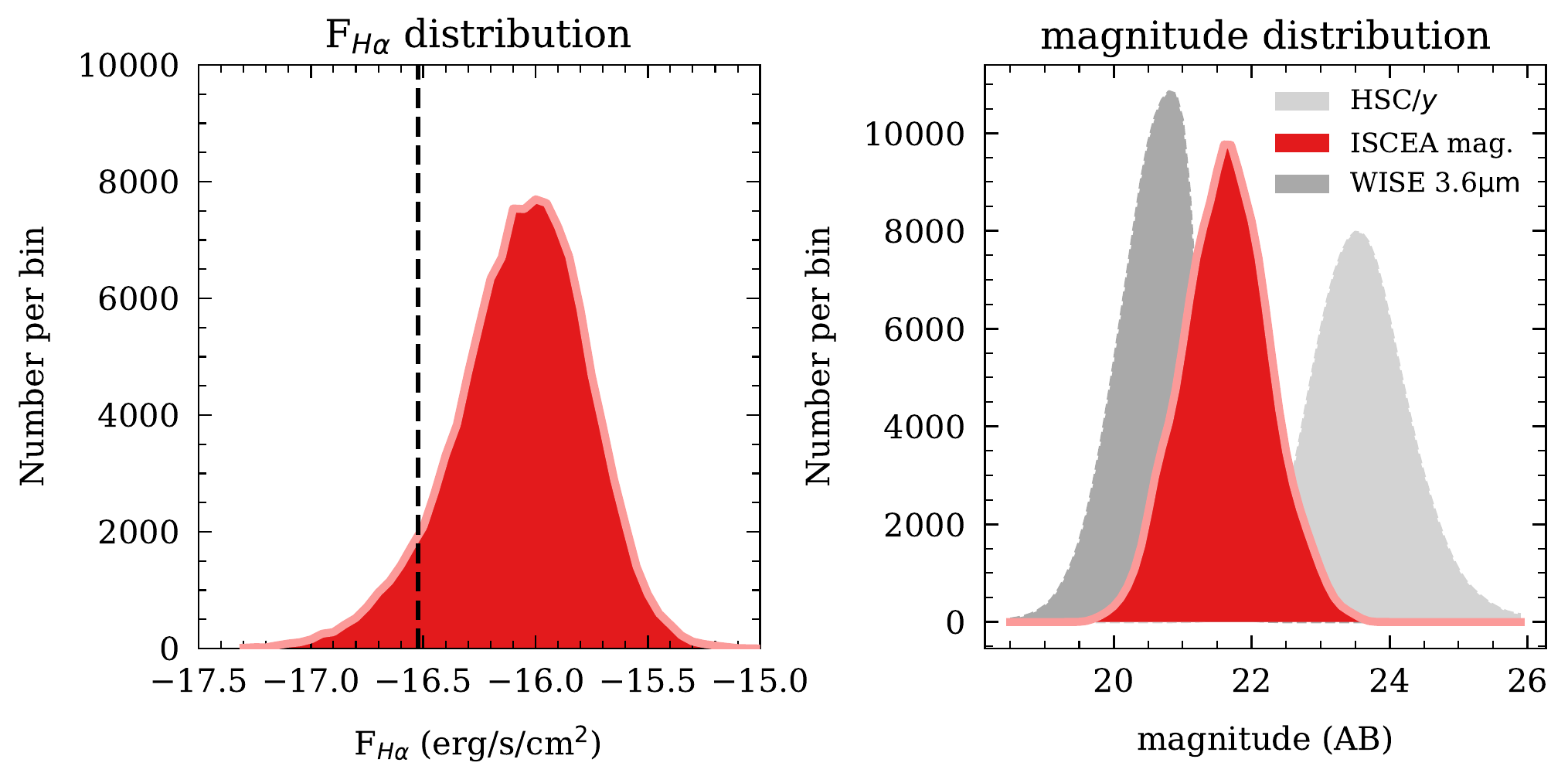}
    \caption{
    The distributions in the estimated H$\alpha$ line fluxes (left) and observed/predicted brightness (right) of the $1000$ galaxies closest to the BCG's photo-$z$ in the \iscea\ FoV for the 100 preliminary protocluster target fields. Shown are the observed brightnesses in \textit{WISE} $3.5\,{\rm \mu m}$ (dark gray) and HSC/$y-$band (light gray) as well as the predicted brightness of the galaxies in the \iscea\ 1.1-2$\mu$m broad $H-$band (red).
    }
    \label{fig:1000nearest-z}
\end{figure}

\begin{figure}
    \centering
      \includegraphics[width = 3in]{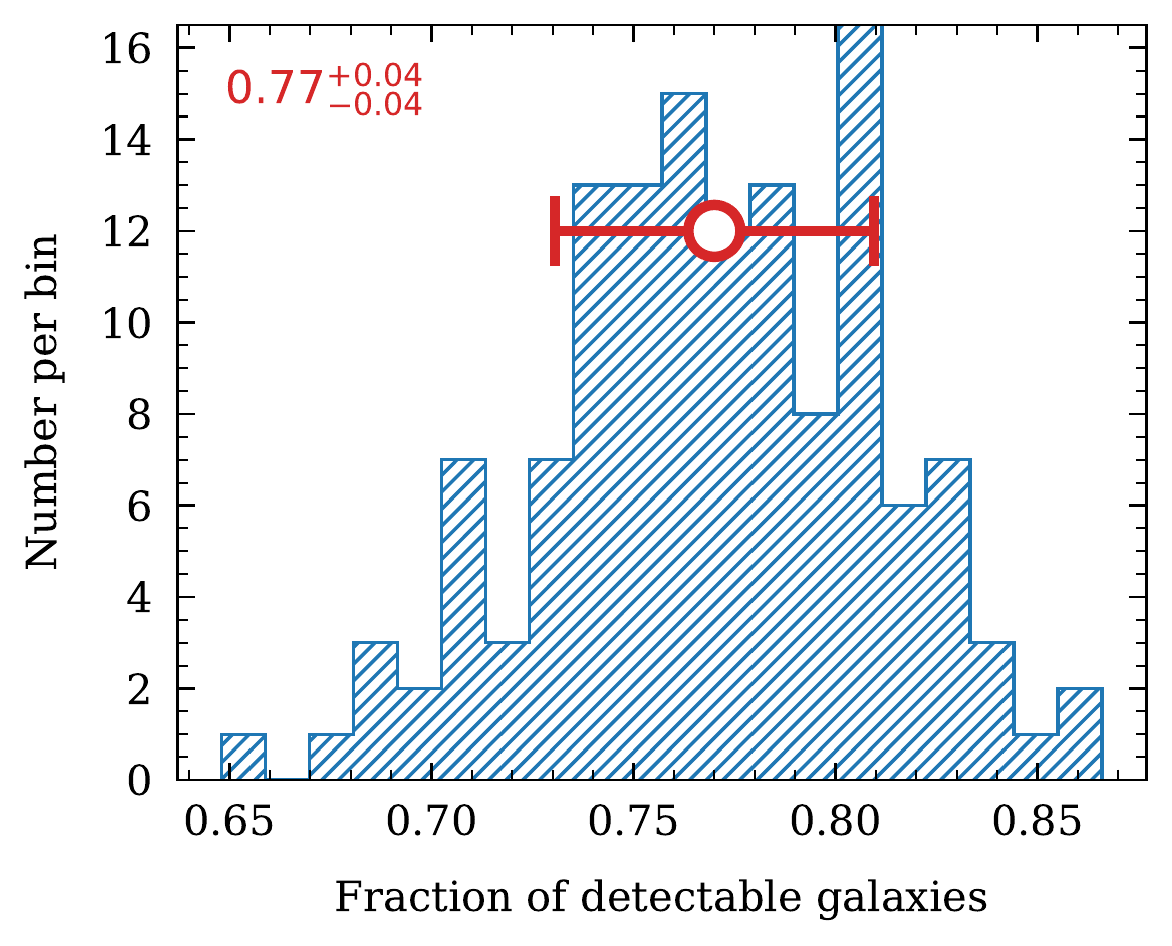}
    \caption{The fraction of \iscea\ H$\alpha$ detections among the $1000$ galaxies closest to the BCG's photo-$z$ in the \iscea\ FoV for the 100 preliminary protocluster target fields. The median fraction is $77\pm4\%$.
}
    \label{fig:1000nearest-z-frac}
\end{figure}

Fig.~\ref{fig:1000nearest-z} shows the distribution in the estimated H$\alpha$ line flux (left panel), and the observed brightness in the \iscea\ $H$-band, WISE 3.5$\mu$m and HSC $y-$band (right panel), for the 1000 galaxies with photo-$z$'s closest to that of the cluster BCG, for the 100 \iscea\  preliminary protocluster target fields (see \S\ref{sec:target_sel}), each the size of the \iscea\  FoV. On average, $\sim$77\% of the 1000 galaxies in each field are estimated to be above the \iscea\  flux limit of $3\times 10^{-17}\esc$ (see Fig.~\ref{fig:1000nearest-z-frac}).
Note that \iscea\  images all galaxies to AB=25 in the \iscea\  broad $H-$band (see \S\ref{sec:obs}), which encompasses all of the 1000 target galaxies in each of the \iscea\  candidate protoclusters from the \cite{Wen2020} catalog (see Fig.~\ref{fig:1000nearest-z}, right panel). The SZ clusters are more massive and expected to contain a larger number of bright galaxies.

\begin{figure}
    \centering
    \includegraphics[width = \textwidth]{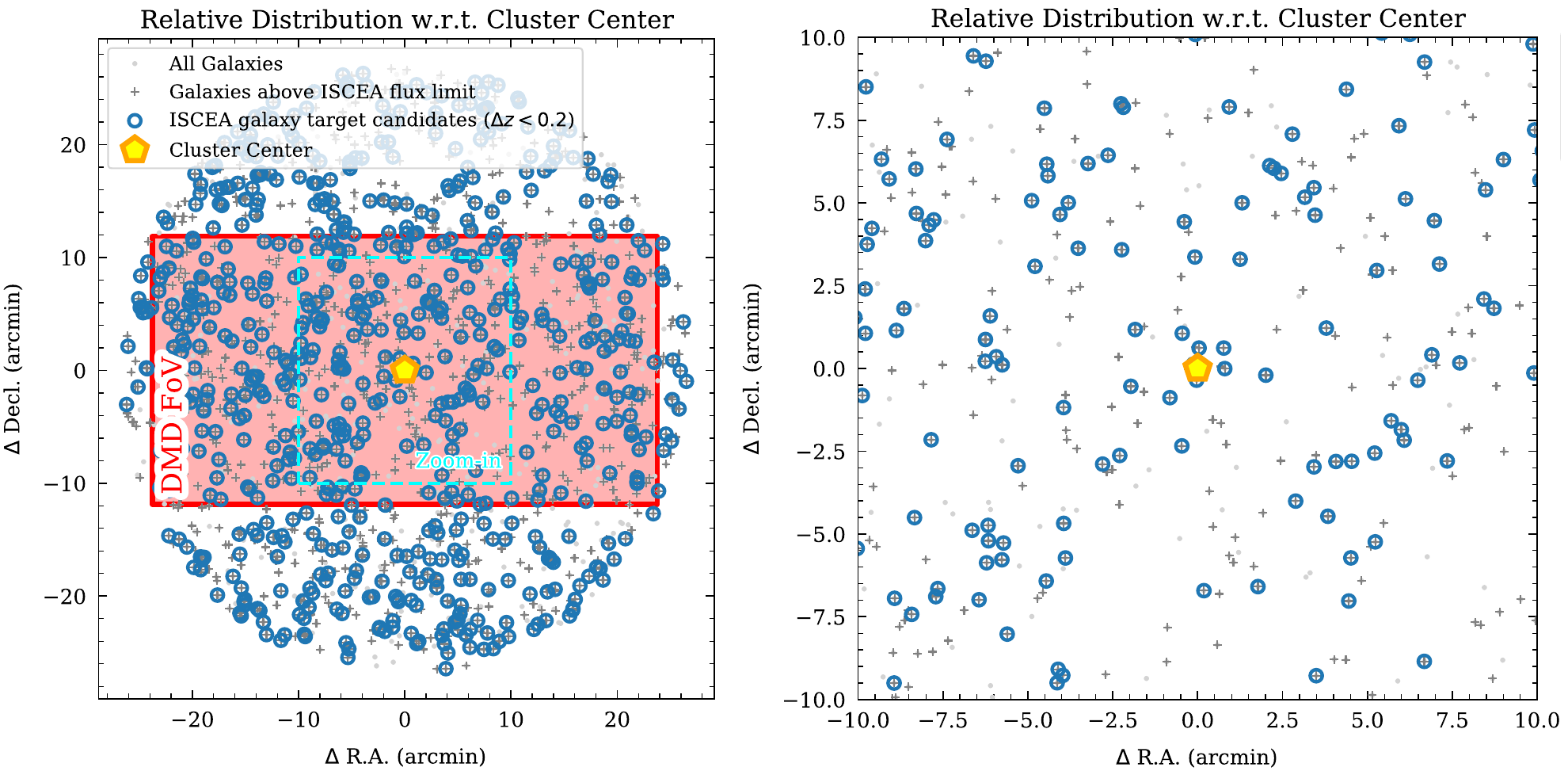}
    \caption{A representative \iscea\  protocluster target field, centered on candidate cluster J123345.3-002945 at $z\sim 2$ from the \cite{Wen2020} catalog. The full \iscea\ DMD FoV (left) as well as a close-up of the central region (right) are shown. Of the 1000 galaxies targeted by \iscea, only the $\sim 300$ within $\Delta z \leq 0.2$ from the BCG are shown to avoid overcrowding the figure.}
    \label{fig:gal_targets}
\end{figure}

Fig.~\ref{fig:gal_targets} shows a representative \iscea\  candidate protocluster target field, centered on candidate cluster J123345.3-002945 at $z\sim 2$ from the \cite{Wen2020} catalog. There are $\sim$ 300 galaxies within $\Delta z_{\rm photo} <0.2$ of the cluster BCG photo-$z$ in this \iscea\  candidate protocluster field, which indicates the rough number of galaxies in an \iscea\ protocluster field belonging to the protocluster, since the photo-$z$ error is $\sim$0.11 \cite{Wen2020}.
Using the \iscea\ mock (see \S\ref{sec:mock}), we can estimate the number of protocluster galaxies expected in each \iscea\ FoV by counting the number of galaxies with H$\alpha$ line flux brighter than $3\times 10^{-17}\esc$ within a volume of 12Mpc$\times$24Mpc$\times$24Mpc at $z\sim 2$ centered on each BCG. Note that at $z\sim 2$, 1 proper Mpc $\sim 2^\prime$.
We find that there are $\sim$210 protocluster galaxies on average with H$\alpha$ line flux brighter than $3\times 10^{-17}\esc$ within an \iscea\ protocluster field, for clusters with the same $M_{500}$ and at the same redshifts as those on the \iscea\  preliminary target list at $z>1.7$ (see \S\ref{sec:count-gal}). Thus $\sim$ 210 galaxies out of $>1000$ observed galaxies in each \iscea\  field are in the cluster or its cosmic web environment, and
70\% of the number of galaxies within $\Delta z_{\rm photo} <0.2$ of the cluster BCG photo-$z$ in a typical \iscea\  candidate protocluster field. This is not surprising, since the photo-$z$ scatter (see Eq.~\ref{eq:zphot}) is $\sim 0.11$ at $z=2$ \citep{Wen2020}, and the \iscea\  flux limit is faint enough to include the majority of H$\alpha$ ELGs in a protocluster.

This photo-z based “blind” selection enables us to target ELGs as well as passive galaxies. Most of the photo-$z$ selected galaxies are expected to be ELGs. The faint H$\alpha$ line flux limit of \iscea, $3\times 10^{-17}\esc$, enables us to achieve high completeness for star-forming galaxies on the main sequence with $S/N\geq 5$ spectra, and derive the fraction of passive galaxies based on the absence of emission lines at the 3$\sigma$ H$\alpha$ line flux limit of $1.8\times 10^{-17}\esc$. The fraction of quiescent galaxies is an important indicator of galaxy evolution (see \S\ref{sec:test-hypo}).

\section{Simulating \iscea\  Data}
\label{sec:sim}

\subsection{\iscea\  mock catalog}
\label{sec:mock}

In order to flow down the \iscea\ Science Objective to requirements, we have created a state-of-the-art galaxy mock catalog by applying a semi-analytical model of galaxy evolution, Galacticus \citep{benson_galacticus:_2012}, to the cosmological N-body simulation UNIT \citep{chuang_unit_2019}, which has a particle mass of $1.2 \times 10^9 h^{-1} \mathrm{M}_\odot$ in a 1~(Gpc/$h$)$^3$ simulation cube, and assumes the \cite{planck_2016} cosmology. Merger trees from the UNIT simulations (generated using the Rockstar halo finder; \citealt{behroozi_rockstar_2013}) are input to the Galacticus semi-analytic model which solves for the properties of galaxies forming inside the halos of each tree, and outputs these galaxies at every snapshot of the UNIT simulation.

We use the same galaxy formation physics in Galacticus as used by \cite{Zhai_2019} to study expected numbers of emission line galaxies for future surveys. The parameters of the Galacticus model physics were calibrated to approximately match available observational data on galaxy evolution, including stellar mass functions and star formation rates. Additionally, \cite{Zhai_2019} calibrated the emission line luminosity functions (and associated dust extinction model) predicted by Galacticus to match the HiZELS \citep{sobral_hizels_2013} observations.

The resulting \iscea\  mock catalog provides physical properties (comoving position, redshift, line-of-sight velocity, halo mass, stellar mass, SFR), as well as observed quantities (H$\alpha$ and \oiii~ line luminosities, and broad-band magnitudes in \iscea, WISE, and HSC bands) for each model galaxy. Additionally, meta-data describing cluster membership and central/satellite status is provided for each galaxy. 

The parameters of the Galacticus model have been tuned using optimization algorithms to find the best match to current observational data. As new observational data become available, including data from \iscea, we will repeat this optimization process (which is fully automated) to refine the model, allowing for the production of new and improved mock catalogs. New data may also highlight limitations of the current model (i.e. mismatches which cannot be resolved by retuning of parameters). These will drive investigations of how to improve the model physics to better understand any such new data.

Of particular interest will be results from \iscea\  which characterize the properties of galaxies in and around protoclusters, across a broad redshift range. The effects of environment within the cosmic web is still poorly constrained observationally and poorly understood theoretically. We expect that \iscea\ data will be invaluable in improving the model of environmental processes in the Galacticus model. These too will be incorporated into future generations of highly realistic galaxy mock catalogs, which will provide a powerful tool for detailed understanding of galaxy evolution in the cosmic web.

\subsection{Photometric Redshift Verification \& \iscea\  Broadband Magnitudes}
\label{sec:photo-z}

It is important to validate the protocluster targets for \iscea\  using the existing multiwavelength, \wise-selected cluster catalogs from \cite{Wen2020}. We identified clusters of interest from the \cite{Wen2020} catalog,  selecting them based upon their predicted redshift. 
We visually inspected the clusters using the \wise\ imaging database to remove cluster targets located near bright stars, and candidates that are detected due to artifacts in the \wise\ data.

We then independently evaluated the photo-$z$'s of the clusters using the photometric data from the \cite{Wen2020} catalog, which includes photometry from Subaru/HSC ($grizY$) and \wise\ W1 (3.4$\mu$m) and W2 (4.6$\mu$m) imaging.   We independently measured photo-$z$'s for the galaxies in clusters at $z_\mathrm{phot} > 1.7$ from the \cite{Wen2020} catalog using EAZY-py\footnote{https://eazy-py.readthedocs.io/}.  EAZY-py fits the photometry using a non-negative linear combination of a set of galaxy spectral templates.  We used the recommended set of templates from \texttt{tweak\_fsps\_QSF\_12\_v3.param} (G. Brammer 2021, private communication), which include a range of galaxy types (star-forming and quiescent) with varying amounts of dust attenuation assuming a  modified dust attenuation law \citep{Kriek2013}.  Compared to the \cite{Wen2020} photo-$z$'s, the values we derive have very small bias, $\mathrm{med}(z)/(1+z)$ = 0.011 for all galaxies.  The scatter is larger, $\sigma_0=0.13$, excluding outliers (where $\sigma_0$ is the normalized absolute deviation, see \citealt{Brammer2008}), but this is reasonable considering we are comparing photo-$z$'s derived from two independent methods, and consistent with the scatter in photo-$z$ estimates from independent codes reported elsewhere (\citealt{Dahlen2013};  \citealt{Wen2020} do not provide uncertainties on individual estimates).    

\begin{figure}
    \centering
    \includegraphics[width = 3in]{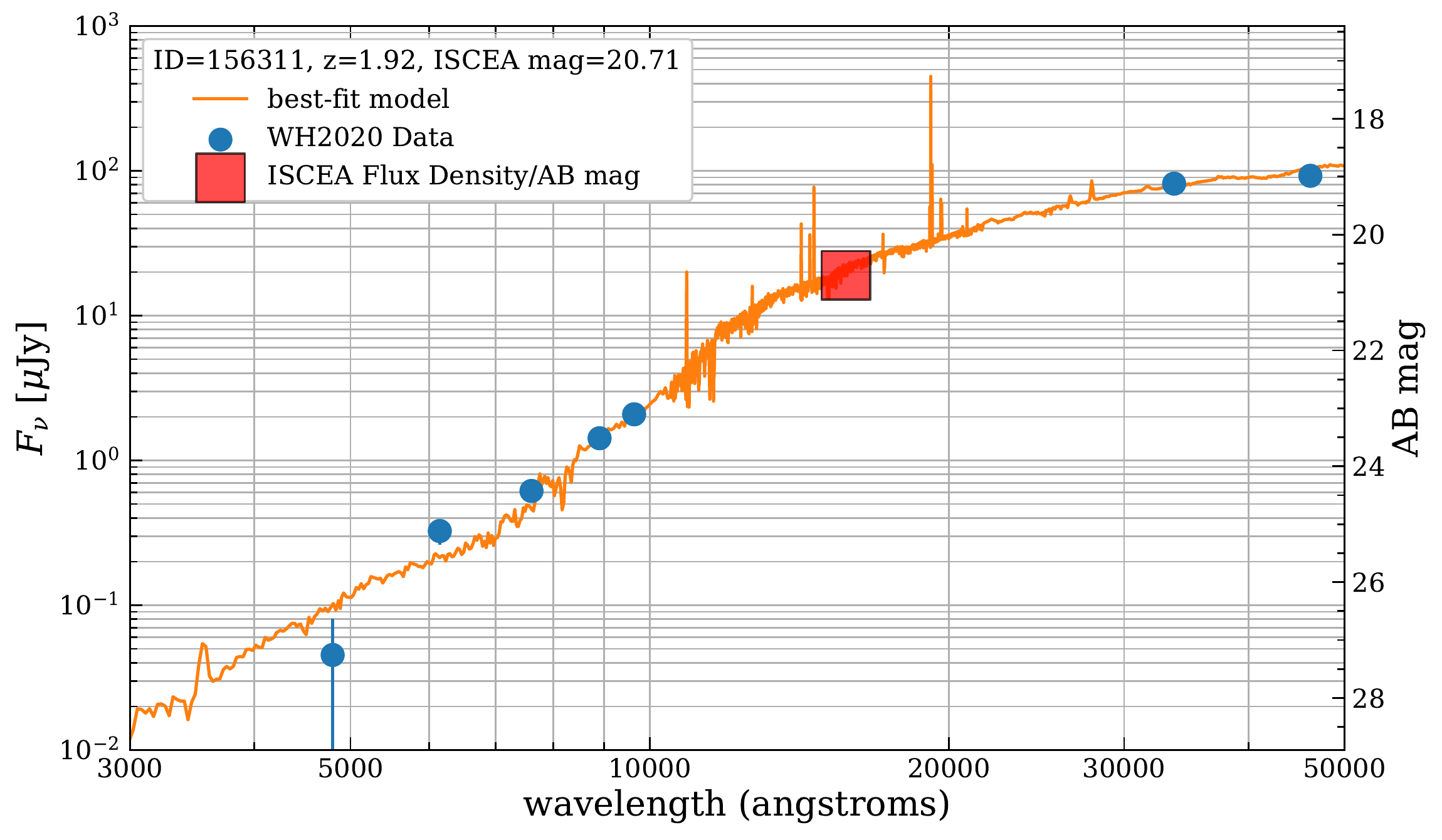}
    \includegraphics[width = 3in]{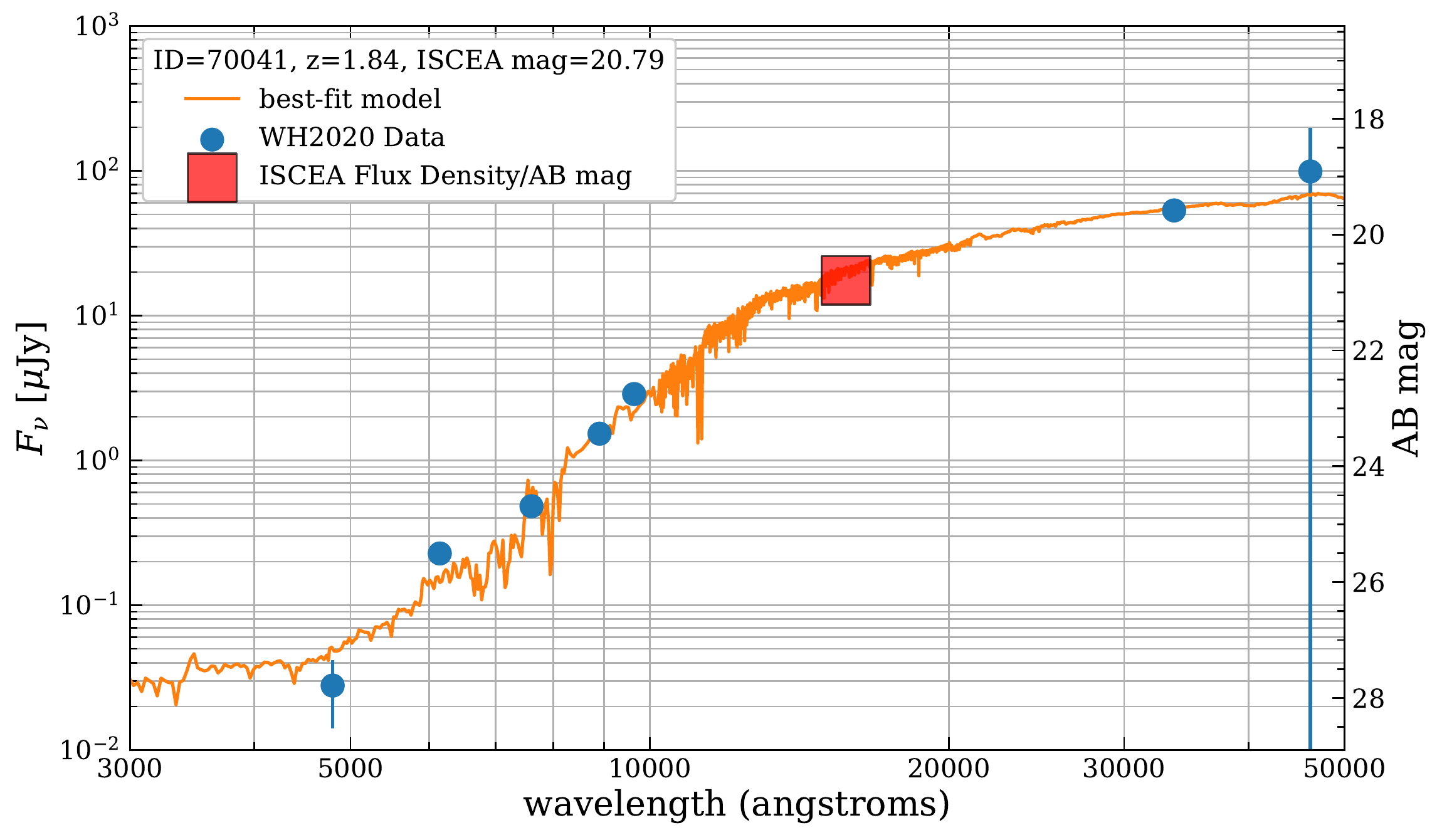}
    \includegraphics[width = 3in]{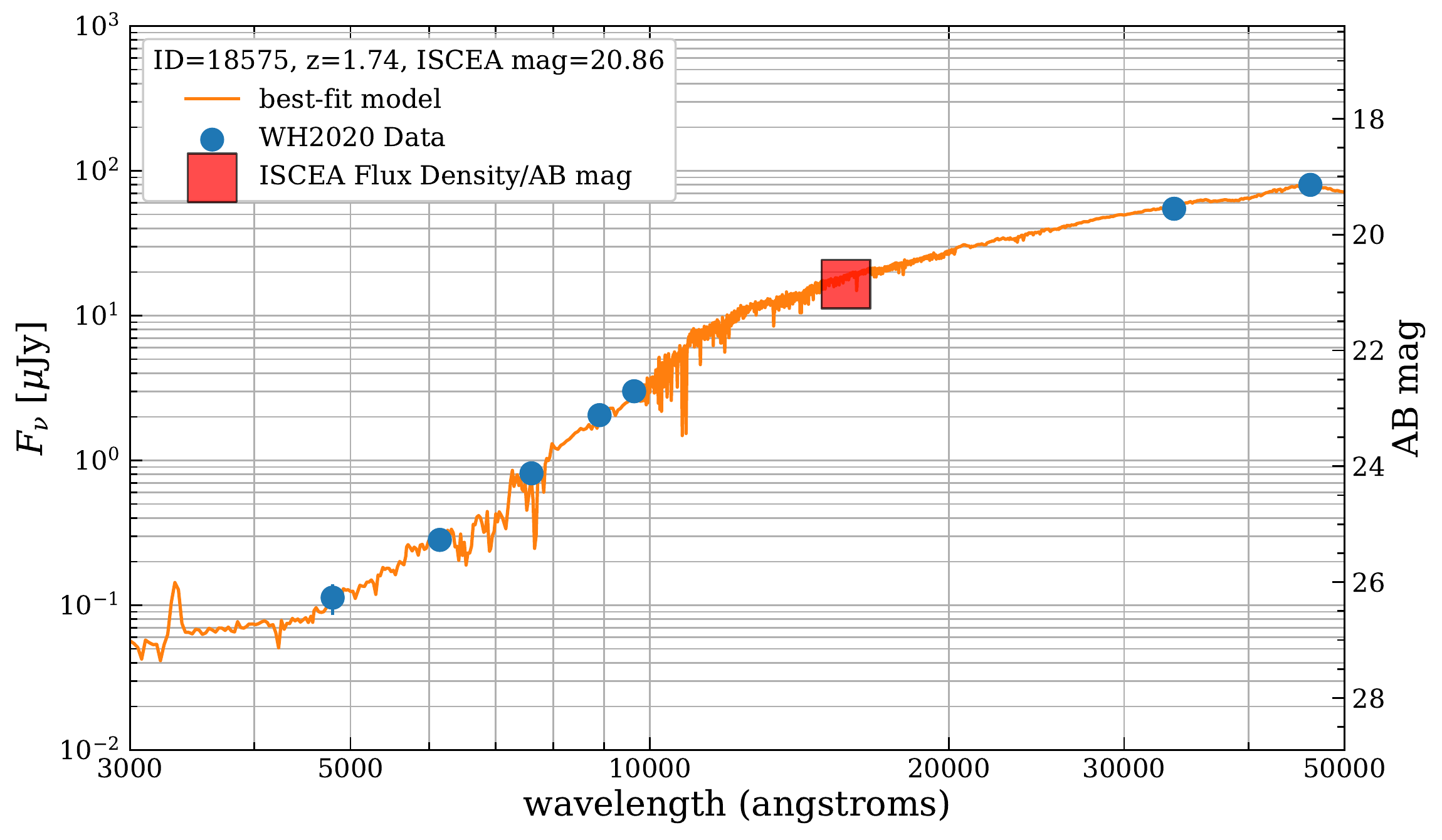}
    \includegraphics[width = 3in]{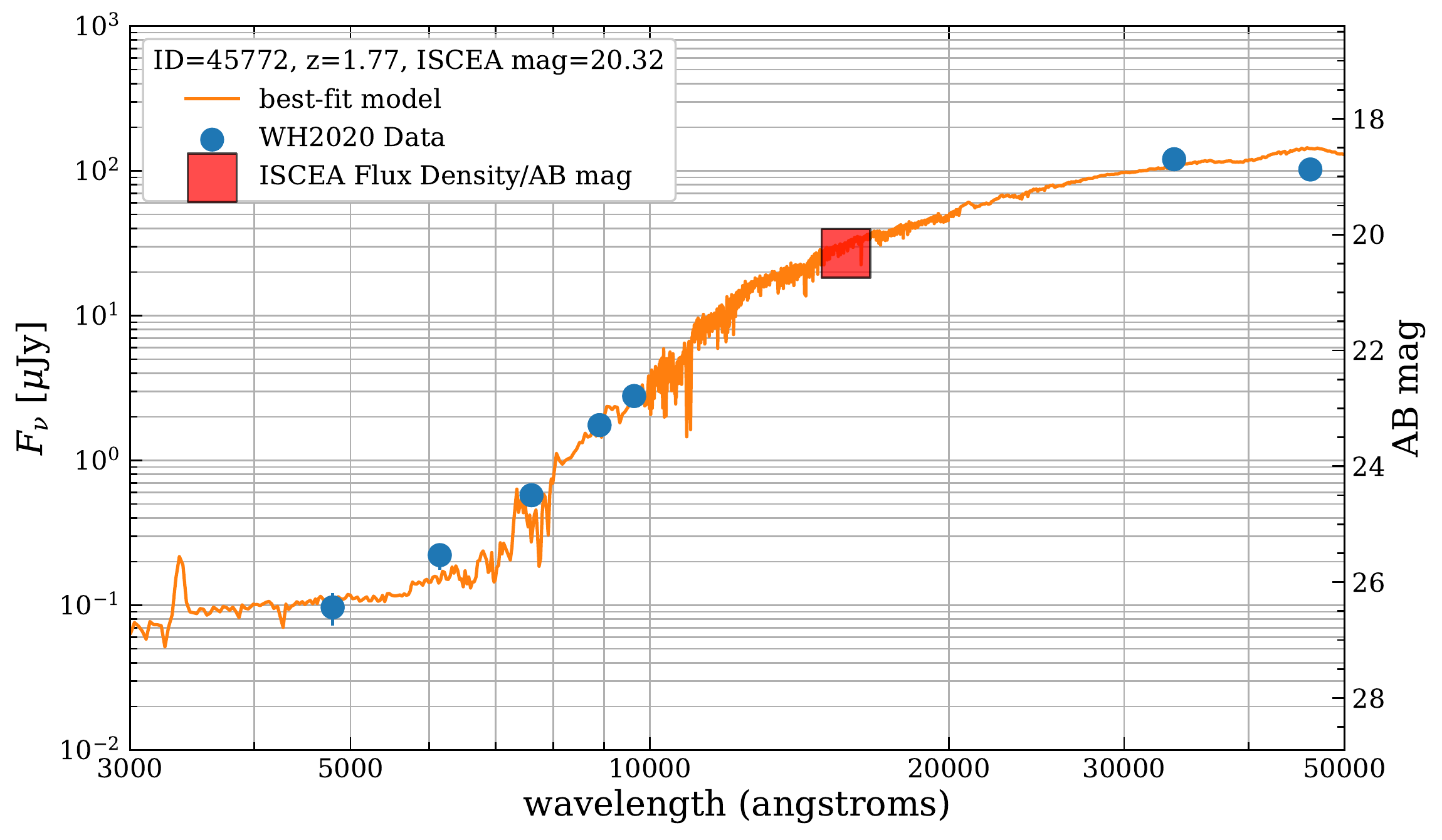}
    \caption{Best-fit spectral fit from the photometric fitting for four example BCGs from the \iscea\ protocluster candidate sample. In each panel, the blue points show the HSC $grizY$ and WISE W1 (3.4$\mu$m) and W2 (4.6$\mu$m) photometry.  The orange curve is the best-fit, non-negative combination of galaxy templates from the photo-$z$ fitting.   The large red square shows the synthesized \iscea\  magnitude derived from the best-fit spectrum assuming \iscea\  passband has uniform transmission over the wavelength range 1.1-2$\mu$m.  Each panel provides the galaxy ID, photo-$z$, and \iscea\  broadband magnitude.}
    \label{fig:SEDfit}
\end{figure}

Fig.~\ref{fig:SEDfit} shows the best-fit photo-$z$ template (constructed from the non-negative linear combination of the EAZY-py templates) for four galaxies in our sample.  These are ``brightest cluster galaxies'' (BCGs) in that they have the brightest \wise\ W1 magnitude of all galaxies associated with their respective clusters from the \citealt{Wen2020} catalog.   

\begin{figure}
    \centering
    \includegraphics[width = 3in]{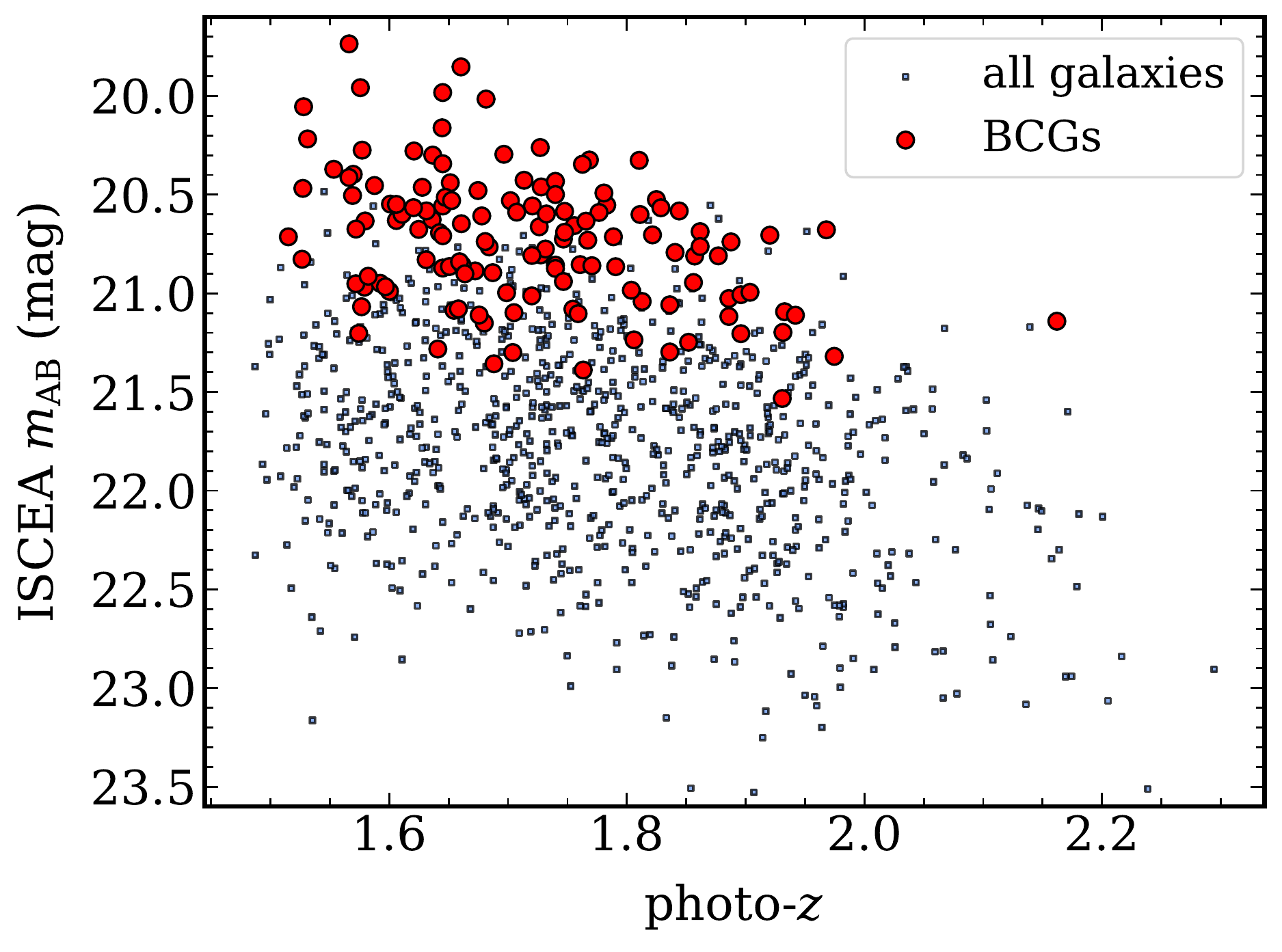}
    \includegraphics[width = 3in]{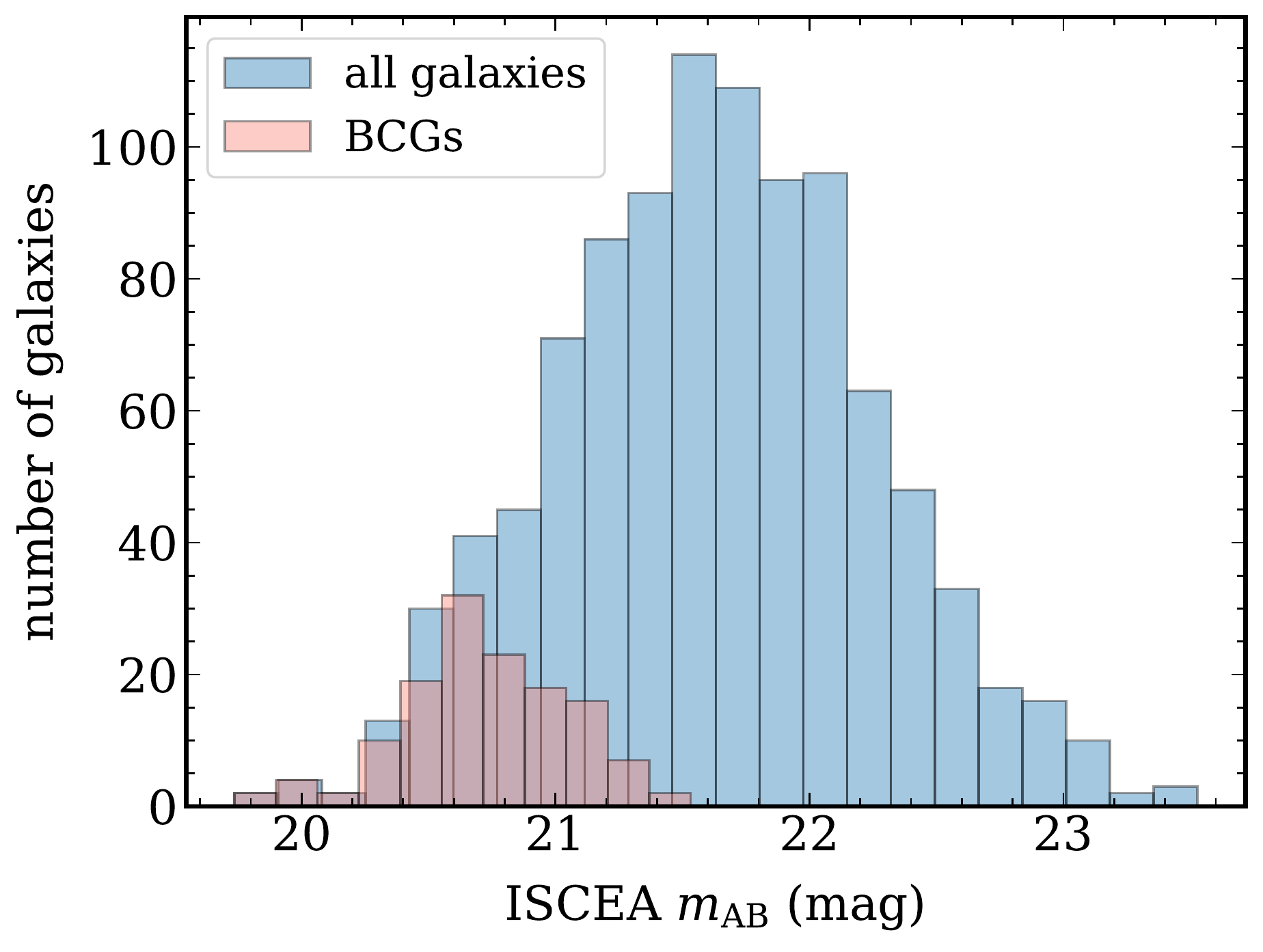}
    \caption{\textit{Left}: Distribution of the predicted \iscea\  broadband magnitudes as a function of photo-$z$ for candidate BCGs (red) and the full cluster population (blue) from \cite{Wen2020}. \textit{Right}: Histograms illustrating the expected magnitude distribution of the same BCGs (red) and galaxies (blue). The distribution of \iscea\  magnitudes peaks around $21.3\,{\rm AB\,mag}$. The decline in the number of galaxies at fainter magnitudes results from the galaxy selection, HSC $i< 26$ and detection in W1 ($\lesssim21.3\,{\rm AB\,mag}$)
    }
    \label{fig:photoz-dist}
\end{figure}

From the best-fit spectral template fit to each galaxy from our photo-$z$ fits, we synthesized (predicted) magnitudes in the \iscea\  passband, \iscea\  $m_\mathrm{AB}$, assuming filter with uniform throughput (i.e., a tophat filter) in the wavelength range $1.1 < \lambda/\mu\mathrm{m} < 2$.  We integrated the best-fit template for each galaxy with this filter (following standard practices, see e.g., \citealt{Fukugita1998} and \citealt{Papovich2001}).  Fig.~\ref{fig:SEDfit} shows the \iscea\  magnitudes for the four example BCGs.   Fig.~\ref{fig:photoz-dist} shows the distribution of \iscea\  magnitudes for the galaxies in the \cite{Wen2020} cluster sample.   The BCGs for the 100 highest redshift clusters are indicated (where the BCG is defined as the brightest galaxy in the \iscea\ $H$-band in each \citealt{Wen2020} cluster). 
The distribution of \iscea\  magnitudes peaks around 21.5 AB mag, with a significant tail below 22nd magnitude (this results from the fact that the \citealt{Wen2020} catalog is selected with HSC $i < 26$~AB and detected in the unWISE catalog with W1 $\la$ 21.3 AB mag).  The median \iscea\  $m_\mathrm{AB}$--W1 colors is 1.4 mag.  

\subsection{Number of \iscea\  Galaxies Per Protocluster Field}
\label{sec:count-gal}

We use the \iscea\  mock (see \S\ref{sec:mock}) to investigate the expected number of galaxies in protoclusters that can be detected by \iscea\  based on their H$\alpha$ line flux.  To do this, we search for clusters with $M_{500} > 0.7\times 10^{14}\,{\rm M_\odot}$ (the same threshold as \citealt{Wen2020}) in the full volume of the simulation. For each cluster, we count the its member galaxies (i.e., satellites in the cluster dark matter halo) that have an H$\alpha$ flux of greater than $3 \times 10^{-17}$ erg s$^{-1}$ cm$^{-2}$ (the \iscea\  5$\sigma$ line flux limit). We then count the number of galaxies above the \iscea\  flux limit in a volume of $\sim$ 12Mpc$\times$24Mpc$\times$24Mpc centered on the cluster, which gives the number of \iscea\  galaxies in a protocluster field.

\begin{figure}
    \centering
    \includegraphics[height = 2.5in]{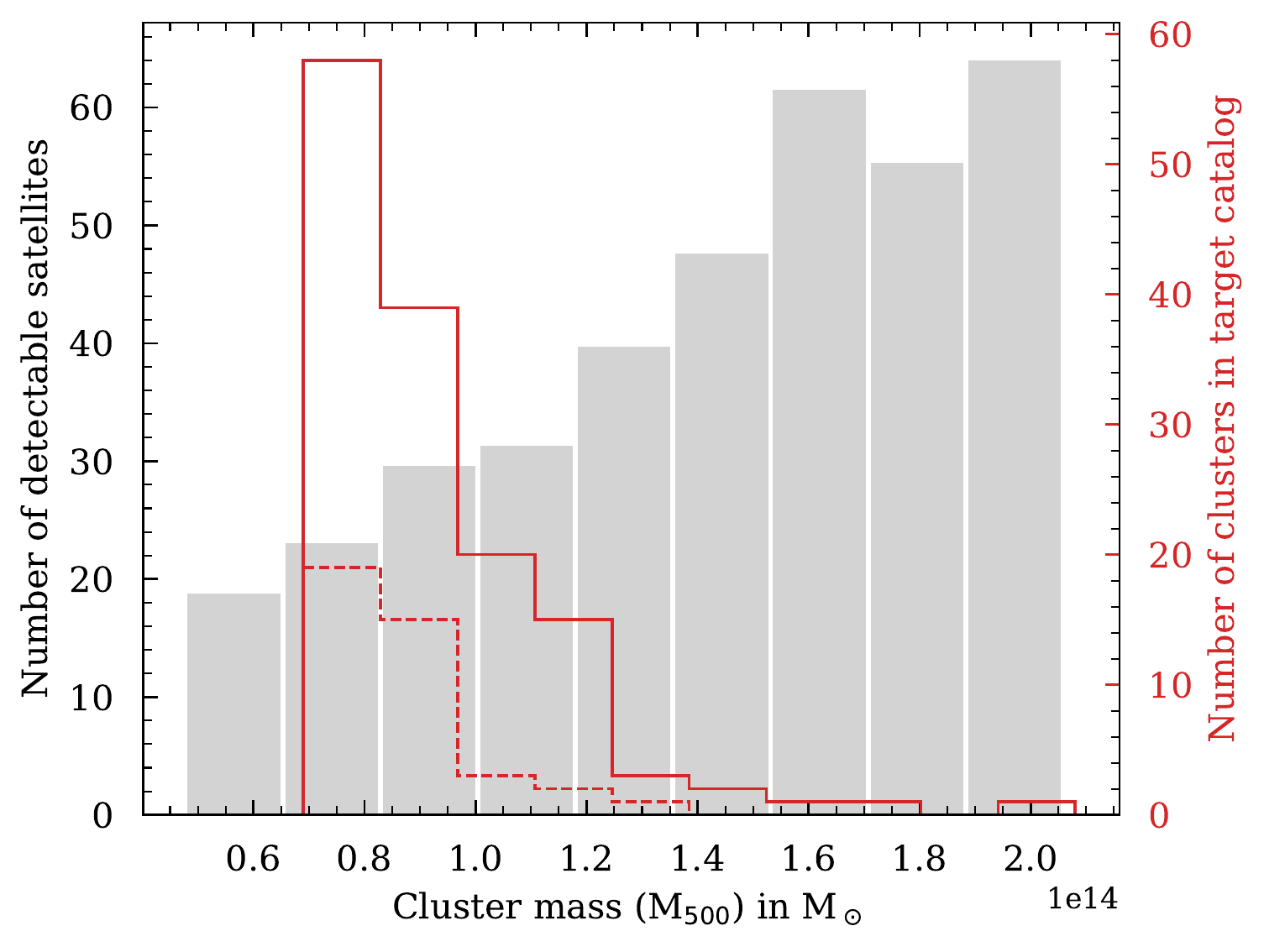}
    \hskip 0.1in
     \includegraphics[height = 2.5in]{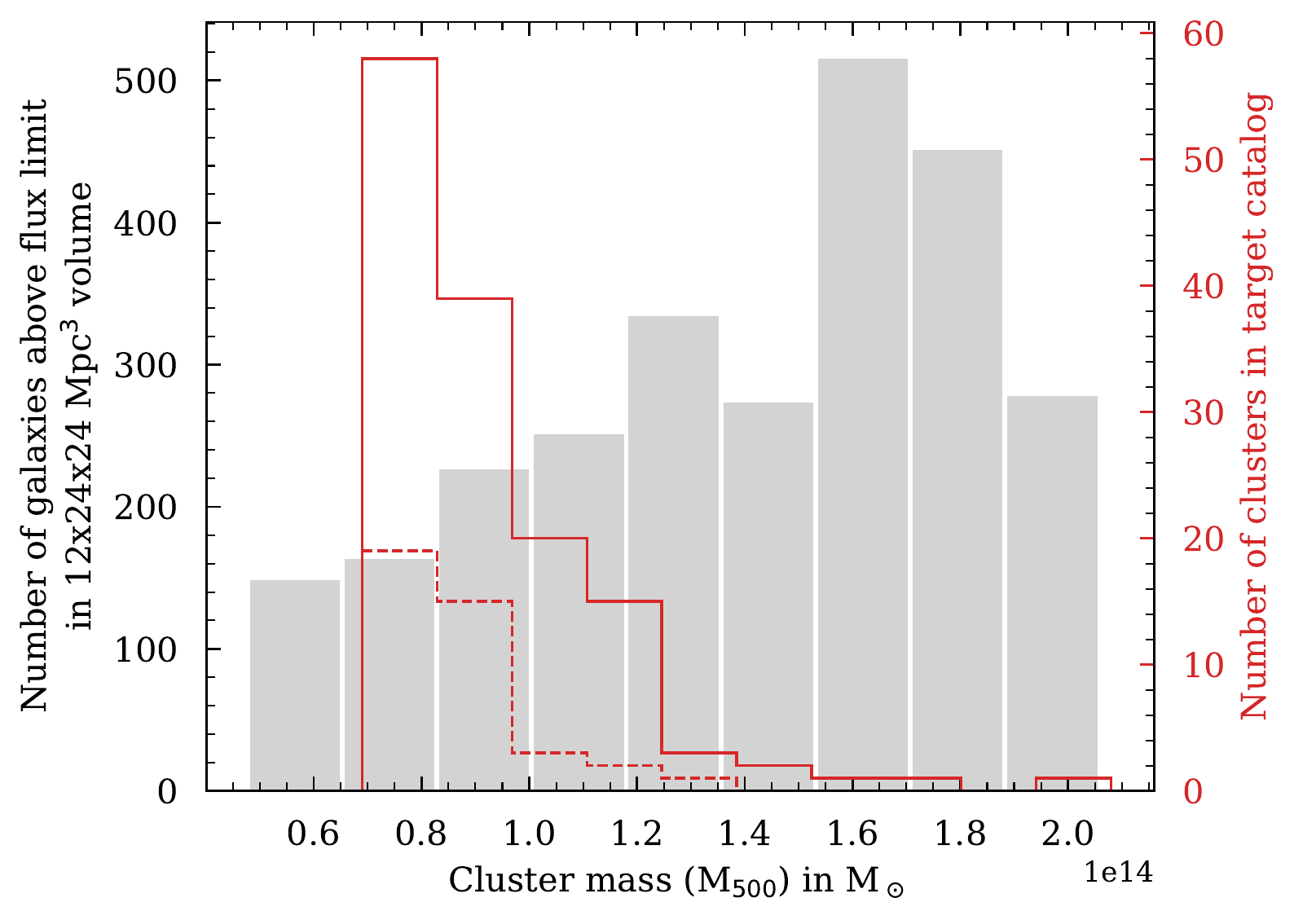}
    \caption{The mean number of cluster member galaxies (left), and the mean number of galaxies within a volume of $12 \times 24\times 24\,$Mpc$^3$ around the cluster central (right), in bins of $M_{500}$ cluster mass. Also shown is the distribution of real cluster masses in red from the \cite{Wen2020} catalog (which all have $M_{500} > 0.7\times 10^{14}\,{\rm M_\odot}$). Specifically, the solid red histogram shows the clusters at $z > 1.7$ and the dashed red histogram shows the clusters at $z > 1.8$ in both panels. 
}
    \label{fig:nsat}
\end{figure}

Here we quantify halo mass in terms of $M_{500}$. We find that in the \iscea\  mock a halo mass of $M_{500} = 0.7\times 10^{14}\,{\rm M_\odot}$ corresponds roughly to $M_{200} = 10^{14}\,{\rm M_\odot}$.
Fig.~\ref{fig:nsat} (left panel) shows the resulting mean number of cluster member galaxies (i.e., mean number of satellites in the cluster halo) in bins of cluster mass, $M_{500}$. Also shown is the distribution of estimated cluster masses of \iscea\  protocluster candidates from the \cite{Wen2020} catalog (in red). Specifically, the solid red histogram shows the clusters at $z > 1.7$ and the dashed red histogram shows the clusters at $z > 1.8$. 
The weighted mean is 33 galaxies per cluster for the clusters in the \cite{Wen2020} catalog at $z>1.7$.

We also use the \iscea\  mock to count the number of galaxies (all, including satellites) in a volume of $12 \times 24\times 24\,$Mpc$^3$ around the cluster central. The $12 \times 24 \,$Mpc$^2$ corresponds to \iscea\  FoV ($24^\prime\times 48^\prime$). The 24 Mpc in the radial direction, chosen to be the same as the protocluster dimension in the transverse direction, corresponds to roughly $\Delta z = 0.02$ over the \iscea\  redshift range of $1.2<z<2.1$.
Fig.~\ref{fig:nsat} (right panel) shows the mean number of galaxies above the \iscea\  H$\alpha$ line flux limit in that volume as a function of $M_{500}$ of the cluster.  Also shown in red is the distribution of estimated cluster masses from the \cite{Wen2020} catalog, with the same line types as in the left panel.
The weighted mean is 210 galaxies with H$\alpha$ line flux $>3\times 10^{-17}\esc$ in the volume of $12 \times 24\times 24\,$Mpc$^3$ around the cluster central, for the clusters in the \cite{Wen2020} catalog at $z>1.7$.
This enables us to estimate the fraction of $\sim 1000$ galaxies observed by \iscea\  in each protocuster field that are in the cluster or its cosmic web environment to be $\sim$21\%.

\subsection{DMD Aperture Loss and Pointing Jitter Modeling and Mitigation}
\label{sec:jitter}

\begin{figure}
    \centering
    \includegraphics[width = 4.5in]{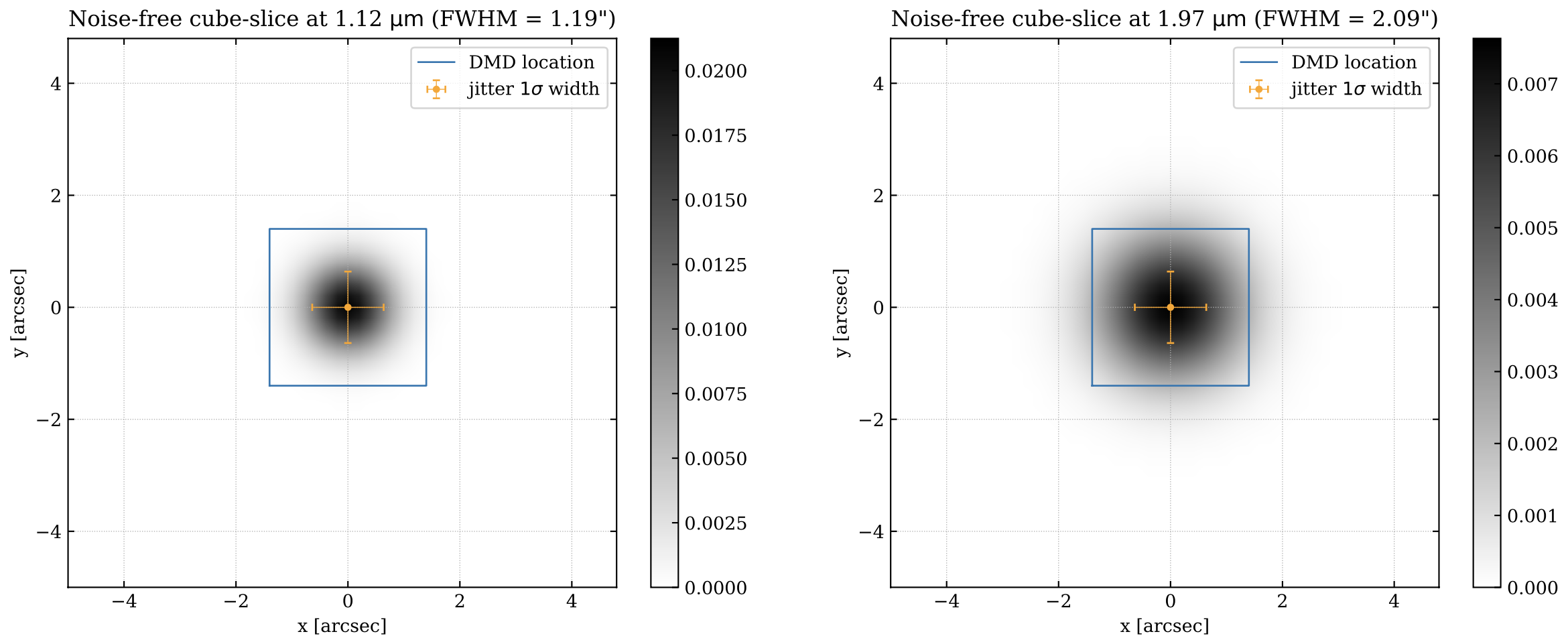} 
    \caption{The PSF-convolved model galaxy at two different wavelengths used to create a simulated observed data cube through a micro-mirror. Note that the source is larger at longer wavelengths because of the increased PSF size. The location of the micro-mirror (i.e., slit) is shown as blue square and the pointing uncertainty (FWHM of 2$^{\prime\prime}$) is indicated as orange cross.
    }
    \label{fig:PSF}
\end{figure}

\begin{figure}
    \centering
   \includegraphics[width = 7in]{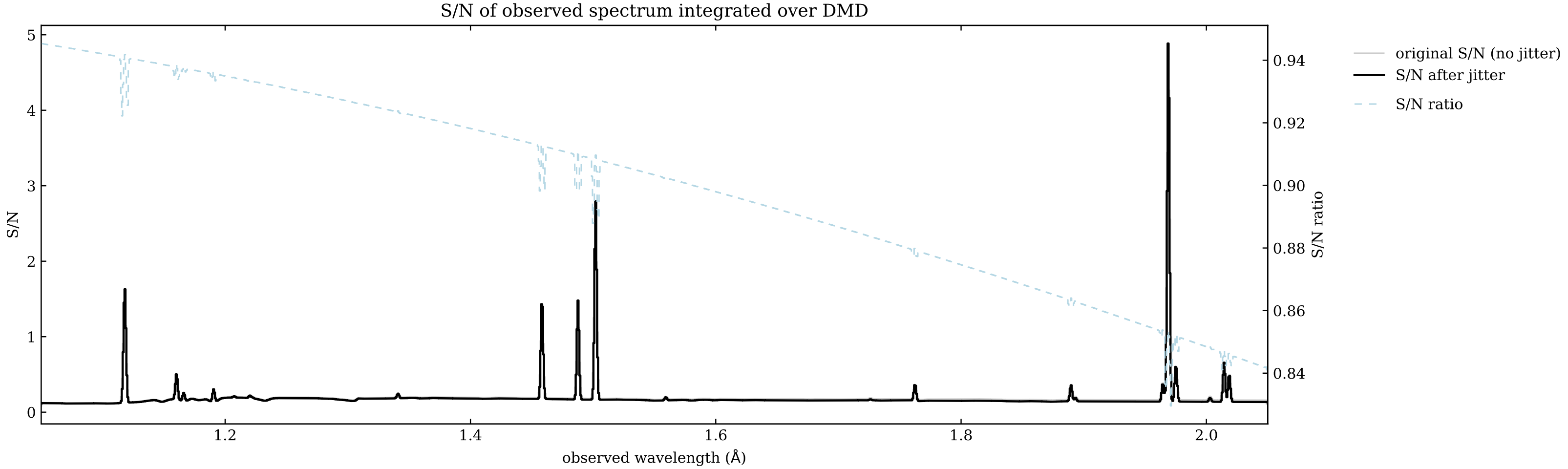}
    \caption{The final output S/N as a function of wavelength including pointing uncertainties (black; left $y-$axis) and the difference in S/N to an observations with no jitter (blue dashed; right $y-$axis). In this example, the S/N drops from $\sim95\%$ (w.r.t. perfect pointing) at the bluest to $\sim85\%$ at the reddest wavelength due to increased DMD micro-mirror aperture loss (i.e., slit loss) at longer wavelengths.
}
    \label{fig:jitter-box}
\end{figure}

We have simulated the decrease in S/N of the observations due to DMD micro-mirror aperture loss (i.e., slit loss) and the pointing uncertainty (a.k.a. ``jitter'') of \iscea. The aperture loss is due to the finite size of a micro-mirror (i.e., slit, $2.8^{\prime\prime} \times 2.8^{\prime\prime}$), and the jitter is caused by pointing instabilities after slews and during exposures, resulting in a shift of the source towards the edges of the micro-mirror.
In order to simulate both of these contributions to the S/N decrease, we have implemented the following model:
\begin{itemize}
    \item[(1)] The source is defined by a Gaussian spatial profile, and an input model spectrum.
    \item[(2)] A spectral cube is created.
    \begin{itemize}
        \item[(i)] For each wavelength, the input source profile is convolved with the corresponding wavelength-dependent optical point spread function (PSF)\footnote{We assume the diffraction limited case in which ${\rm FWHM(PSF)}~\propto \lambda/D$, where $D$ is the diameter of the telescope, see also \S\ref{sec:mission_req}.} to get an image at that wavelength. Fig.~\ref{fig:PSF} shows the convolved source at two different wavelengths (corresponding to \oii and H$\alpha$ wavelengths at $z=2$). The source is bigger at longer wavelengths because of the larger PSF.
        \item[(ii)] The image is then scaled by the flux at the corresponding wavelength.
        \item[(iii)] Noise is added to each of the layers of the cube according to the S/N-wavelength dependence from the ETC (\S\ref{sec:mission_req})
    \end{itemize}
    \item[(3)] The simulated spectrum including jitter is computed following the steps below and iterated over all jitters:
    \begin{itemize}
        \item[(i)] A DMD micro-mirror (i.e., slit)  of $2.8^{\prime\prime} \times 2.8^{\prime\prime}$ is placed on the source. The center of the slit is varied in each iteration according to the pointing uncertainty (see illustration in left panel of Fig.~\ref{fig:jittered}). The source flux is integrated in each of these "jitter boxes" at all wavelengths to obtain a full spectrum.
        \item[(ii)] A wavelength shift is applied to the spectrum as light falls on different detector pixels in the dispersion direction for each jitter position. This simulates the decrease in spectral resolution due to jitter.
        \item[(iii)] At the end of the iterations, the spectra obtained from the individual jitter positions in (i) and (ii) are summed up and combined to form the final spectrum (representing the total integration).
    \end{itemize}
    \item[(4)] The resulting wavelength-dependent S/N of the simulated spectrum including jitter is computed and compared to the S/N obtained without jitter (Fig.~\ref{fig:jitter-box}).
\end{itemize}

We have assumed a jitter characterized by a FWHM of 2$^{\prime\prime}$ and a Gaussian distribution that is sampled every 200 seconds (a 2000 second integration time would therefore have 10 jitters in this simulation). Note that the actual jitter might be closer to a random walk than a Gaussian distribution. More detailed modeling of the satellite's motions will be necessary to estimate the true pattern.
Furthermore, we assumed an ``ideal'' source defined by a Gaussian spatial profile with a FWHM of 0.3$^{\prime\prime}$ and an SED with $F(H\alpha) = 3 \times 10^{-17}\esc$, normalized to an H-band magnitude of 25 AB at $z = 2$. This represents a typical faint emission-line galaxy (galaxy \#6 from Table \ref{tab:spec}).
We assume the established ETC parameters (see \S\ref{sec:mission_req}) and an exposure time of $371\,{\rm ks}$, the required integration time to reach S/N=5 for the H$\alpha$ line flux limit of $3 \times 10^{-17}\esc$ in the absence of pointing jitter (see Table~\ref{tab:obs-time}).

Fig.~\ref{fig:jitter-box} shows an example of the final output S/N (black solid line) and the S/N difference compare to a simulation without jitter (blue dashed line). Note the difference in S/N as a function of wavelength due to the DMD micro-mirror aperture loss and pointing jitter. In this case, the S/N is decreased to about 95\% at bluest and 85\% at reddest wavelengths. The differential decrease in S/N as a function of wavelength is due to the wavelength-dependent size of the PSF, which leads to an increase in aperture loss at the red end of the spectrum.

We find that the decrease in spectral resolution due to pointing jitter is similar at all wavelengths and does not change significantly with the properties of the jitter. Specifically we find a $\sim6\%$ decrease in spectral resolution measured on the H$\alpha$ and \oiii~ lines. In the \iscea\  instrument design, we have included a conservative $10\%$ increase in spectral resolution in order to have margins.

\begin{figure}
    \centering
    \includegraphics[width = 6in]{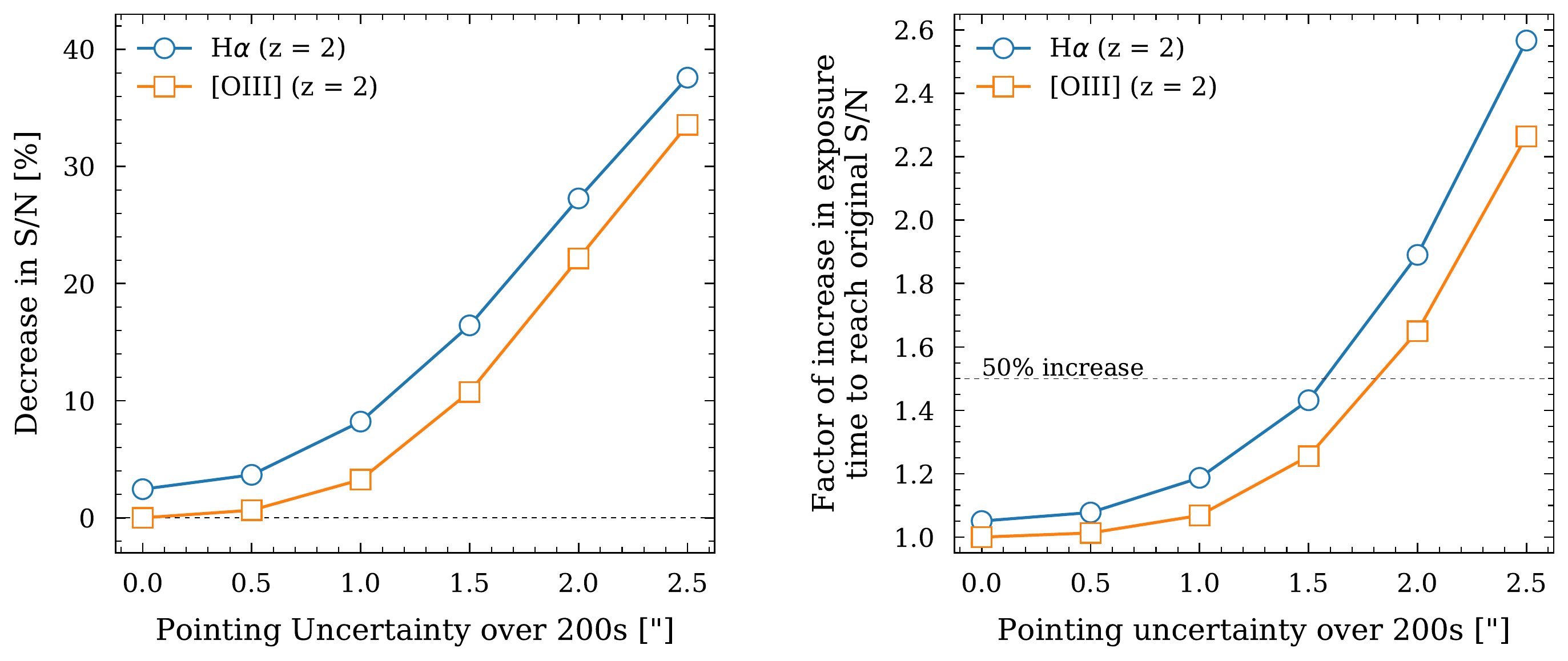}
    \caption{\textit{Left}: Decrease in S/N (in per-cent) as a function of FWHM pointing uncertainty (i.e., jitter) for \oiii~(orange) and H$\alpha$ (blue). \textit{Right}: the factor of increase in exposure time to balance out the effects of pointing uncertainty (i.e., to reach the original S/N without slit loss and pointing uncertainty) as a function of FWHM pointing uncertainty. }
    \label{fig:jittered-SNR}
\end{figure}

\begin{figure}
    \centering
    \includegraphics[width = 5in]{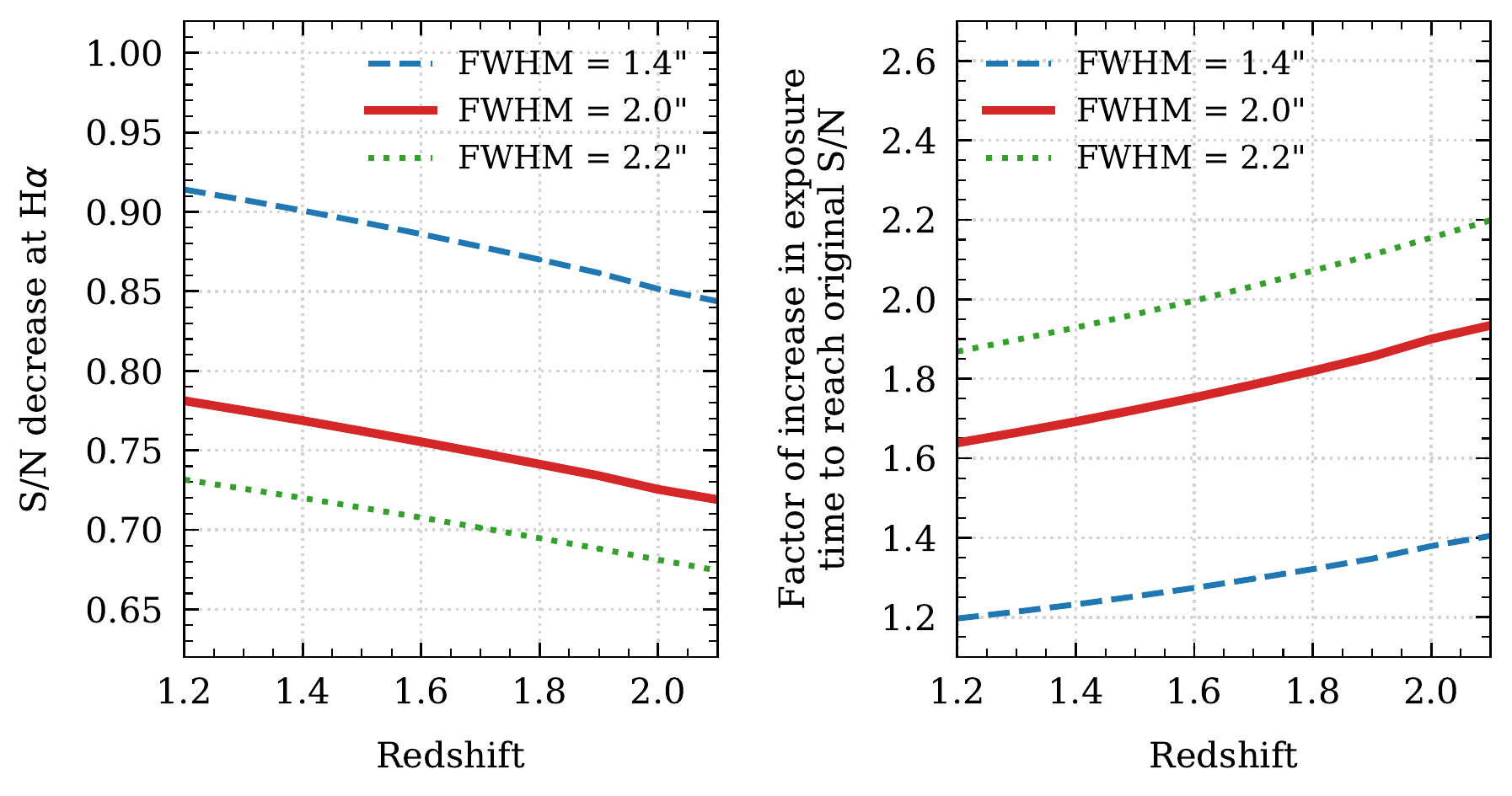}
    \caption{\textit{Left}: S/N decreases due to pointing jitter as a function of target redshift for the H$\alpha$ emission line.
    \textit{Right}: Factor of increase in exposure time to balance out the effects of pointing uncertainty, as a function of target redshift for H$\alpha$. Three different pointing jitter performances (in units of FWHM over 200s) are shown: 1.4$^{\prime\prime}$ (dashed lines), 2$^{\prime\prime}$ (solid lines), and 2.2$^{\prime\prime}$ (dotted lines). For the 100 \iscea\ candidate protoclusters, the mean of the required factor of increase in exposure time is $1.8$ for a 2$^{\prime\prime}$ FWHM pointing jitter over 200s, and $<2.1$ for 2.2$^{\prime\prime}$ FWHM pointing jitter over 200s. }
    \label{fig:jittered-SNR-vs-z}
\end{figure}

Fig.~\ref{fig:jittered-SNR} and Fig.~\ref{fig:jittered-SNR-vs-z} summarize the main results of our study on S/N decrease due to DMD micro-mirror aperture loss and pointing jitter.
The left panel of Fig.~\ref{fig:jittered-SNR} shows the decrease in S/N (in percentage) as a function of pointing uncertainty for \oiii~ (orange) and H$\alpha$ (blue)\footnote{Note that the effect is larger for H$\alpha$ as it is at redder wavelengths thus is affected more by DMD micro-mirror aperture loss.}. The right panel of Fig.~\ref{fig:jittered-SNR} shows the factor of increase in exposure time to balance out the effects of pointing uncertainty (i.e., to reach the original S/N without aperture loss and pointing uncertainty) for both lines. 
The left panel of Fig.~\ref{fig:jittered-SNR-vs-z} shows $S/N$ decreases due to pointing jitter as a function of target redshift for the H$\alpha$ line. The right panel of Fig.~\ref{fig:jittered-SNR-vs-z} shows again the factor of increase in exposure time to balance out the effects of pointing uncertainty for the faintest galaxies to reach the original S/N without jitter, as a function of target redshift for H$\alpha$. Three different pointing jitter performances (in units of FWHM over 200s) are shown: 1.4$^{\prime\prime}$ (dashed lines), 2$^{\prime\prime}$ (solid lines), and 2.2$^{\prime\prime}$ (dotted lines).
\iscea\  baseline mission includes additional observing time to compensate pointing jitter, for the faintest galaxies to reach S/N=5 (see Table \ref{tab:obs-time}). 
{\bf For the 100 \iscea\ candidate protoclusters, the mean of the required factor of increase in exposure time is $1.8$ for a 2$^{\prime\prime}$ FWHM pointing jitter over 200s, and $<2.1$ for 2.2$^{\prime\prime}$ FWHM pointing jitter over 200s. }

Note that the necessary exposure time needed to reach the original S/N is set by the reddest emission line we want to observe. For $z = 2$, the S/N loss in H$\alpha$ is the conservative maximum (at lower redshifts, H$\alpha$ would be bluer, hence less affected by DMD micro-mirror aperture loss due to smaller PSF). 
Note that two contributions to S/N decrease are simulated: 1) DMD micro-mirror aperture loss (i.e., slit loss),  and 2) S/N loss due to pointing jitter. The former results in a constant S/N decrease depending on the size of the source and the PSF. This value for DMD micro-mirror aperture loss can be read off at a jitter FWHM of $0^{\prime\prime}$; it is a $\sim3\%$ S/N loss for H$\alpha$ for our galaxy model (see assumptions above). On the other hand, due to the smaller PSF at bluer wavelength, the \oiii~ emission line is less affected by DMD micro-mirror aperture loss. The second contribution decreases the S/N for increasing pointing jitter.

\subsection{\iscea\  Galaxy Radial Velocity Measurements}
\label{sec:vpec}

Naively, $R=1000$ spectroscopy gives $(1+z)\Delta v/c = \Delta \lambda/\lambda = 0.001$, so $\Delta v = 300/(1+z)\,\kms$. However, each spectroscopic resolution element is sampled by two pixels, and the actual redshift precision of an ELG is determined by how well the emission lines can be centroided on each pixel. If the line is unresolved (but well sampled) and relatively high SNR, 1/5 pixel is a good estimate for line centroiding accuracy. This gives a factor of ten improvement in $\Delta\lambda$ per spectroscopic resolution element for a high SNR line. Allowing for lower SNR, we can expect to measure individual bright lines to $\sim$ 30-50 $(1+z)^{-1} \kms$. 
There will be errors on this because of the brightness of the galaxies themselves, effects due to things other than galaxy motion (e.g. AGN, winds, etc).

Therefore the \iscea\  redshift precision will be $\sim 1-2\times 10^{-4}$. 
The measured redshift 
\begin{equation}
z_{\rm obs}=z_{\rm cos}+z_{\rm pec},
\end{equation}
where $z_{\rm cos}$ is the cosmological redshift, and $z_{\rm pec}$ corresponds to the peculiar radial velocity of an individual galaxy $v_{\rm pec}$.
For clusters embedded within the protocluster environment, the peak of the distribution in $v_{\rm pec}$ corresponds to the mean velocity of the cluster.  
The $v_{\rm pec}$ of galaxies in a protocluster field relative to this mean velocity can be used to identify the members of the cluster, as well as filaments and groups associated with the protocluster.

\subsection{\iscea\  Observing Efficiency}
\label{sec:efficiency}

\iscea's observing time per field of 668ks includes an overhead of 80\% (see Table \ref{tab:obs-time}) to compensate the loss in light from the galaxy targets due to pointing jitter of $2^{\prime\prime}$ FWHM over 200s (see \S\ref{sec:jitter}), so that each measured spectrum has a signal-to-noise ratio $\ge 5$ at the \iscea\  H$\alpha$ line flux limit (see \S.\ref{sec:mission_req}). Therefore the pointing jitter effect is already mitigated, and does not lead to a decrease in \iscea\  observing efficiency. The detailed modeling of subtle effects due to the DMD will be carried out during the Phase A study (see \S\ref{sec:inst_sys}).

We conducted simulations of galaxy target selection by the DMD over multiple visits of a typical \iscea\  target protocluster field, to estimate the fraction of target galaxies that will be observed. For each visit, we prioritize the faintest galaxies (which need to be observed the longest) to be placed on an ``open'' micro-mirror. Galaxies that reach the required S/N$\geq$5 in their spectra (identified via quick data processing and analysis by the SOC, see \S\ref{sec:data_proc}) are then replaced by the next brightest galaxies. This simple observing strategy ensures the most uniform observations.  In the following, we show the results of our pointing (i.e., targeting) simulations for a protocluster field at $z\sim 2$, centered on cluster candidate J123345.3-002945 from the \cite{Wen2020} catalog. The results are very similar for other protocluster fields in our target sample as the density of galaxies is similar in them.

\begin{figure}
    \centering
    \includegraphics[width = 0.52\textwidth]{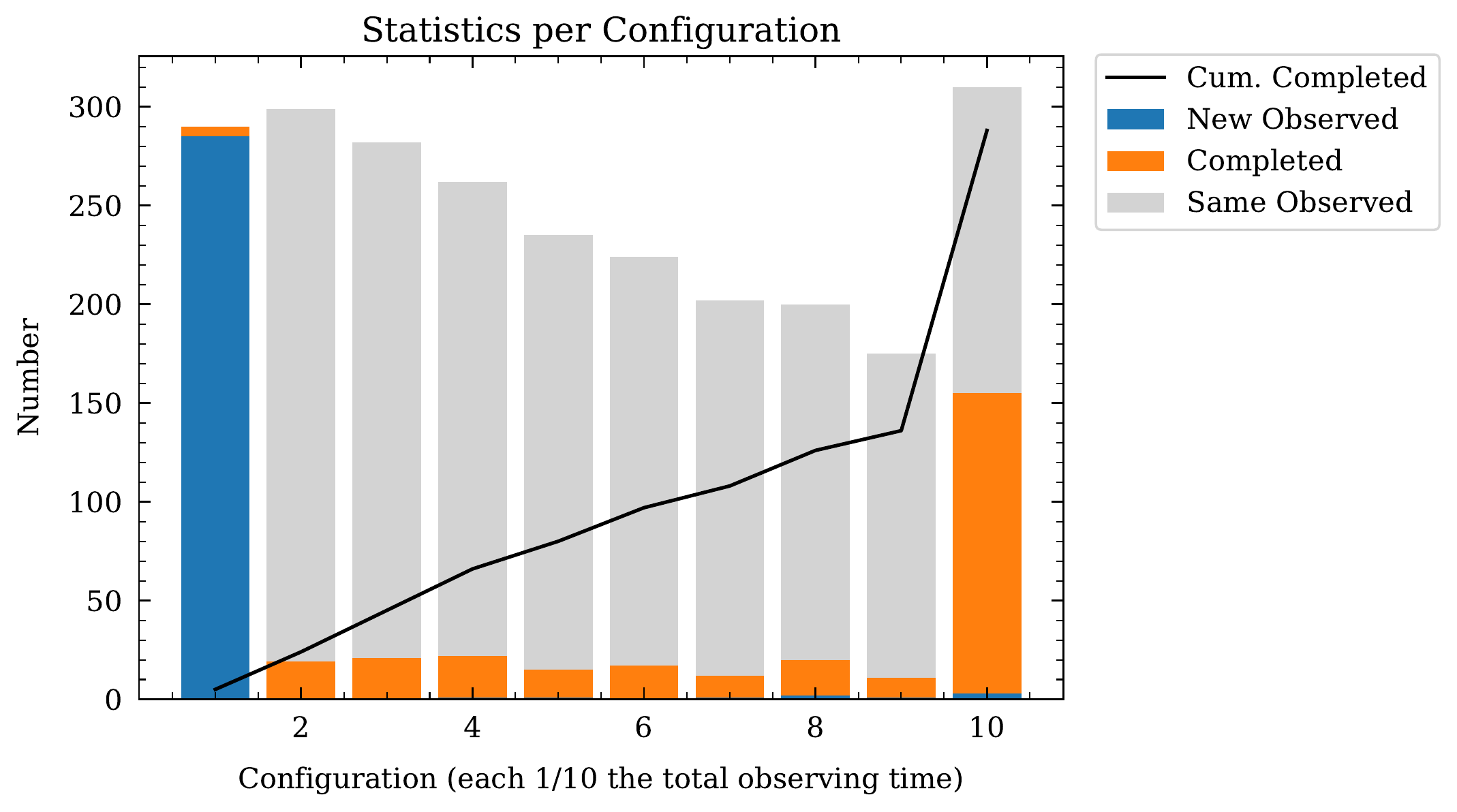}
    \includegraphics[width = 0.44\textwidth]{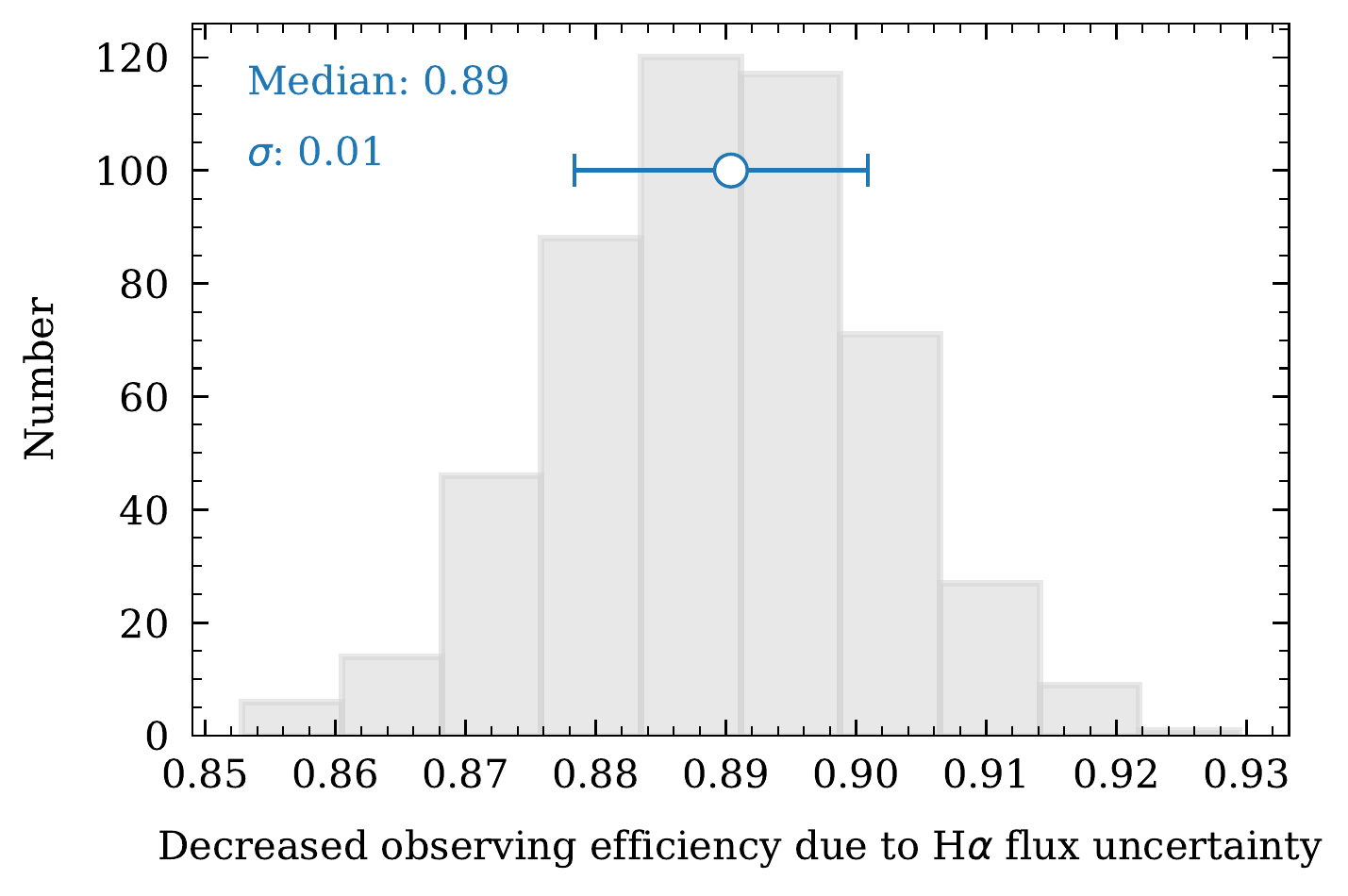}  
    \caption{\textit{Left}: Monte Carlo simulation results of 10 \iscea\  spectroscopic visits to the protocluster field at $z\sim 2$, centered on cluster candidate J123345.3-002945 from the \cite{Wen2020} catalog, 
    the same target field as shown in Fig.~\ref{fig:gal_targets}. Each visit is represented by a bar, with new, continued, and completed observations denoted by blue, gray, and orange, respectively (note that roughly all observations in the 10$^{\rm th}$ visit are completed $-$ hence the similar size of the orange and gray bars). The black curve shows the cumulative number of targets observed as a function of the number of visits. After 10 visits, about $90\%$ of the galaxies are observed.
    \textit{Right}: Decrease factor in the observing efficiency of galaxies above the \iscea\  flux limit due to a factor of 2 uncertainty in the estimated H$\alpha$ line flux, which is used for estimating the \iscea\ observing efficiency.}
    \label{fig:simout}
\end{figure}

By dividing the total observing time per field into 10 visits (each with a total combined exposure time of $66.8\,{\rm ks}$), \iscea\  can observe $\sim 90\%$ of the targeted $\sim 300$ galaxies within $\Delta z_{\rm photo} =0.2$ of the BCG (Fig.\ref{fig:simout}, left panel)\footnote{This number is very similar in the other protoclusters.}. This result is relatively insensitive to the actual number of visits, as long as the number of visits is greater than 3. The main reason for this is because the spatial density of available micro-mirrors (i.e., ``slits'') is significantly higher than the expected number of selected galaxies. The remaining $\sim10\%$ of galaxies are mainly bright galaxies for which less than $\sim600\,{\rm ks}$ of observations are needed to achieve the desired S/N. These galaxies are not observed in favor of completing the observations of fainter galaxies. However, their brightness, hence relatively short observation times, would make them possible targets to follow-up in future observations.

Since \iscea\  can obtain $\sim1000$ spectra simultaneously, we can utilize the unused micro-mirror columns by targeting additional galaxies. Specifically, we will be able to add $\sim700$ galaxies to the target list per field with photo-$z$'s closest to the BCG spectroscopic redshift, but with $\Delta z_{\rm photo} >0.2$, to supplement the list of $\sim$ 300 galaxies with $\Delta z_{\rm photo} \leq 0.2$.
Note that this leads to a completeness in targeting protocluster galaxies of $\sim$100\%. This is because \iscea\  observes $>1000$ galaxies closest in photo-z to the cluster BCG, including $\sim$ 500 within $\Delta z_{\rm phot}=0.33$ of the confirmed cluster BCG, which is $3\sigma$ given the $1\sigma$ photo-$z$ scatter is $\Delta z_{\rm phot} \sim 0.11$ at $z \sim 2$.

Since we do not know the H$\alpha$~flux of the targeted galaxies \textit{a priori}, we compute expected fluxes based on their estimated stellar masses and star formation rates \citep[e.g.,][]{Daddi2007}. The resulting H$\alpha$ flux estimates are uncertain by a factor of $\sim2$ that potentially could cause us to miss faint galaxies, which decreases the true sample completeness. We simulated this effect by a Monte Carlo sampling, thereby changing the obtained fluxes according to this expected uncertainty and re-selecting the sample 500 times. The right panel of Fig.~\ref{fig:simout} shows the resulting distribution of the factor by which the completeness is decreased. We find a median of $89 \pm 1\%$.

In summary, the \iscea\  observing efficiency includes three factors:

\begin{itemize}
    \item[(1)] The fraction of target galaxies observed by dividing the total observing time per field into multiple visits: $\sim 90\%$ (Fig.\ref{fig:simout}, left panel).
    \item[(2)] The decrease factor in the observing efficiency due to the factor of 2 uncertainty in the estimated H$\alpha$ line flux: $\sim89\%$ (Fig.\ref{fig:simout}, right panel).
    \item[(3)] The completeness in targeting protocluster galaxies: $\sim100\%$.
\end{itemize}

Thus the overall \iscea\  efficiency of observing a protocluster field is therefore $\sim80\%$ ($\sim 0.9\times 0.89$).

\section{Data Acquisition}
\label{sec:data_acq}

\subsection{Pointing Correction}
\label{sec:pointing-cor}

\begin{figure}
    \centering
    \includegraphics[width = 3in]{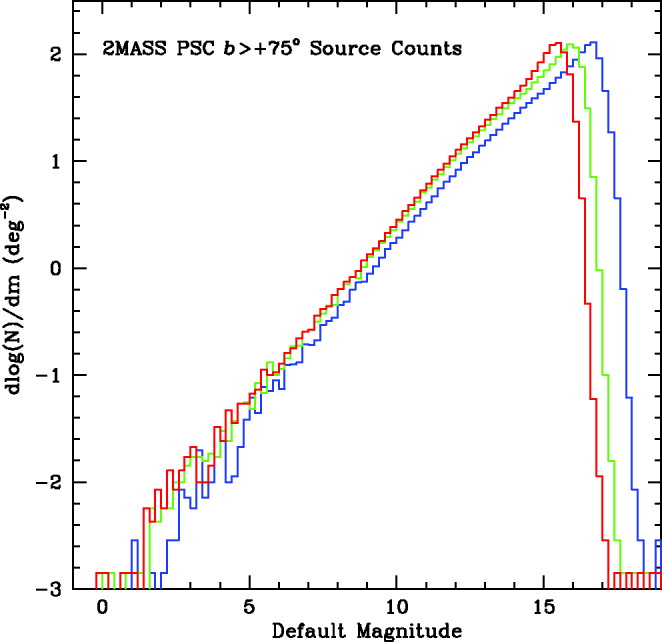}\hskip 0.1in
    \caption{The 2MASS Point Source Catalog source counts (per 0.2 mag bin) for several hundred square degrees near the north Galactic pole \citep{Skrutskie2006}. Blue, green, and red histograms represent $J$, $H$, and $K_s$ Vega mag counts, respectively. There are plenty of bright stars available for \iscea\  pointing correction in each FoV.}
    \label{fig:2mass_stars}
\end{figure} 

To meet the \iscea\  pointing requirement of $\leq 2^{\prime\prime}$ FWHM over 200s, the \iscea\  instrument includes an imaging channel which images bright stars nearly continuously to enable pointing corrections.
Fig.\ref{fig:2mass_stars} shows the differential star counts from the 2 Micron All Sky Survey (2MASS) at $b>+75^\circ$ \citep{Skrutskie2006}, which provides a lower bound since the star counts increase at lower Galactic latitudes.
We expect to find $\ga 10$ stars brighter than Vega H mag 13 in each \iscea\  FoV for the \iscea\  target fields (see Fig.\ref{fig:clusterdist}), which can be used for guiding and pointing corrections.

\subsection{Calibration}
\label{sec:calibration}

The \iscea\  spectral calibration involves several basic steps. These include transforming the relative astrometry between direct and dispersed images (to allow for accurate slit placement of targets from direct images), determination of the spectral shapes (the size and curvature or spectral trace) as a function of position, the wavelength zero-point, the dispersion solution, the line spread function, and the relative and absolute flux calibration of the spectra over the full field of view. With this information, the pipeline software will extract and measure the effective spectrum of each targeted source. Calibration parameters will be applied to the 2D and 1D extracted spectra through the use of a number of calibration reference files in the \iscea\  pipeline.  A full set of initial versions of these files will be created before launch, based on laboratory measurements.  Once in orbit, the files will be updated during initial checkout, and then periodically throughout the mission as needed, based on routine data quality assessment and/or changes to the in-flight system.

The \iscea\  mission calibration plan follows a well-established methodology.
The team calibrates individual detector pixels by targeting open clusters with sufficient stars to fill the \iscea\  FOV, following a strategy similar to that used for the IR channel of WFC3 on HST \citep{Sabbi2010}. Baseline flat-field calibrations will be first obtained on the ground; the
extensive set of data will allow the removal of high-frequency spatial variations due to the detector response (P-flats). Low frequency modulations related to the optical system response (L-flats) will be corrected through pointed on-sky observations, targeting rich stellar clusters like Omega Centauri (NGC 5139) or NGC 2516, to compare the same flux of the same stars at different field positions. To achieve high sigmal-to-noise in relatively short, but typical, individual \iscea\  spectroscopic observations of about 200 sec, requires bright standard stars of 10 - 14 AB mag. 
Multiple, dithered, 100-200 sec exposures of the clusters with sufficient (tens to hundreds) stars per \iscea\  FoV, will allow for precise relative photometric calibration across the FoV.  Absolute flux calibration will target spectrophotometric standard stars used by WFC3/IR, including both white dwarfs and G-type stars \citep{Bohlin2020}. Dispersion solutions will be calibrated on the ground and verified on-orbit, using well characterized NIR spectra of compact planetary nebulae \citep{Lumsden2001} and K and M stars via Hydrogen Paschen, Brackett and other emission and absorption features throughout the \iscea\  wavelength range. Unresolved emission lines will also be used to update the \iscea\  line spread function as a function of wavelength. The spectral trace (curvature and shape vs. wavelength) can be measured by any bright point source, so the star cluster observations will also provide a full mapping of the trace across the FoV, important for accurately extracting the 1D spectra from the 2D spectral images. Fainter stars within each \iscea\  target field will be used to align the slits (i.e., the DMD micro-mirrors), providing the relative astrometry between image and spectra. There are $\sim$ 1000 stars at AB < 17 per square degree that can be used for slit positioning (see Fig.\ref{fig:2mass_stars}).

{\bf Preflight}: The SOC will prepare a list of $\sim 100$ isolated spectrophotometric standards spread across the sky, along with a list of bright PNe and open clusters for use in wavelength and relative photometric and trace calibrations for commissioning and during the science phase of the mission.  In addition, they will select the expected stars in each target field suitable for slit alignment. 

The most critical calibration is the mapping between DMD micro-mirrors and the pixels of the imaging and spectroscopic detectors. This is routinely done creating a regular pattern of bright spots on the DMD and imaging their position on the detectors. The team validates the baseline maps obtained on the ground with on-sky pointing on bright extended regions, e.g., the Orion Nebula, with the same regular DMD patterns. Since there are no moving parts in the optical systems, we expect the solutions to remain stable and with only sporadic checks required. We can easily automate the procedure with minimal impact on the mission efficiency.

Both the spectrometer and imager assemblies are calibrated before integration with the telescope using SwRI facilities in San Antonio to characterize plate scale and vignetting, optical image quality, focus and opto-mechanical alignment, relative spectral response, and pointing dynamic performance. Once integrated with the telescope, end-to-end Instrument Performance Tests (IPT) are performed to verify end-to-end opto-mechanical alignment, and radiometric performance.

{\bf Inflight}: Wavelength and relative and absolute photometric standards will be taken as described above.  In addition, a 100 sec imaging only exposure will be taken for each primary protocluster target, to verify the positions of all galaxies with respect to the stars within the field. This enables the accurate selection of galaxies for spectroscopy. The parallel images of the spectroscopic science data will be downlinked periodically for calibration and verification. On orbit commissioning activities include: 1) dark count measurement before opening the aperture cover, 2) DMD imager to spectrograph mapping, 3) NIR wavelength calibrations using stellar sources, and 4) sensitivity and pointing calibrations by observations of NIR calibration stars.  Suitable NIR photometric and spectroscopic calibration targets are available throughout the sky; calibration observations are made monthly for the first 6 months and bi-monthly thereafter to monitor instrument performance. 

\subsection{Science Observations}
\label{sec:obs}

Upon the completion of initial calibrations in orbit, \iscea\  interleaves target verification and science observations in Year 1. It carries out 1 hour imaging per field over 1.1-2$\mu$m to AB=25 at S/N=5, and 4 hour spectroscopy per field to continuum AB=19.5 at S/N=3 (emission line flux limit of $2.2\times 10^{-16}\esc$ at S/N=5) for 100 candidate protocluster fields, to obtain the redshifts of the already identified Bright Cluster Galaxy (BCG) and $\sim$1000 galaxies closest to it in photo-z in each. Each protocluster is confirmed using spectroscopic redshifts of $>10$ brightest galaxies including the BCG. \iscea’s final target list of 50 confirmed protocluster fields at $1.2<z<2.1$ is based on the imaging and fast spectroscopy of the 100 candidate protoclusters, including 45 confirmed at $1.7<z<2.1$, and 5 at $1.2<z<1.7$. 

All galaxies from \citealt{Wen2020} catalog are brighter than AB mag 24 in the \iscea\  $1.1-2.0\,{\rm \mu m}$ broadband (see Fig.\ref{fig:1000nearest-z}, right panel). \iscea\  broad-H band imaging fills in an important gap in imaging coverage from HSC (optical) to WISE (3.4$\mu$m), and helps improve photo-$z$ accuracy (see \S\ref{sec:photo-z}). 

In each BCG-confirmed protocluster field, we select $\sim$ 1000 galaxies with photo-$z$’s closest to the spectroscopically confirmed BCG redshift as spectroscopic targets. We estimate that in each protocluster target field, there are $\sim$ 210 galaxies with H$\alpha$ line flux above $3\times 10^{-17}\esc$ in the cluster or its cosmic web environment, based on the \iscea\  mock (see \S\ref{sec:count-gal}). The photo-$z$ based “blind” selection enables us to target ELGs as well as passive galaxies. Most of these are expected to be ELGs. The quiescent galaxies can be identified using the non-detection of emission lines at the \iscea\  3$\sigma$ H$\alpha$ line flux limit of $1.8\times 10^{-17}\esc$. The fraction of quiescent galaxy is an important indicator of galaxy evolution. 

The galaxy target list for each \iscea\  protocluster field will be prepared and tested before its scheduled observation by \iscea. The galaxy target lists for the first batch of protocluster targets will be ready before the prime mission begins. The SOC will develop and test the data processing pipelines during the year before launch.

Fig.\ref{fig:DMD_layout} illustrates how \iscea\  will select galaxy targets for multi-slit spectroscopy. The 1020$\times$510 micro-mirrors on the DMD maps to 2040$\times$1020 pixels on the 2048$\times$2048 detector, providing a FoV of $48^\prime \times 24^\prime$, with $2.8^{\prime\prime}\times 2.8^{\prime\prime}$ per DMD micro-mirror. 
Only 2040$\times$2040 of the 2048$\times$2048 pixels are active; the 4 rows of pixels on each edge are reference pixels.
By switching on 1 micro-mirror in each of 1020 micro-mirror columns on the DMD, \iscea\  will measure $\sim$ 1000 non-overlapping spectra simultaneously. 
Since each spectrum is $\sim$ 1161 pixels long on the H2RG detector, the top 70 micro-mirrors correspond to incomplete spectral coverage, with 5.5\% of the spectral coverage missing for the topmost micro-mirror (worst case). The bottom 440 micro-mirrors correspond to complete spectral coverage on the detector. Since both \oiii~and \oii~are captured on all galaxy spectra at $1.2<z<2.1$, this should have no impact on redshift measurements. We expect this to have no impact on the cluster observations, and a very small impact on the observation of the cosmic web filaments. We will mitigate this by tiling the protocluster field for representative protoclusters to achieve uniform spectral coverage, and model the systematic effects from the incomplete spectral coverage for the other protocluster fields. During Phase A, we can carry out a trade study of adding a second H2RG detector in the spectroscopic channel, so that all 1040$\times$768 available micro-mirrors on the DMD correspond to complete spectral coverage, expanding the FoV to $48^\prime \times 36^\prime$. 

\section{Data analysis}
\label{sec:data_ana}

\subsection{Data Processing}
\label{sec:data_proc}

Data will be retrieved from the spacecraft once every 24 hours to enable the updating of the target list. New target lists will be uplinked at least once per week. The raw data volume is 56G bits per 24 hours of observing time without compression, with an overhead of < 10\% for calibration data (both complete and parallel images).  

The \iscea\ data will be processed through two, automated pipelines - one for science data and one for calibration observations. In-orbit calibration data, as described in \S6, will be processed through the calibration pipeline to produce calibration reference files which are used by the science data pipeline. Production and testing of the calibration reference files, along with the design and implementation of the science and calibration pipelines, will be the responsibility of the SOC. An initial set of calibration reference files will be produced by the SOC from the ground test data, updated during in-orbit commissioning and initial calibration. Subsequent updates will occur when changes to the flats, wavelength zero points or dispersion solutions, etc. are determined via regular and periodic calibration observations, as outlined in \S6, and when reprocessing of the \iscea\ science data is necessary or planned as part of regular reprocessing during the mission.  The pipeline executives, responsible for executing and managing data flow, as well as the individual pipeline software modules used to generate L1-L4 data, will be under version control at the SOC during the mission.

\begin{figure}
    \centering
    \includegraphics[height=4in]{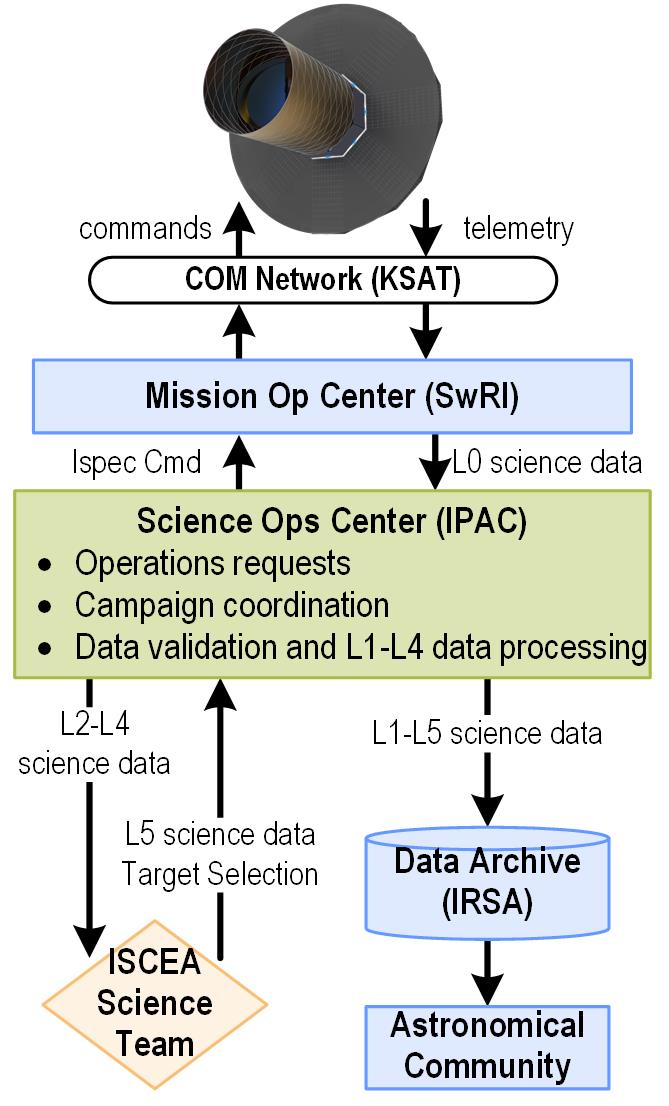}
    \caption{\iscea\  data flow diagram.
    }
    \label{fig:data_flow}
\end{figure}

Basic processing steps such as dark current subtraction, relative and absolute flux calibration, source identification and  extractions will be common to all science and calibration data, although details of how these steps are applied are specific to the type (images or spectra) of data moving through the pipelines. For example, as described in \S6, the \iscea\ spectral calibration involves several steps, such as transforming the relative astrometry between direct and dispersed images, determination of the spectral trace as a function of position on the array, measurement of the wavelength zero-point and dispersion solution, and the line spread function, and  relative and absolute flux calibration of the spectra over the full field of view to produce the L2-L4 data. The necessary calibration parameters will be applied to the 2D and 1D extracted spectra through the use of the calibration reference files (themselves a combination of 2D image files and lists of coefficients) in the \iscea\  science pipeline, necessary for the production of L2 through L4 science data.  In addition to the science and calibration data, the calibration reference files used by the pipeline will be delivered to the archive.  The pipelines will generate regular, automated Data Quality Assessment (DQA) reports for the science data (L1-L4) which will be analyzed by the SOC and \iscea\ science team, and delivered to the IRSA archive. Fig.\ref{fig:data_flow} shows the \iscea\  data flow. Table \ref{tab:data_prod} shows
the data levels and associated processing.

\begin{table}
\begin{tabular}{|l|l|}
\hline
    {\bf Level}	& {\bf Data Level Definitions and Processing}\\ \hline
L0	& Packetized data \\ \hline
L1	& Uncalibrated FITS files. Includes meta and engineering data \\ \hline
L1 to L2 &	Removal of basic detector signatures and artifacts. Application of flux calibration to 2D data. \\ \hline
L2 &	Basic calibrated 2D images and spectral data (detector signatures removed and flux calibrated images) \\ \hline
L2 to L3 &	Extraction of 1D spectra. \\ \hline
L3 &	Extracted 1D spectral data \\ \hline
L3 to L4 &	1D spectral fits to emission and absorption features. \\ \hline
L4 &	Basic spectral fit parameters (redshift, flux, line width) and associated catalogs \\ \hline
L4 to L5 &	Derivation of SFR, ages, SF histories and AGN strengths as a function of mass, local environment and redshift \\ \hline
L5	& High level data products and catalogs \\ \hline
\end{tabular}
\caption{\iscea\  data levels and associated processing.}
\label{tab:data_prod}
\end{table}

\subsection{Modeling Instrument Effects}
\label{sec:inst_sys}

The size of the individual DMD micro-mirrors, about $13.0~\mu$m$\times13.0~\mu$m ($13.7\mu$m$\times13.7~\mu$m center-to-center) is only a few times larger than the wavelengths of interest for \iscea. Therefore, besides reflection, some diffraction and scattering effects will be present. Spreading the reflected light into a larger solid angle than the nominal one (given by the f/\# of the input beam onto the DMD) causes a loss of light, as the collecting optics act as an aperture stop. This may be regarded as equivalent to the slit losses typical of conventional slit spectrographs, with two exceptions: a) some light diffracted from other mirrors may enter the beam, creating an extra background that reduces the "contrast" of the system; b) the diffraction pattern is characterized by lobes and their spacing depends on the wavelength. They modulate the fraction of light captured by the collecting optics, introducing a color term that may be significant affecting both throughput and point spread function.

These effects can be modeled treating the DMD as a blazed diffraction grating and calculating the far-field diffraction pattern. However, an accurate estimate of the DMD efficiency and contrast must take into account the geometry of the system, the degree of coherence of the illuminating source, its position vs. the center of a mirror, polarization effects, etc. Analyzing these effects requires a more advanced treatment based on the exact solution of the Maxwell equations for the specific DMD and \iscea\  configuration. The \iscea\  science team is developing a full theoretical model to quantitatively assess the relevance of these effects, together with an optical system to validate the predictions at visible wavelenghts. The model will be used to optimize the opto-mechanical design of \iscea\  during Phase A, enabling the mission to achieve its required sensitivity and contrast with margins.

\subsection{Science Data Analysis}

\subsubsection{Photometric Redshifts and Stellar Population Properties}

We will derive stellar population parameters (stellar masses, continuum dust attenuation, rest-frame colors) for all galaxies observed by \iscea\ using the measured SEDs from optical/NIR/mid-IR photometry combined with the spectroscopic and/or photometric redshifts.  The photometry for our samples spans HSC $grizY$, \wise\ W1 (3.4~$\mu$m) and W2 (4.6~$\mu$m), with additional coverage from 1.1-2~$\mu$m imagine from \iscea\  $H$-band (see Fig.~\ref{fig:SEDfit}).  In addition, the majority of \iscea\  targets fall within the coverage of LSST, where we can expect $ugrizY$ coverage extending nearly 1 mag deeper than HSC over the 10 year baseline of LSST (the 2 year LSST coverage will be comparable to HSC, providing important calibration and the $u$-band coverage).   

We will first update the photometric redshifts for the galaxies in our sample.  
Prior to \iscea\ launch we will use photometric redshifts based upon the photometry existing at that time (HSC + {\it WISE} + possibly Rubin data) to validate the redshift estimates for the candidate protocluster fields. Following launch we will use updated photometric redshifts that incorporate \iscea\ $H-$band photometry to select protocluster fields for spectroscopy and for slit assignment for galaxies within these fields. Finally, we will use the spectroscopically targeted galaxies to quantify any biases and systematics in the photometric redshifts. 

With the photometric and spectroscopic redshift catalogs, we will then model the photometry using standard SED-fitting practices that include a Bayesian modeling formalism with flexible star-formation histories (e.g., \citealt{Leja2017,Leja2019,Carnall2019}).    Based on previous performance with these methods, with the expected photometry we will achieve typical accuracy of stellar mass of $0.2-0.3$ dex (modulo systematic uncertainties in IMF).  In addition, we will derive continuum dust attenuation values to apply to the \iscea\  line-flux measurements, and we will test these against  measurements from the Balmer decrement for those galaxies in \iscea\  with multiple H-recombination lines detected. 

\subsubsection{Spectroscopic Redshifts and Emission line Analyses}\label{section:emission_line_analysis}

\iscea\  will measure redshifts and H$\alpha$ line fluxes for $>$50,000 galaxies, including an estimated $\sim 210\times50=10,500$ protocluster member galaxies (including $\sim 33\times50=1650$ cluster member galaxies), over the redshift range $1.2 < z < 2.1$.  H$\alpha$ is a fundamental tracer of the SFR as it measures the direct number of ionizing photons from OB associations.   We will convert the H$\alpha$ luminosities to   SFRs following \cite{Kennicutt1998,Kennicutt2012}. We will correct these estimates for dust attenuation using several measures.  First, the \iscea\  spectra will cover H$\beta$ for galaxies with redshifts $z \gtrsim 1.3$ (see Fig~23).  This includes nearly all the protoclusters in our sample. We will measure Balmer decrements, H$\alpha$/H$\beta$, for individual galaxies where H$\beta$ is well detected, comparing the Balmer decrement to the theoretical value expected for Case-B recombination \citep{Osterbrock1989}.  For fainter galaxies, and as a cross check of these corrections for all galaxies, we will stack the spectra of galaxies in bins of stellar mass and redshift to measure the average dust attenuation from the Balmer decrement in the stacks.  As the dust attenuation is expected to decrease with decreasing stellar mass and SFR \citep{Pannella2009,Reddy2015}, the uncertainties (and scatter) in the dust attenuation at the lower end of the mass/SFR function will have less impact on our study.  We will also make the first measurement of how dust attenuation varies as a function of environment (over three orders of magnitude in density) at fixed mass and SFR at these redshifts.    Finally, we will also check the dust reddening from the Balmer decrement to independent estimates from the optical photometry (see above), which will cover the UV rest frame wavelengths (and we can also check if the emission lines suffer more attenuation than the stellar continua, see \citealt{Kashino2013, Reddy2015,Valentino2017}). 

The emission line measurements will also provide estimates of the gas-phase metallicity in the \iscea\  galaxies.   We can derive metallicity estimates using H$\alpha$/[\ion{N}{2}] for galaxies with direct detections (\iscea's spectral resolution is sufficiently high for differentiating H$\alpha$ and [\ion{N}{2}]), and we can derive these values for galaxies in stacks of bins of stellar mass, SFR, and environment.   The H$\alpha$/[\ion{N}{2}] is a reliable tracer of metallicity (e.g., \citealt{Kewley2019}), but prone to uncertainties.  However, by using this ratio for all galaxies in our sample, we remove much of the systematic uncertainties, and this has been attempted previously using smaller samples.   
Furthermore, we can compare these values to estimates from \oiii]/H$\beta$.  This will allow us to study the dependence of metallicity on environment, which is a tracer of galaxy gas flows, feedback and metal enrichment.     We can also test how dust attenuation and metallicity relate in these galaxies by studying the relation between the Balmer decrements and H$\alpha$/[\ion{N}{2}] ratios for galaxies (both direct detections and from the stacked spectra).  

\subsubsection{Quantifying the Environment Using the Local Galaxy Density} 
\label{sec:delta}

We will define environmental densities using multiple methods and metrics.  First, we will use friends-of-friends (FoF) algorithms (e.g., \texttt{pyfof}\footnote{\url{https://github.com/simongibbons/pyfof}}) to statistically identify associated galaxies.
Combined with measurements of the distance to the Nth nearest neighbor (including Bayesian methods using the distance to all Nth neighbors, \citealt{Ivezic2005,Kawinwanichakij2017}) or density measurements, we can derive statistical relations between density and structure mass \citep[e.g.,][]{Peng2010,Muldrew2012,Kovac2014,Kawinwanichakij2017}. For objects detected from the ACT and SPT surveys, we can compare the density measurements here with the measurements derived from the SZ signal.   Simultaneously, we can use these measurements to assign to each galaxy a local density estimate which we will use as a proxy for environment using $\delta = (\rho - \langle\rho\rangle)/\rho$, where $\rho$ is the local environment and $\langle\rho\rangle$ is the average density of all galaxies.  These metrics can achieve relative accuracy in $\log(1+\delta)$ of 0.1--0.15 dex \citep{Kovac2010}. 

Applying these density metrics to protoclusters in the \iscea\  mock catalog (see \S\ref{sec:mock}), we find that with \iscea\  we will probe densities over a large range, from cluster cores ($\log (1 +\delta) \sim 2$) to the field ($\log(1 + \delta) \sim -1$), within each \iscea\  field of view.  With the full sample of \iscea\  observations this will enable us to 
then use the derived properties to study properties of galaxies as a function of SFR, stellar-mass and environment. The measurements will be used to test the \iscea\  Science Hypotheses, which we describe below in \S\ref{sec:test-hypo}.  

\section{Data and Investigation Products}
\label{sec:data-prod}

{\bf Quantity and Quality of  Data:} The \iscea\  investigation will return $> 50\times 1000 $=50,000 galaxy spectra, including
$> 50\times 1000 \times 0.77$=38,500 ELG spectra with $S/N \geq 5$, covering the wavelength range of 1.1-2$\mu$m at a resolution of $R=1000$. Note that 0.77 is the fraction of 1000 \iscea\  galaxy targets per protocluster field expected to have H$\alpha$ line flux above the \iscea\  line flux limit (see Fig.~\ref{fig:1000nearest-z-frac} in \S\ref{sec:sel-gal}). 
 	Of these galaxies, $\sim 210\times 50=10,500$ are in the clusters and their cosmic web environments (see \S\ref{sec:count-gal}).
 	The other galaxies observed by \iscea\  in each protocluster field, $\ga 28,000$, are foreground or background galaxies providing valuable information on the evolution of cosmic large scale structure.
Since we select galaxy targets closest to the cluster BCGs in photo-z to utilize all 1000 DMD "slits" for non-overlapping spectra, most of these are at $z>1.7$, as 90\% of \iscea\  protocluster fields are selected to be at $1.7 < z < 2.1$, and the photo-$z$ uncertainty is $\sim$ 0.11 at $z=2$ \citep{Wen2020}.

\begin{figure}
    \centering
    \includegraphics[width = 6.5in]{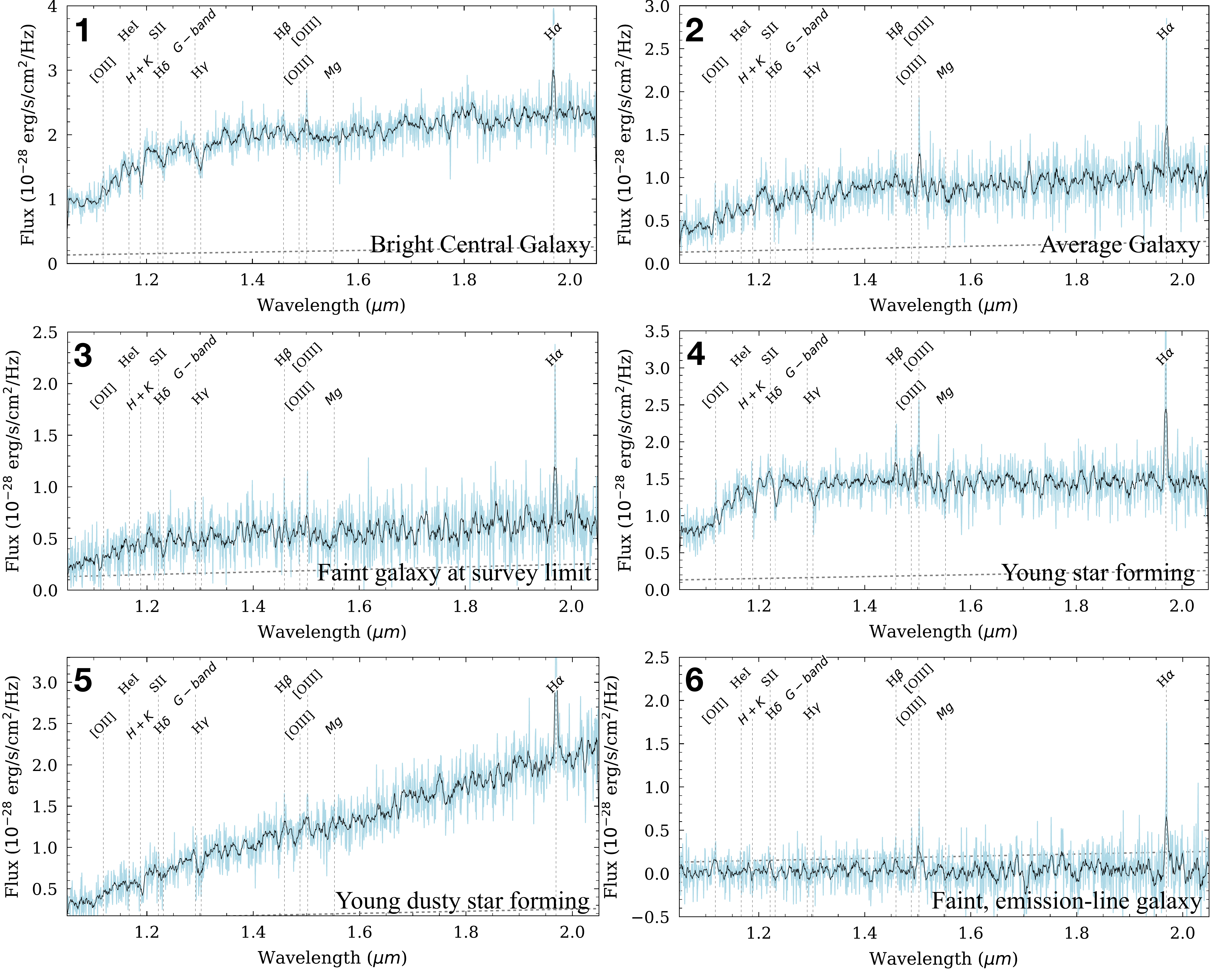}
    \caption{Simulated \iscea\  spectra of different galaxies at redshift $z = 2$ (corresponding to the galaxies listed in Table~\ref{tab:spec}). The different lines show the unbinned spectrum (light blue), the $10\times$ binned spectrum (black), and the sky background (dashed). Prominent spectral features including their names are indicated by the vertical lines. \iscea\  will obtain $>1000$ galaxy spectra per target field.
    }
    \label{fig:jittered-spec}
\end{figure}

Fig.\ref{fig:jittered-spec} shows examples of simulated \iscea\  spectra for $z = 2$ galaxies as tabulated in Table \ref{tab:spec}.  Along with the ELG spectra, \iscea\  will also obtain high $S/N$ spectra of bright elliptical galaxies; both in the cluster and in the cosmic web. These data will provide an invaluable resource for the astronomical community to answer research questions not directly addressed by the \iscea\  science team. 
	
{\bf Investigation Products:} \iscea\  Science Team will fulfill the investigation objectives and requirements by disseminating the results of the investigation by publishing the resultant science papers in refereed scientific journals, and making community presentations. Given the high scientific interest that the astronomical community will have in the data, the high-level products (L5) generated by the science team will be released to the public through IRSA, along with the L1-L4 products (see \S\ref{sec:data_proc}).

The L5 data products will include the data for all spectroscopic targets  with a redshift measurement. For each galaxy we will release the fully calibrated (i.e., wavelength and flux) one dimensional combined spectrum and the corresponding error spectrum. In addition, we will release a catalog including both basic measurements performed on the spectra, and physical parameters computed from these measurements. 
	
Specifically, the spectral measurements will include:
\begin{itemize}
\item{Redshift: for each galaxy we will provide  measured spectroscopic redshift, together with the quality flag describing the reliability of the measurement.} 
\item{Emission line measurements:  we will release the total flux in the emission lines  detected  above 5$\sigma$, and 3$\sigma$ upper limits for the undetected lines  covered by the  observed spectral range. Estimate of line equivalent width (or lower limits) will also be provided.}
\item{Continuum measurements:  for the brightest galaxies with detected continuum, we will provide measurement of the 4000\AA/Balmer  break, and the strength of the detected absorption lines.}
\end{itemize}

Galaxy physical properties released with the catalog will include:
\begin{itemize}
\item{Activity classification: using emission-line-based diagnostic	diagrams (e.g., \oiii/H$\beta$ vs stellar mass M$_*$, see \citealt{Juneau2011}); BPT (\citealt{Baldwin1981}) diagrams (from \oiii/H$\beta$, [\ion{N}{2}]/H$\alpha$, and [\ion{S}{2}]); and other diagnostics from emission line fluxes,  we will classify galaxies activity into star forming and AGN.} 
\item{Stellar population properties: We will provide the stellar mass,  age, star-formation history  and continuum extinction derived from  stellar population synthesis modeling of the broadband data.}
\item{Instantaneous SFR (from H$\alpha$) and nebular extinction from Balmer line ratios, when available.}
\item{Local density:  estimates of the local environmental density described in \S\ref{sec:delta} and \S\ref{sec:test-hypo}.}
\end{itemize}

\begin{table}[]
    \centering
    \begin{tabular}{|l|l|l|l|l|l|l|}
    \hline
    ID & Age & F(H$\alpha$) & $H$-mag & E(B-V) & Metallicity & Description \\\hline
      1   &  3 Gyr & 3 & 20.6 AB & 0 mag & solar & Bright central galaxy \\\hline
       2   &  3 Gyr & 3 & 21.5 AB & 0 mag & solar & Average galaxy \\\hline
        3   &  3 Gyr & 3 & 22 AB & 0 mag & solar & Faint galaxy  \\\hline
         4   &  0.8 Gyr & 5 & 21 AB & 0 mag & $0.5$ solar & Young, star-forming galaxy \\\hline
          5   &  0.8 Gyr & 5 & 21 AB & 0.5 mag & solar & Young, dusty, star-forming galaxy \\\hline
          6   &  3 Gyr & 3 & 25 AB & 0 mag & solar & Faint emission-line galaxy \\\hline
    \end{tabular}
    \caption{List of example galaxies with simulated \iscea\  spectra in Fig.\ref{fig:jittered-spec} at redshift $z = 2$. The H$\alpha$ fluxes in the second column are in units of $10^{-17}\esc$.  \iscea\ will detect the continuum with SNR $>$ 3 for galaxies as faint as $H < 21.5$ in the nominal survey exposure time (668 ks), enabling measurements of galxay redshifts (from continuum breaks and absorption features) and permit stellar population studies. 
    }
    \label{tab:spec}
\end{table}

\section{Modeling and Mitigation of Systematic Effects}
\label{sec:sys}

\subsection{Completeness of Star-Forming Galaxies}

In each of its 50 protocluster fields, \iscea\  selects $>1000$ galaxy targets with photo-$z$'s closest to the spectroscopically confirmed cluster BCG redshift (see \S\ref{sec:sel-gal}).
There are $\sim$ 300 galaxies within $\Delta z_{\rm phot}=0.2$ of the cluster BCG (see \S\ref{sec:sel-gal}) in a typical \iscea\  field, and $\sim$ 210 galaxies above the \iscea\  H$\alpha$ line flux limit in a cluster and its cosmic web environment based on the \iscea\  mock (see \S\ref{sec:count-gal}). Thus \iscea\  is highly complete in observing ELGs above the \iscea\  flux limit in each protocluster.

In the galaxy target list for an average \iscea\  protocluster target field, consisting of $>1000$ galaxies (including all galaxies in the cluster or its adjacent cosmic web), we estimate the missing fraction of ELGs with $f_{\rm line}>3\times 10^{-17}\esc$ to be $\sim$ 10\%, due to the the fraction of target galaxies missed by \iscea\  in its multiple visits to each field (see \S\ref{sec:efficiency}). Significant amount of additional observing time is required to increase the completeness above 90\% (see \S\ref{sec:efficiency}). It would be optimal to carry out additional observations only in a few fields to increase completeness, as a baseline of comparison to mitigate the systematic effect due to H$\alpha$ sample incompleteness.

In addition, we can validate this fraction by carrying out a blind survey of ELGs using the DMD in the IFU mode. IFU mode can be easily implemented by scanning a slit across a spatially extended object. DMDs, however, allow for a more efficient strategy where the spectra are produced opening simultaneously multiple slits with a pre-defined pattern \citep{Fixsen09}. A sequence of such images can pass through an inversion procedure allowing the reconstruction of the datacube with improved signal-to-noise in low signal regime (Hadamard Transform Spectroscopy).
The use of the DMD in the IFU mode is beyond the scope of the \iscea\  Baseline Mission, but will be investigated for possible implementation during the extended mission.

\subsection{Definition of Quiescent Galaxies}

\iscea\  obtains spectra of $>1000$ galaxies in a protocluster field with photo-$z$'s closest to the confirmed BCG spectroscopic redshift (including $\sim 300$ within $\Delta z_{\rm phot}=0.2$), which means that we will have spectra for all of the quiescent galaxies in each of the protoclusters down to the same stellar mass threshold. As such, we will have upper limits on the SFR for these galaxies even if H$\alpha$ is undetected. Since the \iscea\  H$\alpha$ line flux limit of $3\times 10^{-17}\esc$ at 5$\sigma$ ($1.8\times 10^{-17}\esc$ at 3$\sigma$) corresponds to a high completeness of galaxies on the star-forming main sequence at our stellar mass limit, we can define the \iscea\  sample of quiescent galaxies as galaxies which have no detectable line emission (see Fig.\ref{fig:limits} for the 5$\sigma$ limit). We will investigate the implications of this sample definition of quiescent galaxies using the \iscea\  mock and available observational data, to establish a validated metric for comparing \iscea\  data with galaxy evolution models.

The lack of emission lines does post a challenge for redshifts because \iscea\  will only have sufficient sensitivity to obtain continuum redshifts for high mass galaxies. However, the 7 band imaging from HSC ($grizy$) and WISE ($W1$, $W2$), and the broadband NIR imaging  from \iscea\  (1.1-2$\mu$m) can be combined with the spectra to refine the redshift estimates for quiescent galaxies via spectral breaks.  The \iscea\  spectroscopy will also provide important information on the galaxy stellar populations as it probes rest-frame optical features sensitive to ages, star-formation histories, and chemical enrichment. We plan to model jointly the photometry (HSC, WISE, \iscea, and future LSST data, see below) with the \iscea\ spectroscopy. This follows on recent work that combined such datasets in joint analysis \citep[e.g.,][]{Deugenio2020,Estrada-Carpenter2020,Carnall2021,Khullar2021}. This will allow us to test for evolution in continuum-detected galaxies that depend on environment. 

\section{Testing the \iscea\  science hypothesis}
\label{sec:test-hypo}

To meet its Science Objective, \iscea\  is designed to test this Science Hypothesis:
{\bf During the period of cosmic noon, environmental quenching is the dominant quenching mechanism for typical galaxies not only in clusters and groups, but also in the extended cosmic web surrounding these structures.}
The \iscea\  baseline mission will obtain spectroscopic data for a sample of 50 protocluster fields at $1.2 < z < 2.1$ (90\% at $z > 1.7$, and 5 are SZ-detected clusters  at $z>1.5$).  The sample contains galaxies spanning three orders of magnitude in local density (environment), 
enabling quantitative measurements of quenching and star formation as a function of galaxy stellar mass, local density, and proximity of galaxies to filaments during this critical epoch.  The advantage of \iscea\ is that with its large FoV and observing strategy, we can measure spectroscopic redshifts and H$\alpha$-derived SFRs for galaxies in all these structures \textit{within a single dataset}.  Thus, \iscea\ will provide the largest sample of clusters/overdensities along with cosmic web galaxies ever targeted with spectroscopy and is sufficient to place studies of environment at $z\sim 1.7-2.1$ on a comparable footing as studies at $z\sim 1$ (GOGREEN) and $z\sim 0$ (SDSS).
However, \iscea\ works this problem at higher redshift (when the impact of environment first manifests) and it contains substantially higher area coverage (compared to studies using ground-based telescopes, such as Gemini/GMOS) and does not suffer from potential under-sampling of galaxies in high-density environments (e.g., because of fiber-collisions, for SDSS, for example).  \iscea\ is designed to achieve its Science Goal and Objective (see \S\ref{sec:science_req}), and transform our understanding of cosmic evolution at cosmic noon.

\begin{figure}
    \centering
    \includegraphics[width = 4.5in]{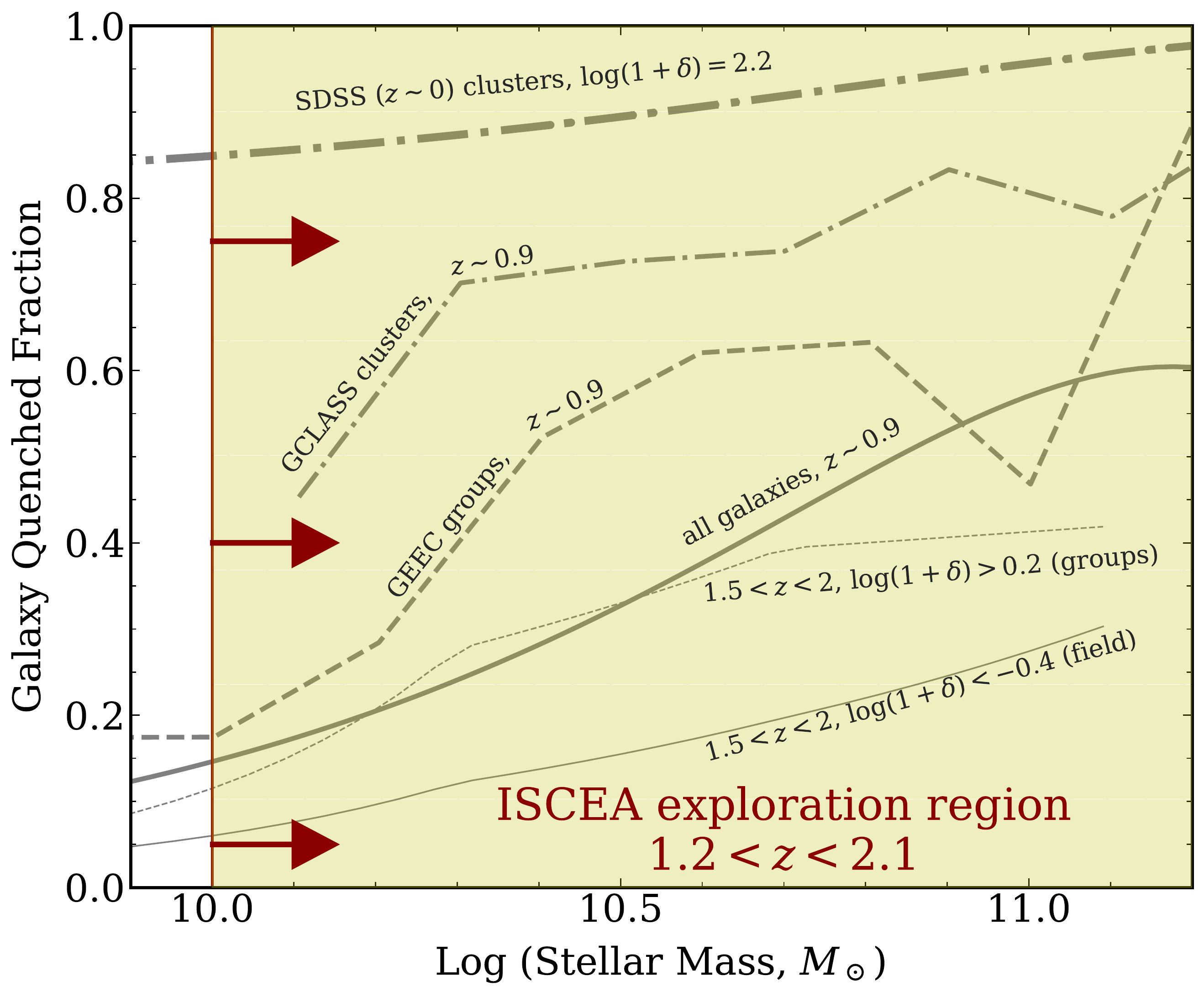}
    \caption{Compilation of the fraction of quenched galaxies as a function of stellar mass, redshift, and environment taken from the literature, and illustrating the mass range probed by \iscea.  Each curve on the plot shows a measurement of the fraction of quenched galaxies taken from datasets in the literature (\citealt{Peng2010,Muzzin2013,Omand2014,Balogh2014,Kawinwanichakij2017}).  Here the ``quenched fraction'' is defined as the ratio of the number of ``quenched'' galaxies to the total number of galaxies (though the definition of ``quenched'' varies somewhat in the literature).  This includes measurements from SDSS at $z\sim 0$ for galaxies in the highest densities (i.e., galaxy clusters, with $\log (1+\delta) \sim 2$), measurements at $z\sim 0.9$ ranging from rich clusters to the field, and values at $z > 1.5$ comparing the effects of group versus the field.  All previous studies at $z > 1.5$ rely on small samples (sometimes including only a single cluster/overdensity) and are highly heterogeneous in their photometric and spectroscopic coverage.   The \iscea\  exploration region augments these studies by enabling the first study  of the galaxy quenched fraction over $1.2 < z < 2.1$, where galaxy structures are collapsing, and spanning the full range of density, from the cores of clusters, cluster outskirts into the cosmic web, groups, and the field with sufficient statistics to test the \iscea\ Science Hypothesis.  
    }
    \label{fig:quenching}
\end{figure}

With \iscea's large FoV of 0.32 deg$^2$, a single pointing covers the full protocluster environment.
It has been shown in recent years that at least at lower redshift the relative importance of environmental and mass quenching is a function of stellar mass (e.g. \citealt{Thomas2005,Peng2010,Davies2016}), and also that even at $z\sim1-1.5$ cluster galaxies show evidence of depleted gas and quenching at least out to the virial radius \citep[e.g.][]{Alberts2016}. 
The latter supports the expectation that the bulk of the quenching occurs at higher redshifts and larger radii than probed by existing studies.  Based on these observations, we can make a hypothesis to test when (at what redshift(s)) and where (in what environments)  quenching occurs and how  these processes depend on stellar mass. 
The \iscea\ hypothesis is that  by $z\sim2$ quenching by ``environment''-related processes is a critical and likely dominant factor driving the subsequent decrease in star formation seen in galaxies not only in clusters and groups but also in the surrounding cosmic web, acting in tandem with mass-related processes (where galaxies quench once their halos achieve a certain mass).
The key to testing the above hypothesis is to measure SFRs and stellar masses for a statistical sample of galaxies spanning the full range of local densities at $z\sim2$.  Because our hypothesis is that the environmental and mass-quenching depends on galaxy stellar mass, structure (halo) mass, and local density, we must measure the differential effects for each \textit{within the same structures}.  This is achieved because of \iscea's large FoV which enables us to trace galaxy SFRs and stellar masses from high-density environments into the cosmic web and to the field in the same connected structures.   Fig.~\ref{fig:quenching} shows the region of the parameter space of ``quenched fraction'' of galaxies probed by \iscea\ as a function of redshift, stellar mass, and environment, in comparison to other studies of galaxies.  \iscea\ is therefore the first of its kind, and will  measure these metrics at $1.2 < z < 2.1$ with the lowest systematics of any current (or planned) other projects.

\iscea\ will test its Science Hypothesis by quantifying several key metrics of the galaxy population. This includes the galaxy ``red'' fraction (the quenched population), the ``green'' fraction (galaxies in transition), and the ``blue'' fraction (star-forming galaxies).   The galaxies will be selected based on their colors (e.g., rest-frame $UVJ$, see \citealt{williams2009,Whitaker2011,vanderburg2013,Foltz2018}), their specific SFRs (sSFR $\equiv$ SFR / $M_\ast$, see \citealt{Barro2017}), and/or based on spectroscopic line indices \citep[see, e.g.,][]{Haines2017}.   We will confirm the vast majority of these galaxies using the \iscea\ spectroscopy.    
Assuming a conservative total observing efficiency for \iscea\ of 80\% (see \S\ref{sec:efficiency}),
the 45 protocluster fields at $1.7<z<2.1$ provide a sample of 
(45 structures) $\times$ (210 galaxies per structure) $\times$ (80\% efficiency) $\approx$ 7560 galaxies
(including $45\times33 \approx 1485$ galaxies in the densest areas, e.g., cores of clusters, see \S\ref{sec:count-gal}).  All of the \iscea\ galaxies will have stellar masses, $\log (M/{\rm M_\odot})\geq 10$, with which to measure the dependence of sSFR upon local density.  With this sample size we can divide the sample into five bins in stellar mass even within the highest density regions and within nine bins in local density (extending from the densest ``cores'' to the sparser ``field'') 
while still maintaining sufficient statistics for 10\% statistical uncertainty per bin. To extract maximal information we will also fit a parametric model for the quenching function as shown in Fig.~\ref{fig:quenching}, with only the slopes of the halo and environmental quenching lines and their amplitudes as free parameters, which provides a sufficient level of constraint to determine the redshift evolution of the quenching function over the full redshift range probed by \iscea. 

\begin{figure}
    \centering
    \includegraphics[width = 0.67\textwidth]{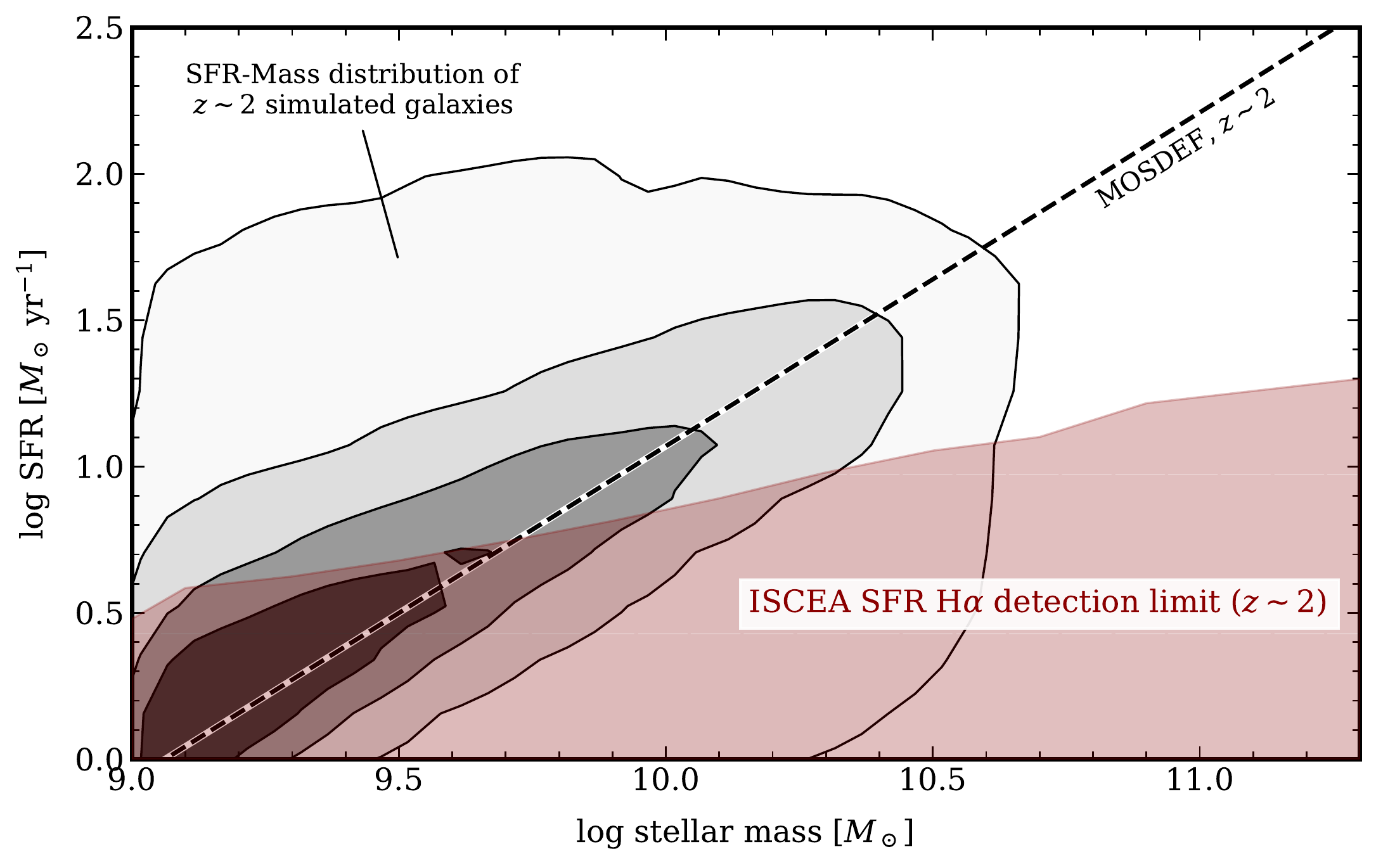}
    \caption{The SFR detection limit for \iscea\  derived from mapping the H$\alpha$ emission to the star formation main sequence (SFMS). The contours show the distribution (25\%, 50\%, 75\%, and 95\%-tiles) of galaxy SFRs and stellar masses for galaxies at $1.9 < z < 2.0$ from the \iscea\  mock.  The dashed line shows the empirical relation derived from galaxy H$\alpha$ emission at $z\sim 2$ (from MOSDEF, \citealt{Shivaei2015}). The red-shaded area shows the region below the \iscea\  detection limit at $z=2$ assuming the $5\sigma$ detection limit for H$\alpha$ ($3\times 10^{-17}$ erg s$^{-1}$ cm$^{-2}$) and an average dust attenuation that increases with stellar mass following \citet{Pannella2009}.  \iscea\  will be sensitive to the H$\alpha$ emission from all galaxies above $\log M_\ast/{\rm M_\odot} \gtrsim 10$ at $z=2$ (and will be sensitive to H$\alpha$ emission from fainter / lower-mass galaxies at lower redshifts).  
    \label{fig:limits}}
\end{figure}

\begin{figure}
    \centering
    \includegraphics[width=7in]{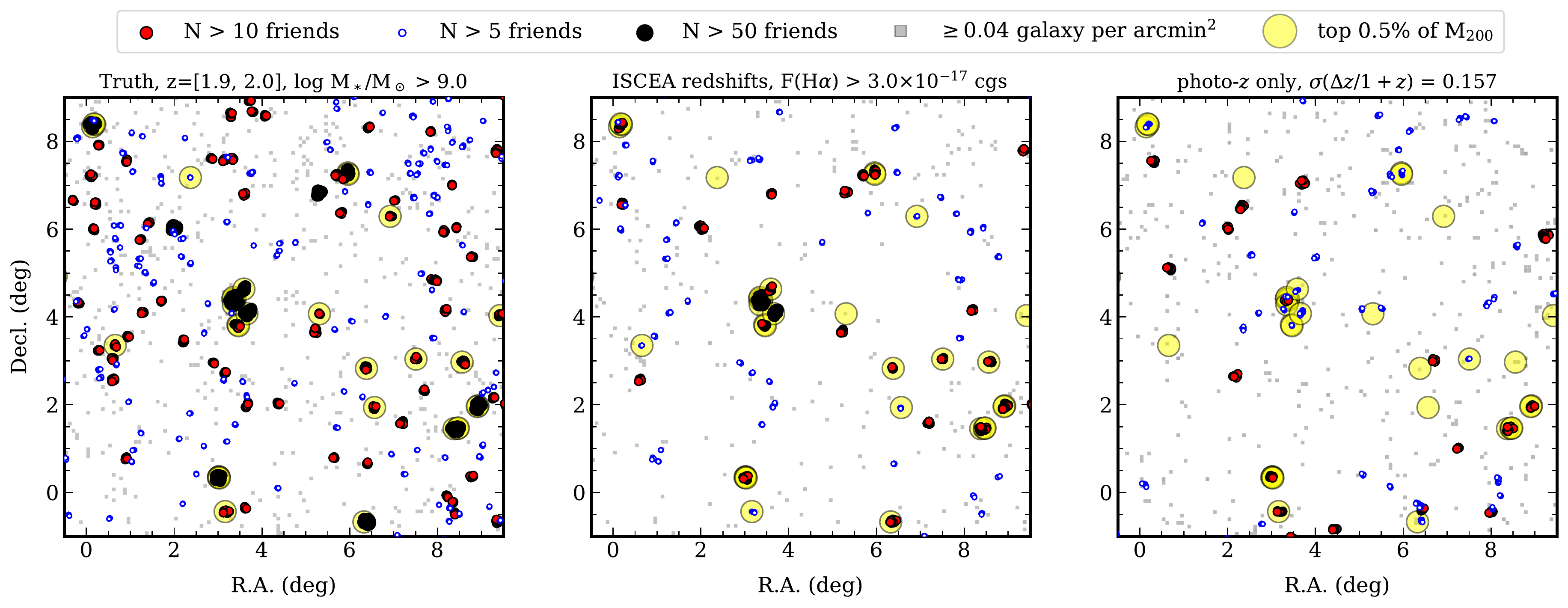}
    \caption{A $10^\circ \times 10^\circ$ slice at $1.9 < z < 2.0$ from the \iscea\  mock.  The left is "truth" using a mass-limited galaxy sample from our mock catalog.  The middle panel shows the results for galaxies detected in H$\alpha$-emission by \iscea\ .  The right panel shows results using only photo-$z$'s. 
    In each panel, the gray squares show a density map of $5^\prime \times 5^\prime$ regions with $\ge 1$ galaxy in them (a surface density of $\Sigma > 0.04$ arcmin$^{-2}$).  The circles show galaxies with $>5$, $>10$, $>50$ ``friends'' (from the FoF algorithm) from each sample, as labeled. The large yellow circles show the top (most massive) 0.5\% $M_{200}$ objects (from the $M_{200}$ distribution in the mock).
    }
    \label{fig:density1}
\end{figure}

In what follows we describe the primary methods we will use to estimate important quantities to test the \iscea\ Science Hypothesis. 
\begin{itemize}
    \item \textit{\underline{Estimating SFRs}}:  The \iscea\ spectroscopy will cover H$\alpha$ luminosities for all objects in the redshift range of the samples.  This will be the primary means by which we will estimate galaxy SFRs.  We will use standard conversions between H$\alpha$ luminosities and the SFR \citep{Kennicutt1998,Kennicutt2012}, adjusted to our assumed IMF.  We will apply dust attenuation corrections using multiple methods as discussed above (\S\ref{section:emission_line_analysis}).  Fig.~\ref{fig:limits} shows the SFR detection limit for \iscea\  estimated by mapping the H$\alpha$ emission to the star formation main sequence (SFMS). The contours show the distribution of galaxy SFRs and stellar masses for galaxies at $1.9 < z < 2.0$ from the \iscea\  mock (see \S\ref{sec:mock}).  The dashed line shows the empirical relation derived from galaxy H$\alpha$ emission at $z\sim 2$ (from MOSDEF, \citealt{Shivaei2015}). The red-shaded area shows the region below the \iscea\  detection limit at $z=2$ assuming the $5\sigma$ detection limit for H$\alpha$ and an average dust attenuation that increases with stellar mass following \citet{Pannella2009}. \iscea\  will be sensitive to the H$\alpha$ emission from all galaxies above $\log M_\ast/{\rm M_\odot} \gtrsim 10$ at $z=2$ (and will be sensitive to H$\alpha$ emission from fainter / lower-mass galaxies at lower redshifts).

More importantly, the \iscea\ spectroscopy provides accurate redshifts for galaxies, which is otherwise a major limiting factor for deriving the quiescent-galaxy fraction.  \iscea\  will be able to measure redshifts in the case of passive galaxies (with no detectable emission features) to a continuum magnitude of {$\sim$ 21.5 (AB) at 3$\sigma$} in the nominal exposure time ($668\,{\rm ks}$) in the \iscea-$H$ band (see Fig.~\ref{fig:jittered-spec}).  This corresponds to stellar mass limit of $\log M_\ast/{\rm M_\odot} \gtrsim 11-11.3$ for passive galaxies from $z=1.5-2.1$, modulo assumptions about the star-formation history. For galaxies below this mass with no detectable emission lines, we will assign probabilistic association with the protocluster volume based on photometric redshifts.  
We will also model \iscea\  data quantitatively using the \iscea\  mock, which will be continuously modified and improved for application to \iscea\  data.

\begin{figure}[ht]
    \centering
    \includegraphics[width=\textwidth]{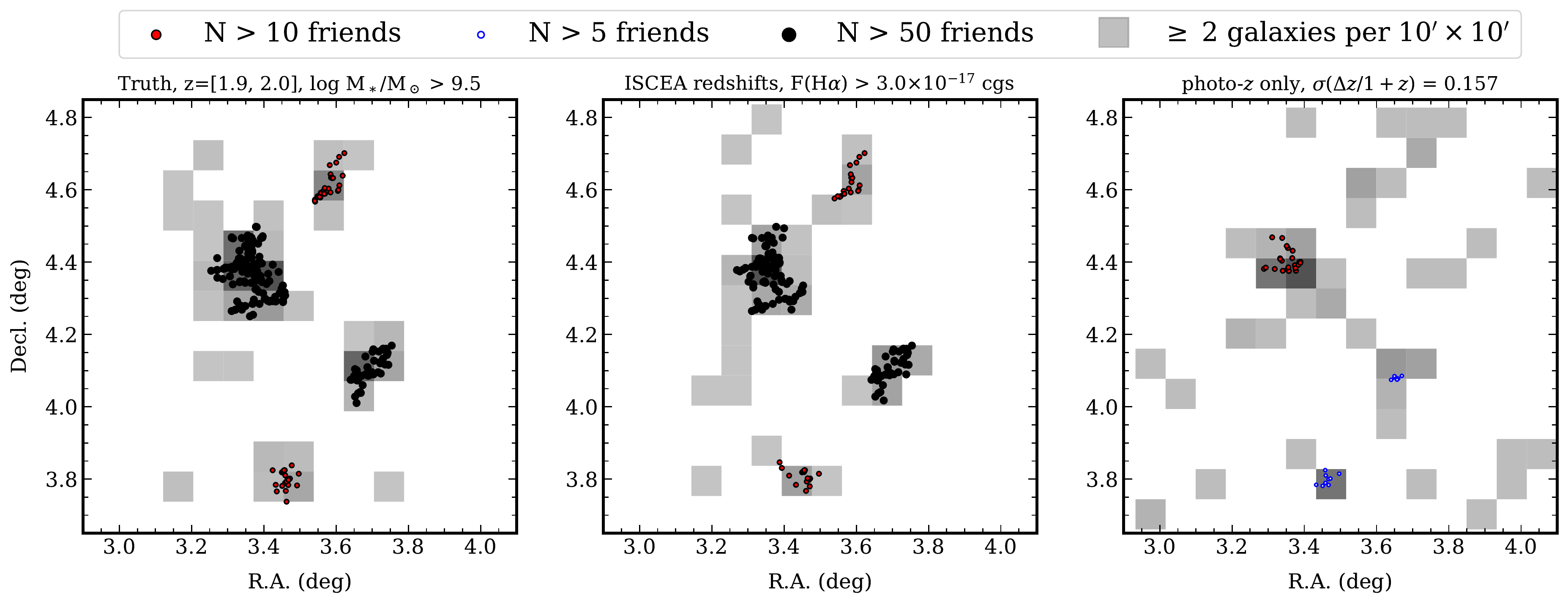}
    \caption{Example of how \iscea\ traces the true large scale structure. The panels show a zoom-in from the \iscea\ mock catalog, centered on several fiducial structures identified from our friends-of-friends algorithm for galaxies in the redshift slice, $1.9 < z < 2.0$ (each panel is 1.2$^\circ \times 1.2^\circ$).     Each panel shows the galaxy density distributions selected by different means, including the ``True'' distribution (left panel, showing results for all galaxies with stellar mass $\log M_\ast/{\rm M_\odot} > 9.5$), the \iscea-selected measurement (center panel, showing results for all galaxies with H$\alpha$ flux above the \iscea\ detection limit), and results based solely on photometric redshifts (right panel).  Circles show objects with a high number of ``friends'' (as labeled).  The gray scale shows the local galaxy density in 10$\arcmin \times 10\arcmin$ cells $>$2 galaxies per cell.  Note that the \iscea--measured galaxies trace the structures with high fidelity as evidenced by the ``true'' surface density from the mock out to tens of arcminutes from the centers of the structures. This allows use to use the \iscea\ dataset to quantify galaxy properties over a wide range of environment.   This is not achievable with the photometric-redshift measurements (right panel) where the measured galaxy density is noisy because of redshift errors.
    }
    \label{fig:density3}
\end{figure}

\item \textit{\underline{Measuring Environment / Local Density}}:  We will measure the environment of galaxies using multiple methods that both identify the structures themselves and the local densities.  First, we will identify the structures based on galaxies physically associated on the sky and redshift.  For galaxies with measured spectroscopic redshifts, we will assign galaxies as ``associated'' if they fall within some range of velocity, typically $\Delta v \lesssim 2000$~km s$^{-1}$.  This is sufficient to distinguish galaxies from the Hubble flow based on their peculiar velocities.  For galaxies with photometric redshifts, we will use a probabilistic approach where some fraction of their probability density function ($\mathcal{P}$) is contained within a specified $\Delta z$ (where typically we will require that $\mathcal{P} = \int\ P(z)\ dz \gtrsim 70$\% when integrated over a range of redshift of $\Delta z/(1+z) \approx $10\% about the mean redshift of the structure as defined by the \iscea\ spectroscopy, but this may depend on galaxy type and color, see \citealt{Papovich2010}).    

We will then assign galaxies to ``structures'' based on a friends-of-friends algorithm (see \S\ref{section:emission_line_analysis}).  The power of the friends-of-friends algorithm is that it associates galaxies into structures based on a linking length that we can adjust:  based on the \iscea\ mock, we find that a linking length of $\approx2^\prime$ (1 proper Mpc at $z=2$) accurately identifies these galaxies into their ``true'' structures.  We will update these simulations based on the final \iscea\ performance specifications and using updated mocks, in order to calibrate these methods. Fig.~\ref{fig:density1} shows how the friends-of-friends algorithm applied to \iscea--like data from the mock are able to recover structures (over fields of 10 $\times$ 10 deg$^2$).   Fig.~\ref{fig:density3} shows examples of regions comparable to the size of the \iscea\ FoV, where the algorithm recovers the galaxies associated with structures.   
Once we have identified galaxies in each structure, we will also measure the local density using either the number of galaxies per surface area ($\Sigma$ measured in \# of galaxies per arcmin$^2$ or per Mpc$^2$) and/or the number per unit volume ($\rho$, measured in units of galaxies per Mpc$^2$).   The local density, is then defined as the ``overdensity'', $\delta$, relative to the mean density, $\bar \Sigma$ or $\bar\rho$, in the usual way (e.g., \citealt{Peebles1980}),
\begin{equation}
    \delta \equiv \frac{\Sigma - \bar{\Sigma}}{\bar{\Sigma}}, 
\end{equation}
such that $\delta > -1$ because $\Sigma > 0$ (and a similar relation exists for $\rho$). See also \S\ref{sec:delta}.

Fig.~\ref{fig:groups} shows that using this method we will be able to identify both the structures and recover the densities accurately with \iscea\ data, which is not possible using photo-$z$'s.  This provides confidence that we will recover the ``richness'' of structures accurately with \iscea.  For comparison, at this redshift structures with $>$100 ``friends'' have halo masses, $\log M_{200}/{\rm M_\odot} > 13$.  This correspond to progenitors of Fornax-sized clusters (or larger) at $z=0$ \citep{Moster2012,Behroozi2019}.   Fig.~\ref{fig:density3} shows the surface density of galaxies measured from \iscea\ compared to that from ``truth'' for several structures identified from our friends-of-friends algorithm applied to the \iscea\ mock at $z=1.9-2.0$.  The surface density recovers the structure accurately, which means we will be able to trace the local density of galaxies in structures at $z\sim 2$ from the central regions (e.g., cluster and group ``cores'') to the filaments and the field.   This enables us to achieve the measurement to test the \iscea\ Science Hypothesis.  Note that Fig.~\ref{fig:density3} shows that this is not true for photometric-redshift surveys, where the accuracy is insufficient to measure the galaxy density accurately (nor is this sufficient to identify structures accurate from ``friends-of-friends'', see Fig~\ref{fig:density1}).

\item \textit{\underline{Stellar Masses and specific SFRs}}:
Stellar masses will be computed using the combination of deep $grizy$ photometry from the HSC Survey \citep{Aihara2018} and \iscea\  NIR photometry.  In addition, we will include the \wise\ imaging (which will detect the massive, red galaxies in our sample), and LSST optical imaging ($ugrizY$) which will provide important coverage in the $u$-band (we expect to have $\sim$2 year LSST imaging by the start of \iscea.\footnote{We assume that the 2 year depth of LSST will be approximately $\approx0.9\,{\rm mag}$ shallower than the full 10-year depth \url{https://www.lsst.org/scientists/keynumbers}.  This is comparable to the current HSC depth in $grizY$ which effectively doubles the existing exposure time (and LSST will ultimately be nearly 1 mag deeper than the HSC data).} \citet{Lower2020} have recently shown that non-parametric models of the star-formation history yield much more robust and unbiased stellar mass determinations than standard parametric approaches (see their Figure 3). We plan to use their non-parametric approach with Prospector \citep[][]{Johnson2021} to determine stellar masses for the galaxies in our sample.  Using these methods we expect to achieve stellar masses with an accuracy of $0.2-0.3\,{\rm dex}$ \citep[for fixed IMF, see][]{Conroy2013}.    

We will then estimate \textit{specific SFRs}, using the measurements of the SFRs and stellar masses for all galaxies in the sample, following sSFR $\equiv$ SFR/M$_\ast$.   We expect the errors on the sSFR to propagate from both the SFRs and stellar masses, and they may be as high as $\delta(\log~\mathrm{sSFR}) \sim 0.5$ dex for some objects.  However, the measurements we need require only accuracy of $\sim$1 dex, which is achieved by our measurements.  

\end{itemize}

Using the measurements of galaxy stellar masses, SFRs (and specific SFRs), and local density measurements, we will then be in a position to test the \iscea\ Science Hypothesis.   For example, Fig.~\ref{fig:ssFR-quenching-timescale} shows the expected performance of \iscea.  The top panel shows the accuracy in measuring specific SFRs as a function of galaxy density, using customary units of $\log(1 + \delta)$.   Using a sample of 45 clusters at $1.7 < z < 2.1$, \iscea\ will achieve relative accuracy of $0.2-0.3$ dex in bins of density.   This is a significant improvement over current studies at $z\sim 2$ that focus on measurements from individual clusters (e.g., \citealt{Koyama2013} using observations of MAHALO fields).  \iscea\ enables us to measure the density in many more bins of density compared to the current ``state of the field'' (the top panel of Fig.~\ref{fig:ssFR-quenching-timescale} shows we expect to triple the number of bins in density while improving the uncertainties by more than a factor of two).   

\begin{figure}
    \centering
    \includegraphics[width = 0.5\textwidth]{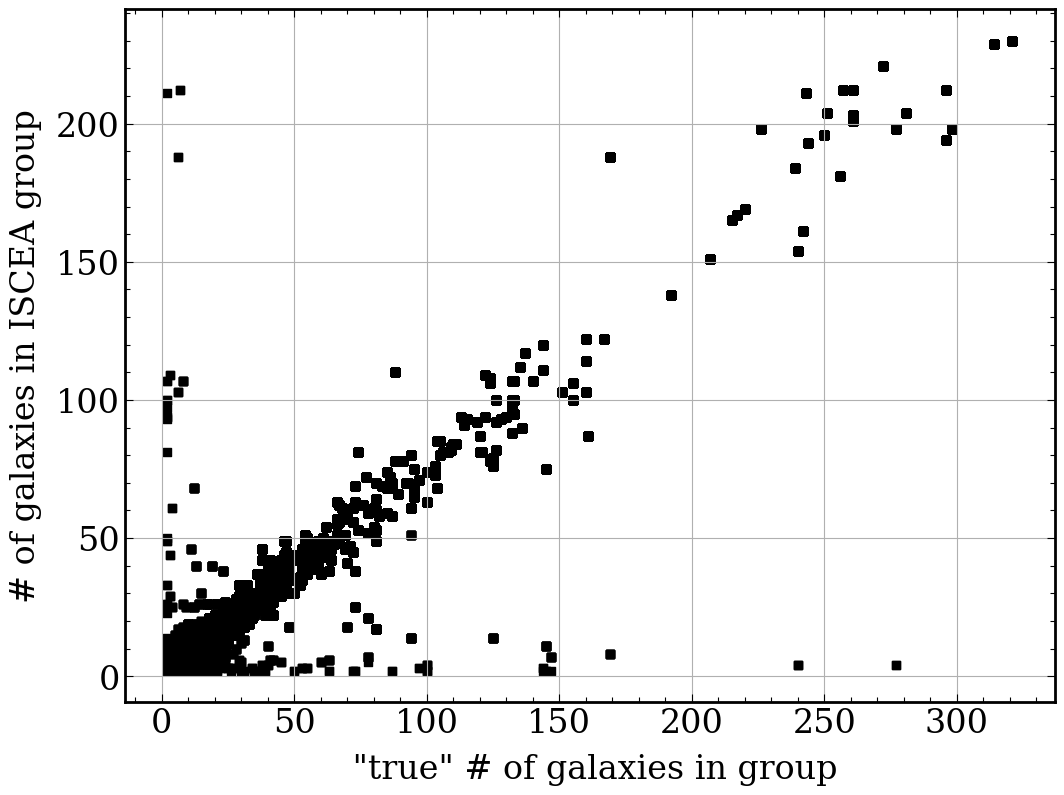} 
    \caption{The number of galaxies in a group (truth) against the number of galaxies in the group identified from our analysis of the \iscea\ mock catalog for structures at $z=1.9-2.0$. This provides confidence that we will recover the ``richness'' of structures accurately with \iscea, particularly for the higher-mass structures.   For comparison, at this redshift the structures with $>$100 ``friends'' have halo masses $\log (M_{200}/{\rm M_\odot}) > 13$ in the \iscea\ simulation.  These all correspond to progenitors of Fornax-sized clusters or larger at $z=0$ \citep{Moster2012,Behroozi2019}.  }
    \label{fig:groups}
\end{figure}

\begin{figure}
    \centering
    \includegraphics[width=4.5in]{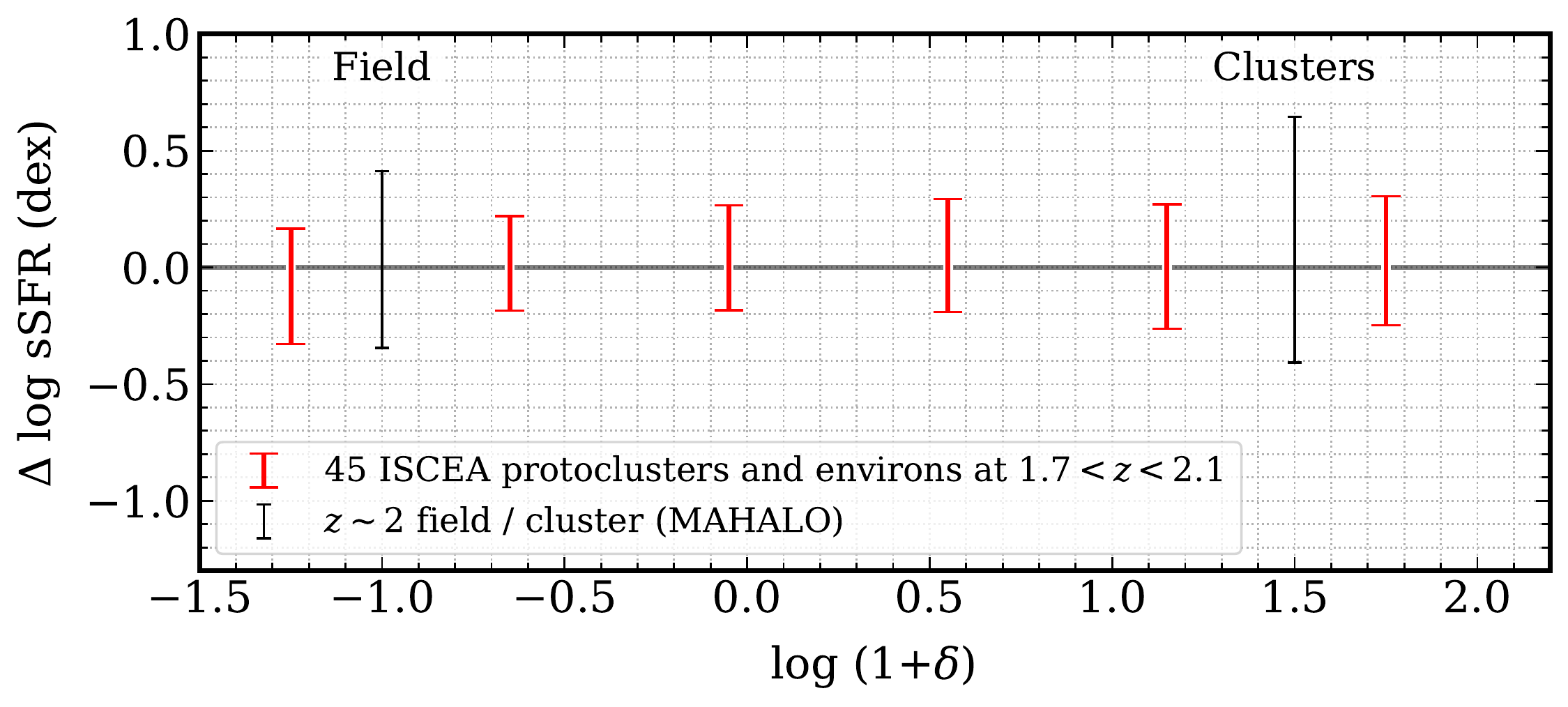} \\
    \includegraphics[width = 4.5in]{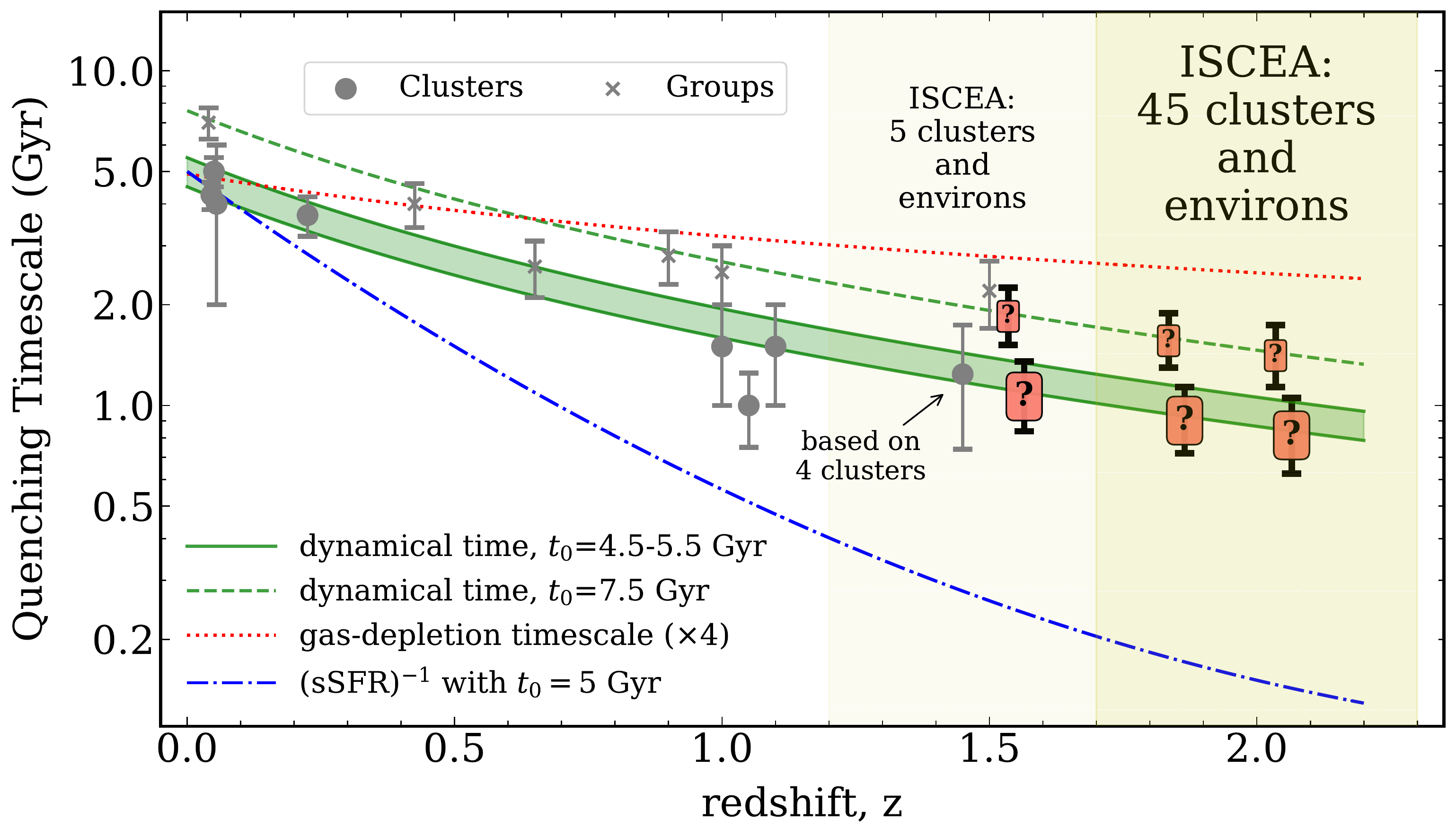}
    \caption{\textit{Top}: 
    The 1$\sigma$ uncertainty in the specific SFR measurements as a function of local density expected for the \iscea\  sample of 45 protoclusters at $1.7<z<2.1$.  The density measures log$(1+\delta)$ are derived from the local galaxy density for galaxies associated with each protocluster in the \iscea\  mock.  Contemporary surveys (such as MAHALO, \citealt{Koyama2013}) cover smaller volumes (including 4 clusters). 
    \textit{Bottom}: Evolution of Quenching timescales derived for clusters and groups at $z \la 1.6$ compared to theoretical predictions (adapted from \citealt{Foltz2018}).  \iscea\  will measure the quenching timescales at $z\sim 1.6$ with improved statistics, and make the first decisive measurement at $1.7 < z < 2.1$. The gray data points in the bottom panel are taken from the compilation of \cite{Foltz2018}, which consist of data from
\cite{Wetzel2013,Muzzin2014,Taranu2014,Haines2015,Balogh2016,Foltz2018}.
The red points for \iscea\  are predictions assuming the a quenching timescale of 5 Gyr for the \iscea\  galaxies in cluster environments, and 7 Gyr for the group environments. We have assumed the error bars scale from \cite{Foltz2018}, assuming three bins with redshift $(z=1.55, 1.85, 2.05)$ with number of targets of 5, 23, 22 in each redshift bin. We have also scaled the SPT/ACT clusters to have 3$\times$ the number of objects as those in \cite{Foltz2018} given the richness of the former are $\sim 3\times$ higher.
 }
    \label{fig:ssFR-quenching-timescale}
\end{figure}

\begin{figure}
    \centering
    \begin{tabular}{cc}
    \includegraphics[width = 3in]{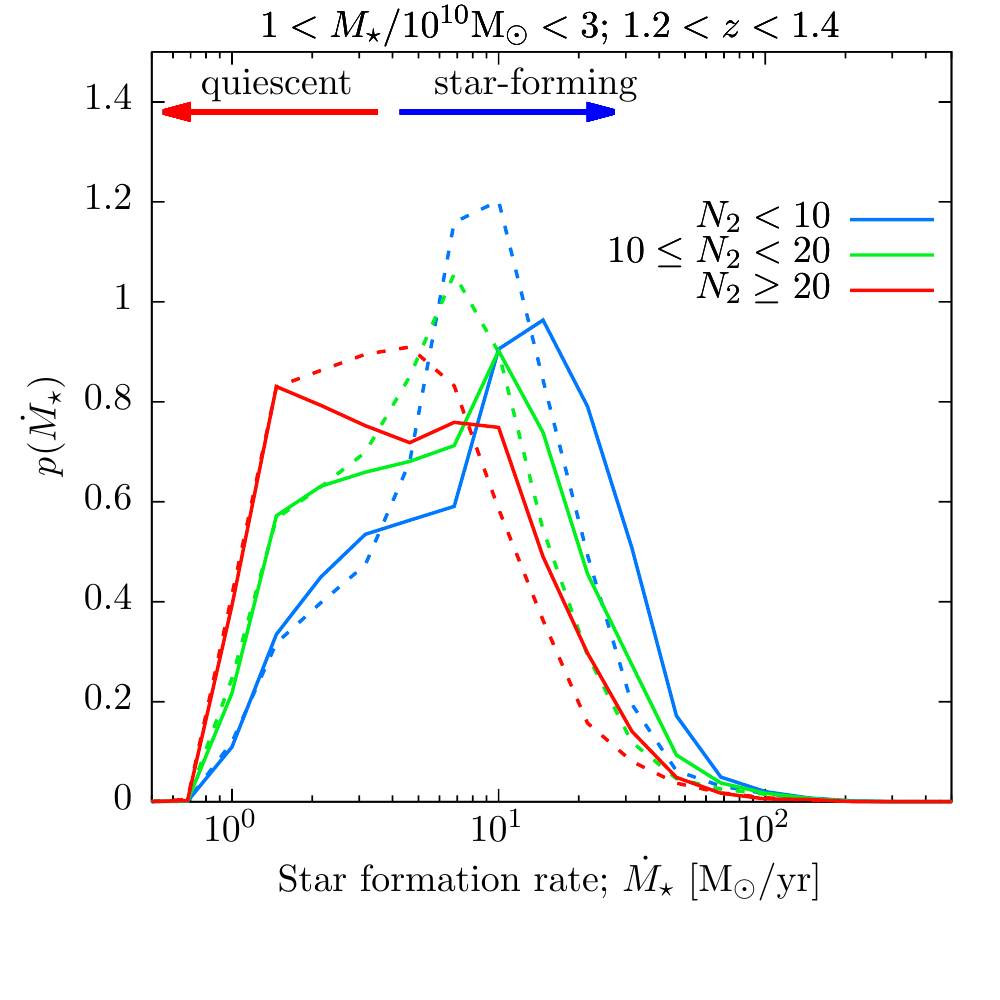} &
    \includegraphics[width = 3in]{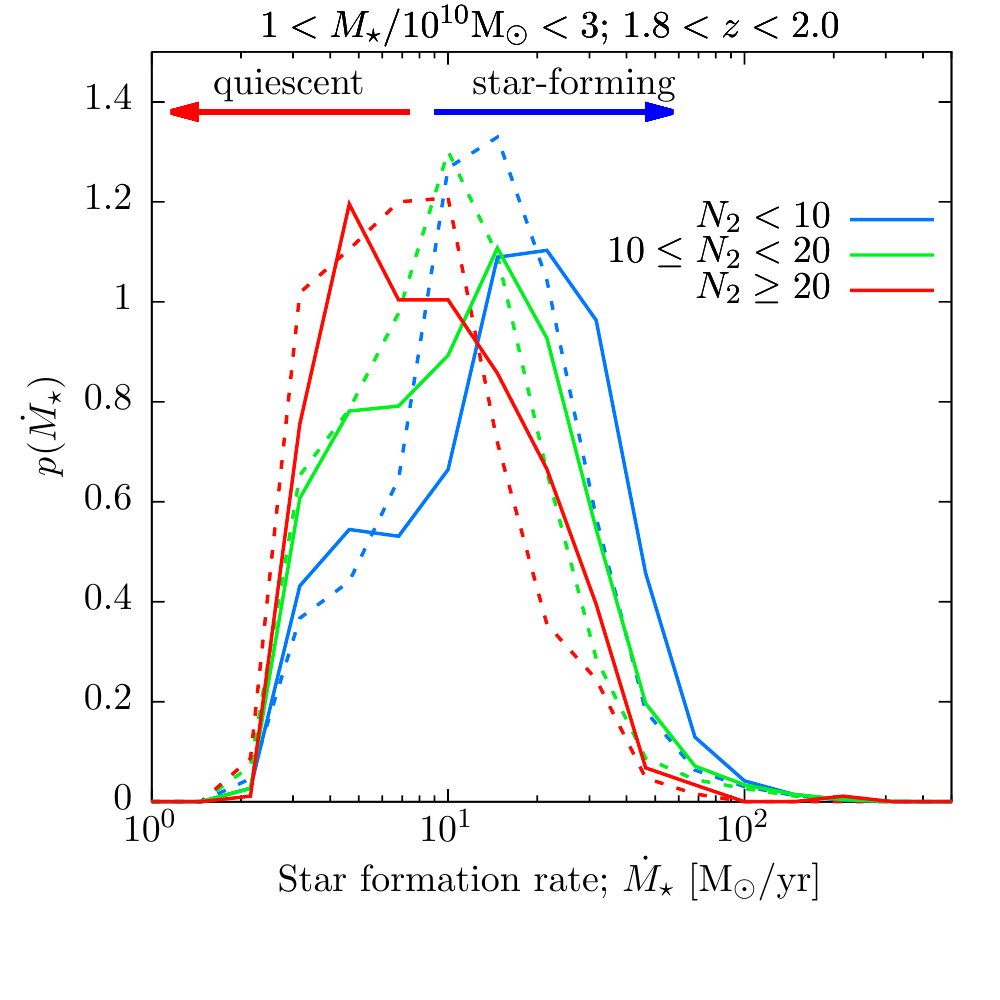}
    \end{tabular}
    \caption{The distribution of SFR for galaxies in the \iscea\  mock selected to have H$\alpha$ fluxes of $f_\mathrm{H\alpha} > 3 \times 10^{-17}$ergs/cm$^2$/s, and stellar masses of $2 < M_\star/10^{10}\mathrm{M}_\odot < 3$ (solid lines) and $1 < M_\star/10^{10}\mathrm{M}_\odot < 2$ (dashed lines), in redshift slices of $1.2 < z < 1.4$ (left) and $1.8 < z < 2.0$ (right). Red and blue arrows illustrate the division between quiescent and star forming galaxies. Galaxies are split by a measure of their local environment, $N_2$, defined as the number of neighboring galaxies within a sphere of radius 2~Mpc (comoving).  The distributions show that while the  SFR in the quiescent range is insensitive to stellar mass, it is very sensitive to local mass density, consistent with the \iscea\ Science Hypothesis, and fully testable with the data \iscea\ will provide.  }
    \label{fig:sfrDistribution}
\end{figure}

 \iscea\ data quality will also enable us to study the quenching timescales for galaxies in different environments. 
 Using the \iscea\ data, we will measure quenching timescales, which can be constrained by solving a continuity equation between star-forming galaxies (blue galaxies), which transition to quenched galaxies (these are so-called, ``green'' or ``green-valley''), and then become fully quenched galaxies (red galaxies), all of which must occur with some transition timescale and quenching timescale ($t_Q$). The quenching timescale is then derived from the solution to these differential equation (e.g., see \citealt{Foltz2018}).   The bottom panel of Fig.~\ref{fig:ssFR-quenching-timescale} shows the estimated accuracy of the quenching timescales for galaxies in \iscea\ in three bins of redshift and two bins of density.  This accuracy will enable us to rule out certain physical processes because such processes predict quenching timescales (e.g., if the example in Fig.~\ref{fig:ssFR-quenching-timescale} is correct, then we can rule out physical models such as ``quintesscence'' or those that scale with gas-depletion timescales in favor of others, such as processes that scale with dynamical time [e.g., ``overconsumption'', see \citealt{McGee2014}]).  It is crucial to make this measurement at $z\sim 2$ where the difference in quenching timescales is largest as such processes have less time to act.  This enables us to differentiate between the importance of competing physical processes, which is most pronounced at high redshift when the timescales are shorter (compared to the Hubble time).   Thus, we will test the \iscea\  Science Hypothesis of the environmental quenching as the main quenching mechanism during cosmic noon.

\iscea\  will provide data for measuring the star-forming fraction and quiescent fractions of galaxies as a function of local mass density (over three orders of magnitude) and stellar mass (over more than an order of magnitude). This will allow us to differentiate between mass quenching and environmental quenching as the dominating mechanism in typical (satellite versus central) galaxies.  
Fig.~\ref{fig:sfrDistribution} shows the distribution of SFR for galaxies in the \iscea\  mock catalog in two bins of stellar mass, $1 < M_\ast/10^{10}\,{\rm M_\odot} < 2$ and $2 < M_\ast/10^{10}\,{\rm M_\odot} < 3$, selected to have H$\alpha$ fluxes of $> 3 \times 10^{-17}\esc$, in redshift slices of $1.2 < z < 1.4$ (left panel) and $1.8 < z < 2.0$ (right panel).  As expected, the SFR increases with redshift from $z\sim 1.3$ to $z\sim 1.9$. Furthermore, while the  distribution of the SFR in the quiescent range is insensitive to stellar mass, it is very sensitive to local mass density (measured by $N_2$, the number of neighboring galaxies within a sphere of comoving radius of 2~Mpc). In the Galacticus model used to construct the \iscea\  mock, this quenching is driven primarily by ``starvation'', in which the circumgalactic medium of a galaxy is rapidly stripped away by ram pressure forces when it becomes a satellite in a larger system. This removes the supply of new, cooling gas for the galaxy, causing it to run out of fuel for new star formation and, eventually, to be quenched.
This clearly shows that current galaxy formation theory predicts the dominance of environmental quenching over mass quenching for typical galaxies at cosmic noon ($1.2<z<2.1$), consistent with the \iscea\  Science Hypothesis. This will be tested by the \iscea\  measurement of the SFR distributions.

\section{\iscea\  Secondary Science Program}
\label{sec:stars}
During two months each winter, the \iscea\ protocluster targets have limited visibility due to spacecraft orbital constraints. This opens windows of opportunity to carry out a productive secondary science program on \iscea\ during the Baseline Mission. 

\begin{figure}
    \centering
    \includegraphics[width = 3.5in]{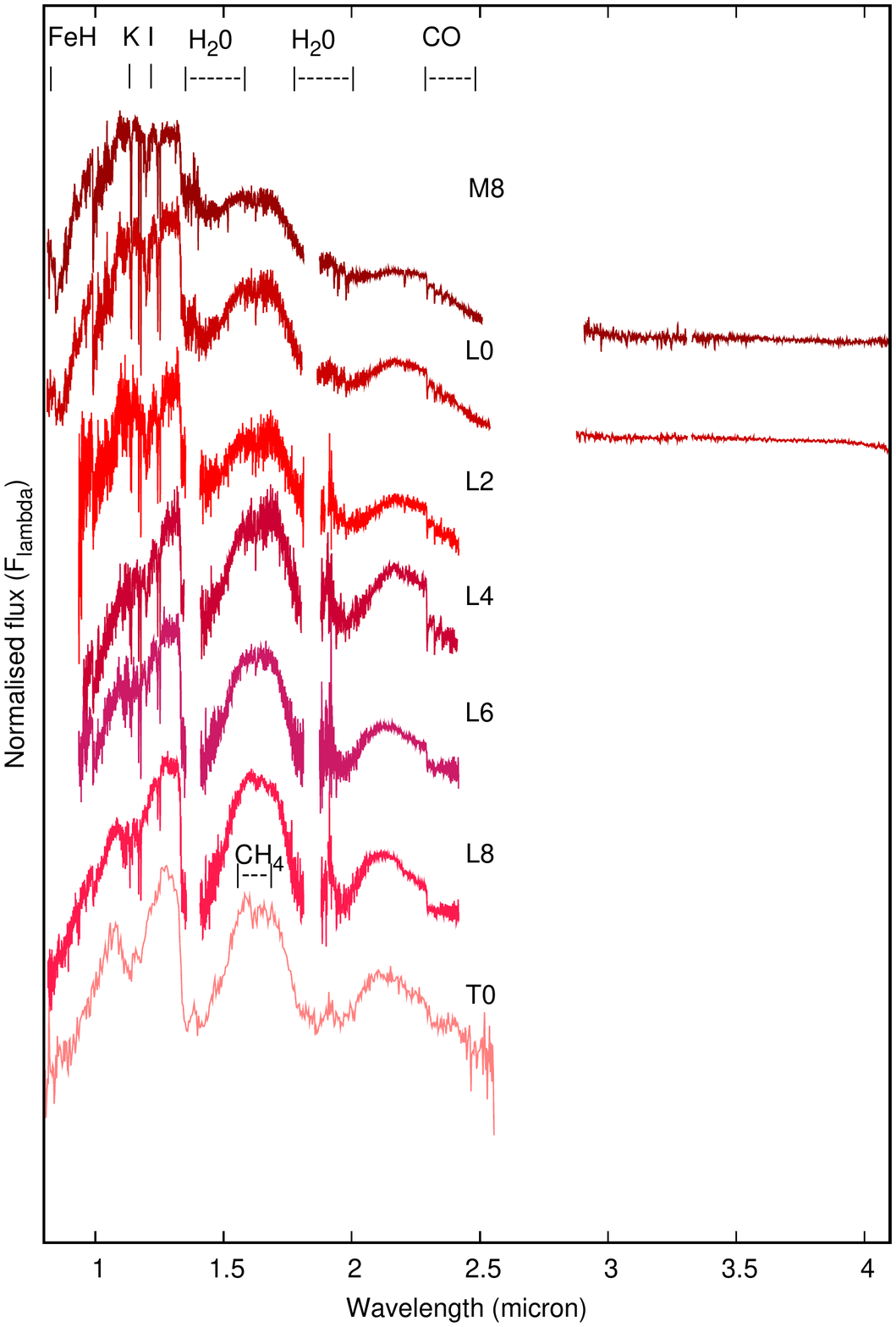}
    \includegraphics[width = 3.5in]{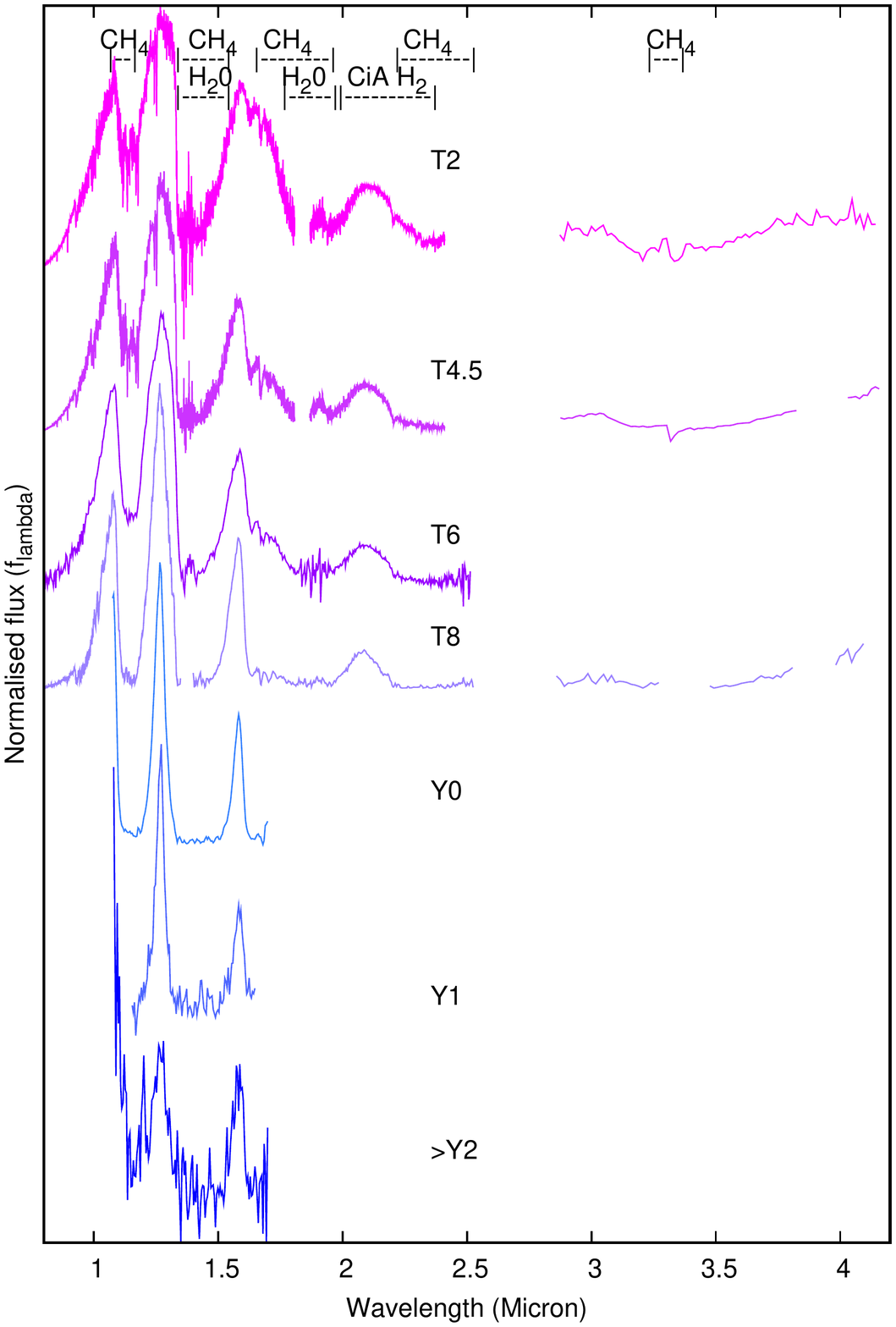}
    \caption{The near-infrared spectral sequence from early-M dwarfs to early-Y brown dwarfs. All spectra are normalized to 1 at their peak and then a flux offset is applied \citep{Helling2014}. }
    \label{fig:BDspectra}
\end{figure}

Complementary to our main extragalactic fields, the Milky Way provides a multitude of targets for compelling observations that uniquely exploit \iscea\  capabilities. In particular, the spectral range of \iscea\  is ideally tailored to the discovery and characterization of low mass stars and brown dwarfs, down to planetary masses, characterized by deep H$_2$O absorption bands (see Fig.~\ref{fig:BDspectra}). A spectroscopic survey of young stellar clusters representative of a variety of environments (source density, age, metallicity, etc...) can thus provide fundamental information on the bottom-end of the stellar Initial Mass Function, the environmental dependence of the minimum-mass for opacity limited fragmentation, and the origin and frequency of free-floating planetary objects, together with a multitude of spectra that can be used as proxies for exoplanet atmospheres.

\section{Summary}
\label{sec:summary}

We have presented a comprehensive discussion of the science program on the Infrared Satellite for Cosmic Evolution Astrophysics (\iscea), a small astrophysics mission with the Science Goal to discover how galaxies have formed and evolved in the cosmic web of dark matter at cosmic noon. \iscea's Science Objective is to determine the history of star formation and its quenching in galaxies as a function of local density and stellar mass when the Universe was 3-5 Gyrs old ($1.2<z<2.1$). \iscea\  tests the Science Hypothesis that environment is the dominant cause of quenching in typical galaxies during cosmic noon by observing  clusters, groups, and the extended cosmic web surrounding these structures at $1.2<z<2.1$.

\iscea\  meets its Science Objective by making a 10\% shot noise measurement of star formation rate (SFR) down to $6\,{\rm M_\odot\,yr^{-1}}$ using H$\alpha$ out to a radius of 10 Mpc in each of 50 protocluster (cluster + cosmic web) fields at $1.2<z<2.1$. \iscea\  measures the star formation quenching factor (the fraction of quiescent galaxies) in those fields, and galaxy velocity with a precision $<50\,$km/s to deduce the true, 3D distribution of galaxies in each protocluster field.

\iscea\ meets its science requirements with significant resiliency to pointing performance and thermal control (see Table \ref{tab:obs-time}), with $>30$\% margin on observing time.
If the \iscea\  instrument and mission requirements are fully met, these margins will be used to enhance \iscea\  science.
\iscea\  can observe 24 additional confirmed protoclusters at $1.7<z<2.1$ to the H$\alpha$ line flux limit of $3\times 10^{-17}\esc$ at 5$\sigma$ (for a total of 69 at $z>1.7$), using 568ks for additional target selection if needed. 
Alternatively, \iscea\  can observe the same 50 protoclusters to the fainter H$\alpha$ line flux limit of $\sim 2.4\times 10^{-17}\esc$ (corresponding to the SFR of $\sim 4.8\,{\rm M_\odot\,yr^{-1}}$). 

\iscea\  provides ground-breaking data: galaxy spectra with both H$\alpha$ and \oiii~ emission lines, or absorption features, to measure key galaxy properties over three orders of magnitude in local density over 50 fields, covering a wide range of environments, from crowded cluster cores to field galaxies, mapping the cosmic web surrounding each cluster directly tying the dense nodes to the surrounding filaments (see Fig.\ref{fig:cosmic-web}). \iscea\  transforms our understanding of cosmic evolution by providing for the first time, robust measurements of how density quenches star formation in galaxies at $z\sim 2$ (see Figs.~\ref{fig:quenching}, \ref{fig:ssFR-quenching-timescale} and \ref{fig:sfrDistribution}).

\iscea\  is a small satellite observatory with a 27cm$\times$27cm square (equivalent to 30cm diameter) aperture telescope, with a FoV of 0.32 deg$^2$.
Adopting the approach pioneered by the ground-based multi-object spectrographs IRMOS \citep{MacKenty2006}, RITMOS \citep{Travinsky2018}, and SAMOS \citep{Robberto2016}, \iscea\  uses a DMD as its programmable slit mask to obtain spectra of $\sim$1,000 galaxies simultaneously. It has an effective resolving power of $R=\lambda /\Delta\lambda=1000$ with $2.8^{\prime\prime}\times 2.8^{\prime\prime}$ slits over the near-infrared (NIR) wavelength range of 1.1 to 2$\mu$m, a regime not accessible from the ground without large gaps in coverage and strong contamination from airglow emission (see Fig.\ref{fig:air-trans}). \iscea\  has a pointing accuracy of $\leq 2^{\prime\prime}$ FWHM over 200 sec. This enables deep spectroscopy of hundreds of galaxies per protocluster field over a wide FoV. \iscea\  has a Sun synchronous Low Earth Orbit (LEO) and a prime mission of 2.5 years. Given its unique capability for multi-slit spectroscopy in crowded fields, \iscea\  can have a vigorous astrophysics Guest Observer program in its extended mission, thereby offering the community valuable opportunities to advance a broad range of science topics: from galaxy evolution to the Milky Way structure, from the spectroscopic follow-up of NIR counter-parts of gravity wave events, to the characterization of Outer Solar System objects. 
\iscea's space-qualification of DMDs opens a new window for spectroscopy from space, enabling revolutionary advances in astrophysics (see, e.g., \citealt{Wang2019}).



\end{document}